%% file: m31_lp_accepted.tex
\documentclass{aa}

\usepackage{graphicx}
\usepackage{txfonts}
\usepackage{lscape}
\usepackage{booktabs}
\usepackage{longtable}
\usepackage{caption}

\def\xmm{{\it XMM-Newton}}

\def\chandra{{\it Chandra}}
\def\einstein{{\it Einstein}}
\def\hubble{{\it Hubble}}
\def\spitzer{{\it Spitzer}}

\def\mth{{M\,31}}

\def\nh{\hbox{$N_{\rm H}$}}

\def\hii{H\,{\sc ii}}
\def\ha{H$\alpha$}
\def\sii{S\,{\sc ii}}

\defcitealias{2005A&A...434..483P}{PFH05\ }
\defcitealias{2011A&A...534A..55S}{SPH11\ }
\defcitealias{2016A&A...590A...1R}{3XMM\ }

\newcommand{\pcat}{\citetalias{2005A&A...434..483P}}
\newcommand{\scat}{\citetalias{2011A&A...534A..55S}}
\newcommand{\txmm}{\citetalias{2016A&A...590A...1R}}

\begin{document}

   \title{Deep XMM-Newton Observations of the Northern Disc of M31\\ I: Source Catalogue\thanks{Based on observations obtained with \xmm, an ESA science mission with instruments and contributions directly funded by ESA Member States and NASA.}}

   \titlerunning{Northern Disc of M31, I: Source Catalogue}

   \author{Manami~Sasaki\inst{1} 
           \and
           Frank~Haberl\inst{2}
           \and
           Martin~Henze\inst{3,4}
           \and
           Sara~Saeedi\inst{5} 
           \and
           Benjamin~F.~Williams\inst{6}
           \and
           Paul~P.~Plucinsky\inst{7}
           \and
           Despina~Hatzidimitriou\inst{8,9}
           \and 
           Antonios~Karampelas\inst{9,10}
           \and 
           Kirill~V.~Sokolovsky\inst{9,11,12}
           \and 
           Dieter~Breitschwerdt\inst{13}
           \and
           Miguel~A.~de~Avillez\inst{14}
           \and
           Miroslav~D.~Filipovi\'c\inst{15}
           \and
           Timothy~Galvin\inst{15,16}
           \and
           Patrick~J.~Kavanagh\inst{17} 
           \and
           Knox~S.~Long\inst{18}
          }

   \authorrunning{Sasaki et al.}

\institute{
Dr Karl Remeis Observatory and ECAP, Universit\"{a}t Erlangen-N\"{u}rnberg, Sternwartstr. 7, D-96049 Bamberg, Germany, \email{manami.sasaki@fau.de}
\and
Max-Planck-Institut f\"{u}r extraterrestrische Physik, Giessenbachstra\ss e, D-85748 Garching, Germany
\and
Department of Astronomy, San Diego State University, San Diego, CA 92182, USA
\and
Institute of Space Sciences (IEEC-CSIC), Campus UAB, Carrer de Can Magrans, s/n 08193 Barcelona, Spain
\and
Institut f\"ur Astronomie und Astrophysik, Universit\"at T\"ubingen, Sand 1, D-72076 T\"ubingen, Germany
\and
Astronomy Department, University of Washington, Box 351580, Seattle, WA 98195, USA
\and
Harvard-Smithsonian Center for Astrophysics, 60 Garden Street, Cambridge, MA 02138, USA
\and
Department of Astrophysics, Astronomy \& Mechanics, Faculty of Physics, University of Athens, 15783 Athens, Greece
\and
IAASARS, National Observatory of Athens, Vas. Pavlou \& I. Metaxa, 15236 Penteli, Greece
\and
American Community Schools of Athens, 129 Aghias Paraskevis Ave. \& Kazantzaki Street, Halandri, 15234 Athens, Greece
\and
Sternberg Astronomical Institute, Moscow State University, Universitetskii~pr.~13, 119992~Moscow, Russia
\and
Astro Space Center of Lebedev Physical Institute, Profsoyuznaya~St.~84/32, 117997~Moscow, Russia
\and
Zentrum f\"{u}r Astronomie und Astrophysik, Technische Universit\"{a}t Berlin, Hardenbergstrasse 36, D-10623 Berlin, Germany
\and 
Department of Mathematics, University of \'Evora, R. Rom\~ao Ramalho 59, 7000 \'Evora, Portugal
\and
Western Sydney University, Locked Bag 1791, Penrith, NSW 2751, Australia
\and
CSIRO Astronomy and Space Science, PO Box 1130, Bentley Delivery Centre, WA 6983, Australia
\and
School of Cosmic Physics, Dublin Institute for Advanced Studies, 31 Fitzwillam Place, Dublin 2, Ireland
\and 
Space Telescope Science Institute, 3700 San Martin Drive, Baltimore, MD 21218, USA 
}

   \date{Received ; accepted }


  \abstract
   {We carried out new observations of two fields in the star-forming northern ring of \mth\ with \xmm\ with two exposures of about 100 ks each. A previous \xmm\ survey of the entire \mth\ galaxy revealed extended diffuse X-ray emission in these regions.}
   {We study the population of X-ray sources in the northern disc of \mth\ by compiling a complete list of X-ray sources down to a sensitivity limit of $\sim7\times10^{34}$ erg s$^{-1}$ (0.5 -- 2.0 keV) and improve the identification of the X-ray sources. The major objective of the observing programme was the study of the hot phase of the interstellar medium (ISM) in \mth. The analysis of the diffuse emission and the study of the ISM is presented in a separate paper.}
   {We analysed the spectral properties of all detected sources using hardness ratios and spectra if the statistics were high enough. We also checked for variability. In order to classify the sources detected in the new deep \xmm\ observations, we cross-correlated the source list with the source catalogue of a new survey of the northern disc of \mth\ carried out with the \chandra\ X-ray Observatory and the \hubble\ Space Telescope (Panchromatic \hubble\ Andromeda Treasury, PHAT) as well as with other existing catalogues.}
   {We detected a total of 389 sources in the two fields of the northern disc 
of \mth\ observed with \xmm. We identified 43 foreground stars and candidates
and 50 background sources. Based on the comparison to the results of the
Chandra/PHAT survey, we classify 24 hard X-ray sources as new candidates for 
X-ray binaries (XRBs). In total, we identified 34 X-ray binaries and candidates
and 18 supernova remnants (SNRs) and candidates.
We studied the spectral properties of the four brightest 
SNRs and confirmed five new X-ray SNRs. 
Three of the four SNRs, for which a spectral analysis was performed,
show emission mainly below 2 keV, which is consistent with shocked ISM.
The spectra of two of them also require an additional component with a higher
temperature.
The SNR [SPH11] 1535 has a harder spectrum and might 
suggest that there is  a pulsar-wind nebula inside the SNR. 
For all SNRs in the observed fields,
we measured the X-ray flux or calculated upper limits. 
We also carried out short-term and long-term variability studies of the X-ray 
sources and found five new sources showing clear variability. 
In addition, we studied the spectral properties of the transient source 
SWIFT J004420.1+413702, which shows significant variation in flux over a period
of seven months (June 2015 to January 2016) and associated change in absorption.
Based on the likely optical counterpart detected in the Chandra/PHAT survey, 
the source is classified as a low-mass X-ray binary.}
   {}


   \maketitle
%

\section{Introduction}\label{intro}

The Andromeda galaxy (M\,31) is the largest galaxy in the Local
Group and the nearest spiral galaxy to the Milky Way, located at a
distance of 783~kpc \citep{2016MNRAS.458.3282C}.
With a similar mass and metallicity to those of our Galaxy, it
is also the closest example of the type of galaxy that dominates
redshift surveys.
This archetypal spiral galaxy thus provides a unique opportunity to
study and understand the nature and the evolution of a galaxy similar
to our own.

The star-formation history in M\,31 has been
studied in detail in observations with both the {\it Hubble} Space
Telescope and large ground-based telescopes
\citep[e.g., the Local group Galaxies Survey, LGS,][]{2002AAS...20110407M,2006AJ....131.2478M}.
Deep {\sl HST} photometry
has shown that the mean age of the disc of M\,31 is $6 - 8$~Gyr
\citep{2006ApJ...652..323B}.
\citet{2003AJ....126.1312W}
measured a mean star-formation rate of about 1~$M_{\sun}$~yr$^{-1}$
in the full disc of M\,31 and produced maps of star-formation rate in
different age ranges.
In addition, the northern disc was observed with the \hubble\ Space Telescope 
over a period of 4 years in the Panchromatic \hubble\ Andromeda Treasury (PHAT) 
survey \citep{2012ApJS..200...18D}.
Each field was observed with the Advanced Camera for Surveys (ACS) and the 
Wide Field Camera 3 (WFC3) 
yielding photometry of over 100 million stars in M\,31
from the near-infrared (NIR) to the ultraviolet (UV)
\citep{2014ApJS..215....9W}. Based on these data
the star-formation history has been measured on a few 100 pc
scale \citep{2017ApJ...846..145W}.

At longer wavelengths,
\citet{2006ApJ...638L..87G} showed by using data taken with the
\spitzer\ Space Telescope that 
M\,31 has spiral-arm structures merged with the prominent star-forming ring
at a radius of $\sim$10~kpc.
The newest images of the {\it Herschel} Space Observatory 
show a radial gradient in the gas-to-dust 
ratio and indicate that there are two distinct regions in M\,31 with different 
dust properties inside and outside $R \approx 3.1$~kpc
\citep{2012ApJ...756...40S,2012A&A...546A..34F}.
In addition to the well-known dust ring at a radius of $\sim$10~kpc
\citep{1984A&AS...55..179B,1993ApJ...418..730D}
with enhanced star formation,
\citet{2006Natur.443..832B} 
found a dust ring with
a radius of 1 -- 1.5~kpc, which had apparently been created in
an encounter with a companion galaxy, most likely M\,32.

First observations of individual sources in M\,31 in X-rays were
performed with the {\sl Einstein} Observatory
in the energy band
of 0.2 -- 4.5~keV and yielded the first catalogues of X-ray sources in the
field of M\,31 
\citep{1979ApJ...234L..45V,1991ApJ...382...82T}.
In the 1990s, \mth\ was observed with the R\"ontgen Satellite ROSAT
in the 0.1 -- 2.4~keV band. A total of 560 sources were confirmed
in the field of M\,31 
\citep{1997A&A...317..328S,2001A&A...373...63S}.
Newer observations with the next generation X-ray satellites
\chandra\ X-ray Observatory and X-ray Multi-Mirror Mission (\xmm) 
have provided a comprehensive
list of X-ray sources and allowed the study of individual objects
\citep[e.g.,][]{2001A&A...378..800O,2002ApJ...577..738K,
2002ApJ...578..114K,2004ApJ...615..720W,2005A&A...434..483P,
2005ApJ...634..314T,2006ApJ...643..356W,
2008A&A...480..599S,2008ApJ...689.1215B,2016MNRAS.461.3443V}
in selected regions
as well as in a survey of the entire M\,31 galaxy performed with \xmm\
between June 2006 and February 2008.
The catalogue of the first \xmm\ survey of \mth\ with 856 sources
was published by Pietsch et al.\ (\citeyear{2005A&A...434..483P},
PFH05 hereafter) and an updated \xmm\ catalogue with 1948 sources by 
Stiele et al.\ (\citeyear{2011A&A...534A..55S}, SPH11, hereafter).

We performed new deep observations of the northern star-forming disc of \mth\ 
with \xmm\ in order to
study the morphology and properties of the hot interstellar medium (ISM)
and to study the star-formation history of \mth\ by achieving a much lower
flux limit than before and thus obtaining a more complete 
sample of X-ray sources in M\,31. 
First results about transient sources have been reported 
by \citet{2015ATel.8227....1H,2015ATel.8228....1H,2016ATel.8825....1H,2016ATel.8826....1H,2016ATel.8827....1H}.
In this paper,
we present the point source catalogue of the new deep \xmm\ observations
of the northern disc of \mth. 
The study of the ISM using the new \xmm\ data is presented in a dedicated
paper by Kavanagh et al. (in prep.).

\section{Observations and data analysis}\label{data}

We have observed two fields in the northern disc of \mth\ in a large 
programme (LP) of \xmm\ (PI: M.\ Sasaki). 
The data were taken with the European Photon Imaging Cameras
\citep[EPICs,][]{2001A&A...365L..18S,2001A&A...365L..27T} in full-frame mode
using the thin filter for the pn camera (EPIC-pn) and the medium filter for 
the two MOS cameras (EPIC-MOS1/2),
in order to minimise the contamination by background but to maximise the
sensitivity for soft diffuse emission.
Each of the two fields was observed twice. 
The observation IDs and effective exposure times are given in Table 
\ref{obslist}. We had requested a total exposure of 200 ks for each field
split into two observations separated by 6 months to study the variability
of the sources.
While the first three observations (ObsIDs 0763120101, 0763120301, and 
0763120401) were carried out with exposure times of $\sim$90 to 100 ks,
observation 0763120201 was affected by high background flares.
The data were analysed using the \xmm\ Science Analysis System (SAS) 
ver.\,15.0.0.
Data processing and analysis were performed in the same manner as in
\citet{2013A&A...558A...3S}.

\begin{table*}
\caption{
\label{obslist}
List of \xmm\ LP observations of the northern disc of \mth.}
\centering
\begin{tabular}{ccccccrrrrr}
\hline\hline
OBS$^*$ & ObsID & Field & RA J2000.0& Dec J2000.0& Date & \multicolumn{3}{c}{EPIC net exposure [ks]}\\
 & & & (hms) & (dms) & & pn & MOS1 &MOS2 & $\Delta$RA$^\dagger$ & $\Delta$Dec$^\dagger$ \\
\hline
1 & 0763120101 & 1 & 00:44:21.99 & +41:31:16.6 & 2015-06-28 & 93.2 & 102.1 & 100.8 &  1.30\arcsec\  &  --1.66\arcsec\\
2 & 0763120201 & 1 & 00:44:21.99 & +41:31:16.6 & 2016-01-21 & 54.8 & 80.3  & 79.9  &  0.88\arcsec\ &   2.13\arcsec\\ 
3 & 0763120301 & 2 & 00:44:49.38 & +41:49:31.1 & 2015-08-11 & 99.0 & 105.2 & 104.9 &  2.03\arcsec\ &  --0.01\arcsec\\
4 & 0763120401 & 2 & 00:44:54.38 & +41:49:12.7 & 2016-01-01 & 61.0 & 91.3  & 91.8  & --0.87\arcsec\ &   1.32\arcsec\\ 
\hline\hline
\multicolumn{11}{l}{\footnotesize $^*$: These numbers will be used in the entire manuscript to refer to the observations.}\\
\multicolumn{11}{l}{\footnotesize $^\dagger$: Attitude shift applied to the data.}
\end{tabular}
\end{table*}

\subsection{Source detection}

In order to perform source detection,
images of each observation were created for each one of the three cameras
in the energy bands of $B_1$ = 0.2 -- 0.5 keV, 
$B_2$ = 0.5 -- 1.0 keV, $B_3$ = 1.0 -- 2.0 keV, $B_4$ = 2.0 -- 4.5 keV, 
and $B_5$ = 4.5 -- 12.0 keV. Source detection 
was performed simultaneously on all 15 images for each observation.
Details about the source detection routine and possible uncertainties
can be found in \citet{2013A&A...558A...3S}.
We thus obtained one final source list for
each observation. This source list includes information such as the 
source position, detection likelihood, count rate, and hardness ratios 
(see Sect.\,\ref{hardnessratios}) for each detection.

\subsection{Astrometry correction}

In order to improve the astrometry of the \xmm\ sources, we cross-correlated
the source list obtained from the source detection routine with
the USNO-B1.0 catalogue 
\citep{2003AJ....125..984M}
in the optical and the 2MASS catalogue 
\citep{2006AJ....131.1163S}
in the near-infrared (NIR) for each observation. The USNO-B1.0 catalogue also
contains proper motion information for the stars, which was taken into
account when comparing the X-ray position with the optical position.
For each cross-correlation result, the optical/NIR position at the 
epoch of the \xmm\ observation was plotted on optical $I, V, U$ images 
of the Local Group Survey 
\citep[LGS,][]{2002AAS...20110407M,2006AJ....131.2478M} 
together with the \xmm\ positions and verified by eye.
In addition, the list was 
cross-correlated with the \xmm\ catalogue of SPH11, in which the X-ray 
sources had been classified. Of particular interest for the astrometric 
correction are the foreground stars and the background active galactic nuclei 
(AGNs). Only firm correlations between the optical/NIR and the X-ray positions
of sources which had been identified as a foreground star or an AGN were
used to calculate the weighted mean of the linear offset between the
optical/NIR and X-ray positions for RA and Dec separately,
using the inverse of the positional error of the X-ray detections as weights.
The weighted means of the offsets in RA or Dec for the four observations were
in the range of 0.01\arcsec\ -- 2.13\arcsec.
These weighted means were then used to correct the attitude
information for each observation (see Table \ref{obslist}). 
The entire source detection routine and
the following cross-correlation of the source lists with the catalogues have
been carried out once more using the astrometrically aligned data. The
linear offset to the optical and NIR positions for X-ray sources identified
as foreground stars or AGNs was reduced to $<$0.5\arcsec.
After the correction, the offsets between coordinates had a mean dispersion of 
$\sim$1.2\arcsec, which is comparable to the mean positional error of the
detections (1.25\arcsec).

\subsection{Artefacts}

The raw source list produced by the standard detection algorithm includes some 
false detections, which have to be removed. False detections arise for a number 
of reasons:  
out-of-time (OOT) events, 
chip gaps, edges of the field of view (FOV), or wings of 
the point-spread function (PSF) of brighter sources. 
Therefore, not all the significant detections resulting from the source 
detection routine
are real astrophysical sources. In addition, part of a diffuse emission in a 
source-crowded region can also be listed as a source.

Out-of-time  events are caused when bright sources are observed with
EPIC-pn. In this case many photons are registered while the CCD is read out,
resulting in incorrect values for the position along the read-out (RAWY) and
thus causing the emission spread over the entire column. 
Detections of OOT events as sources can be identified and removed by comparing 
the EPIC-pn image to the EPIC-MOS1/2 images of the same observation.

In order to remove the false detections,
we inspected all images one by one and verified the detections in our source
list. Also the detections on different cameras and different observations were
compared with each other.
False detections caused by 
OOT events, chip gaps, FOV edges, and wings of the PSF were 
marked as such and removed from the final source list.
Sources detected inside a crowded region with supernova remnants (SNRs) or 
\hii\ regions were marked as 'diffuse'. Such a detection was kept in the
source list if it could be clearly identified with a source after the
comparison with other catalogues and data (see Sect.\,\ref{catalogs}).
The number of detected sources in the four observations originally
was 187, 125, 206, and 153, from which 9, 2, 10, and 9 sources were
removed, respectively (sorted by ObsID).

\subsection{Final source list}

\begin{figure*}
\centering
\includegraphics[trim=15 10 15 10,width=.49\textwidth]{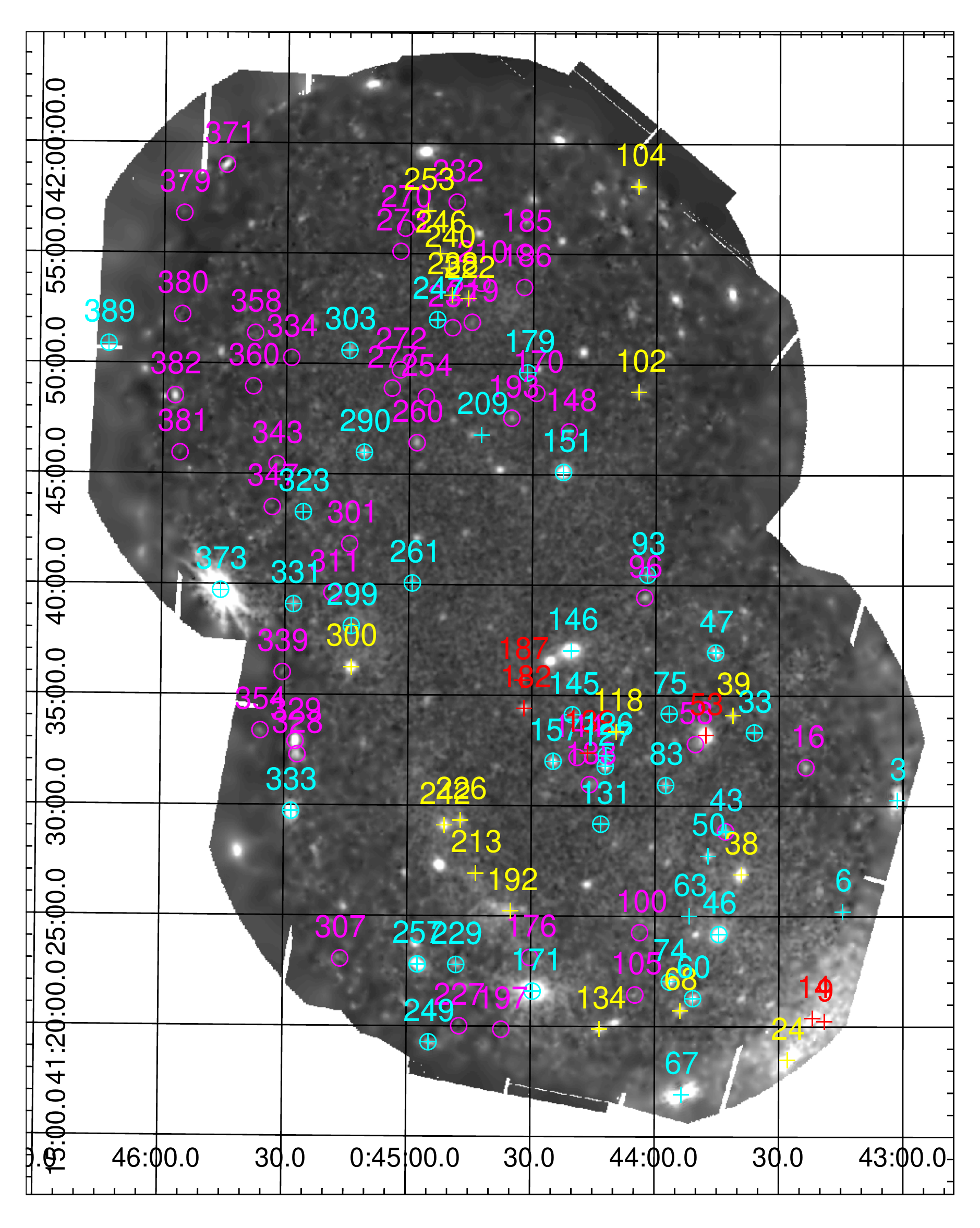}
\includegraphics[trim=15 10 15 10,width=.49\textwidth]{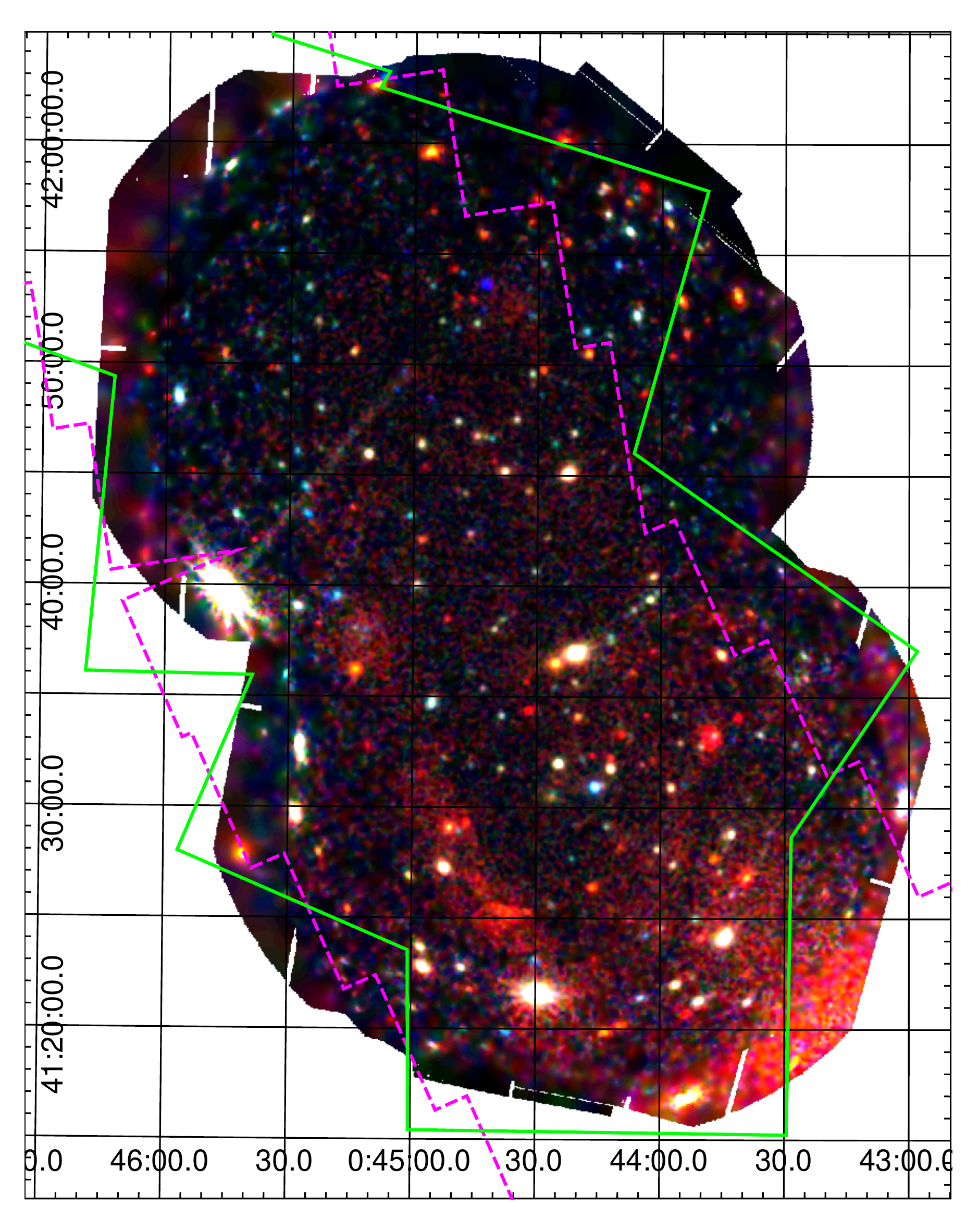}
\caption{
Left:
Exposure-corrected mosaic image 
in the energy band 0.2 -- 1.0 keV. Marked sources are: cyan 
(XRBs and candidates), yellow (SNRs and candidates), red (SSSs and candidates).
XRBs and candidates, for which a spectral analysis was performed, are 
additionally marked with a circle (see Sect.\,\ref{spectra}). Background sources
with fitted spectra are marked with magenta circles.
Right:
Exposure-corrected mosaic images 
of the \xmm\ LP observations of the northern disc of \mth\ 
in three colours
(red: 0.2 -- 1.0 keV, green: 1.0 -- 2.0 keV, blue = 2.0 -- 12.0 keV)
with the footprints of the new \chandra\ survey and 
the PHAT survey shown by 
solid green line and dashed magenta line, respectively.
All images are shown in log-scale.
}
\label{mosaic}
\end{figure*}

Once the artefacts were removed, 
the four detection lists of the four new \xmm\ LP
observations have been combined into one final source list
with 389 sources. Since each of the
two positions in the northern disc of \mth\ was observed twice, most of
the sources were detected in two observations. In overlapping regions, 
there are also sources which are detected three or four times.
In these cases of multiple detections, the detection with the highest detection
likelihood and thus the best statistics was selected for the final
source list. Typically, this
also corresponds to the detection with the best positional accuracy.
The final source list can be found in Table \ref{soutab}. This table also
includes the classification of the sources, either from Stiele et al.\ 
(\citeyear{2012A&A...544A.144S}, see Sect.\,\ref{catalogs}) or newly
defined in this work. Candidates of a source class
are given in $<$ $>$ brackets.
A mosaic image of the four observations is shown in Fig.\,\ref{mosaic}.
More SNRs and
XRBs are found closer to the galactic centre than in the other parts of the 
galactic disc (see Fig.\,\ref{mosaic}, left). 
In addition, SNRs are likely located in the ring of \mth.
Supersoft sources are found closer to the nuclear region.

\subsection{Cross-correlation with other catalogues}\label{catalogs}

The \xmm\ LP source list was cross-correlated with the
catalogue of the \xmm\ survey of \mth\ by SPH11 by searching for
sources with positions that are consistent in the two catalogues
within the 3$\sigma$ positional errors.
The source list was also compared with the revised list of SNRs and candidates 
compiled by Sasaki et al.\ (\citeyear{2012A&A...544A.144S}, SPH12 hereafter).
In addition, the source list was compared to other catalogues and publications
in order to classify the sources detected in the new \xmm\ LP survey of the
northern disc of \mth, again by identifying correlations within 3 $\sigma$ 
errors for the positions:
list of radio sources \citep{2012SerAJ.184...41G,2014SerAJ.189...15G},
catalogue of optical SNRs \citep{2014ApJ...786..130L},
studies of XRBs by, e.g., 
\citet{2008ApJ...689.1215B,2012ApJ...757...40B,2014ApJ...791...33B,2014MNRAS.443.2499W}, 
and the study of globular cluster (GlC) sources using 
the Nuclear Spectroscopic Telescope Array (NuSTAR)  by
\citet{2016MNRAS.458.3633M}.
We used the source classifications  by SPH11 as a basis and used the
newer studies to confirm or revise the classifications.

\subsection{Comparison to \chandra\ and PHAT data}

Williams et al. (\citeyear{2018arXiv180810487W}, accepted)
performed a new survey of the 
northern disc of \mth\  with \chandra\ and present a combined analysis
of the \chandra\ data with the PHAT results.
They have also performed a detailed study of the
stellar population at the position of the X-ray sources in the new
\chandra\ catalogue. 
We cross-correlated the new \xmm\ LP source list with the source list of 
the new \chandra\ survey of \mth.
Even though the areas of the sky, which have been observed with \chandra\ 
and with \xmm\ do not agree perfectly (see Fig.\,\ref{mosaic}, right), 
there are 197 sources which have been detected both with
\xmm\ and with \chandra.
For these sources  we thus
obtained more accurate X-ray positions from the \chandra\ data, while the 
\xmm\ data allow us to study the spectra and time variability of the sources 
(see Sect.\,\ref{spectra}, Sect.\,\ref{variability}).
Based on the comparison to the results of \chandra\ and \hubble\ surveys 
we identified X-ray sources
that coincide with foreground stars in the Milky Way, 
stars, globular clusters or 
young clusters in \mth, or background galaxies or galaxy clusters.

Nine out of the 43 foreground stars and candidates are also detected
with \chandra\ and were confirmed as foreground stars based on the PHAT data.
In addition, we identified 24 hard X-ray sources coinciding with stars or 
stellar
clusters in \mth, which makes them new likely candidates for XRBs.

\section{Analysis of spectral properties}

\subsection{Hardness ratios}
\label{hardnessratios}

\begin{figure*}
\centering
\includegraphics[width=0.49\textwidth,trim=40 0 60 20,clip=]{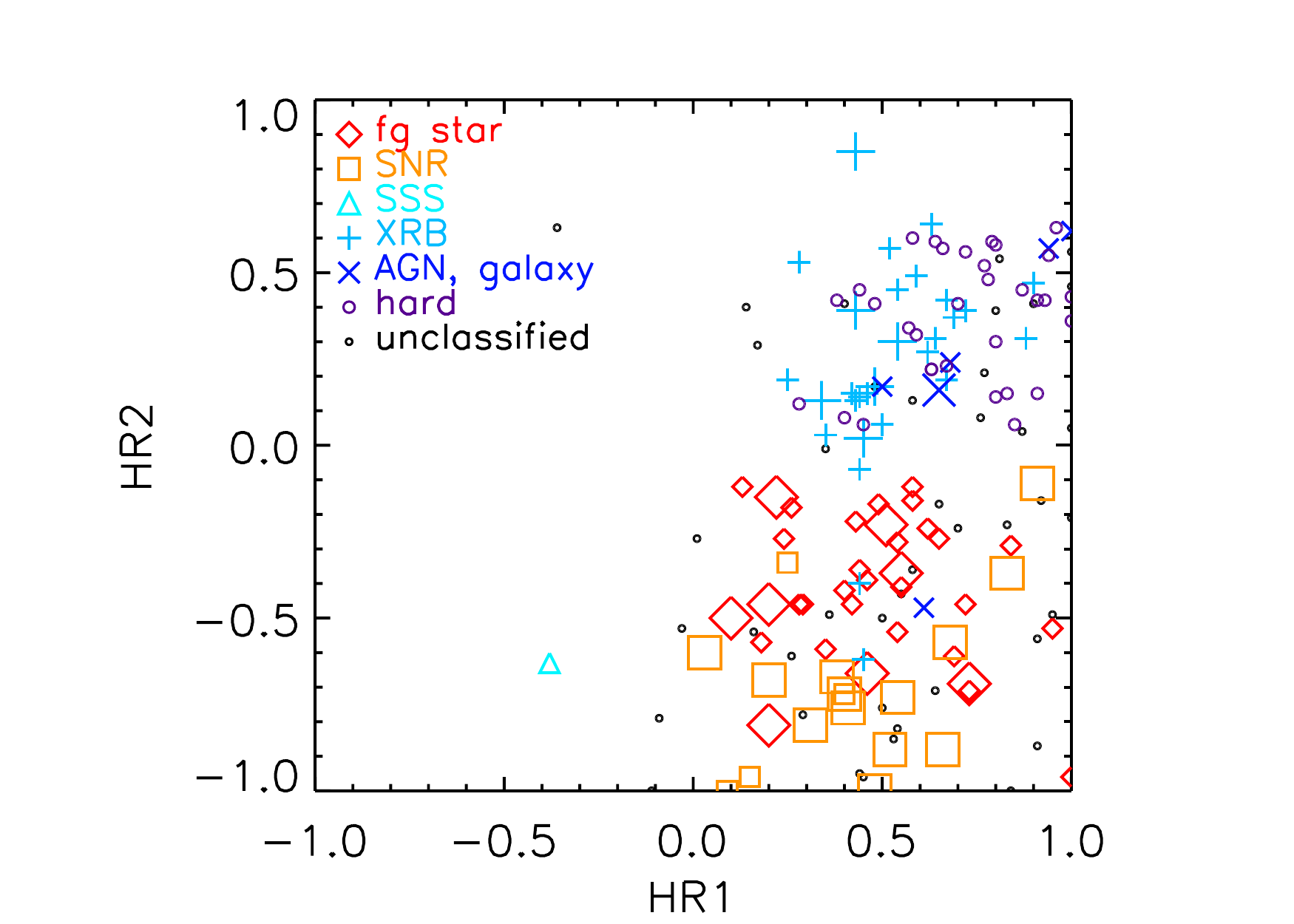}
\includegraphics[width=0.49\textwidth,trim=40 0 60 20,clip=]{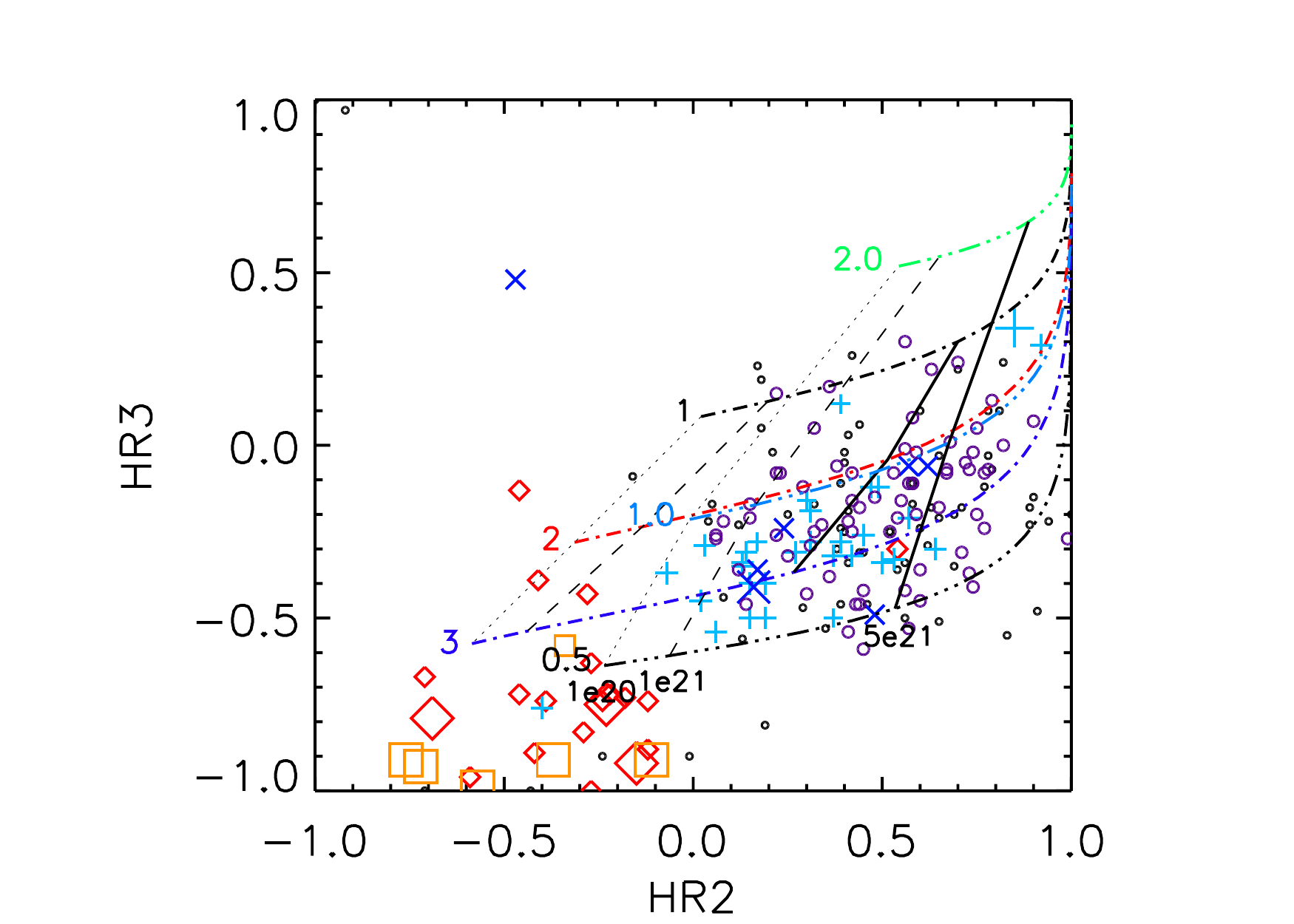}\\
\includegraphics[width=0.49\textwidth,trim=40 0 60 20,clip=]{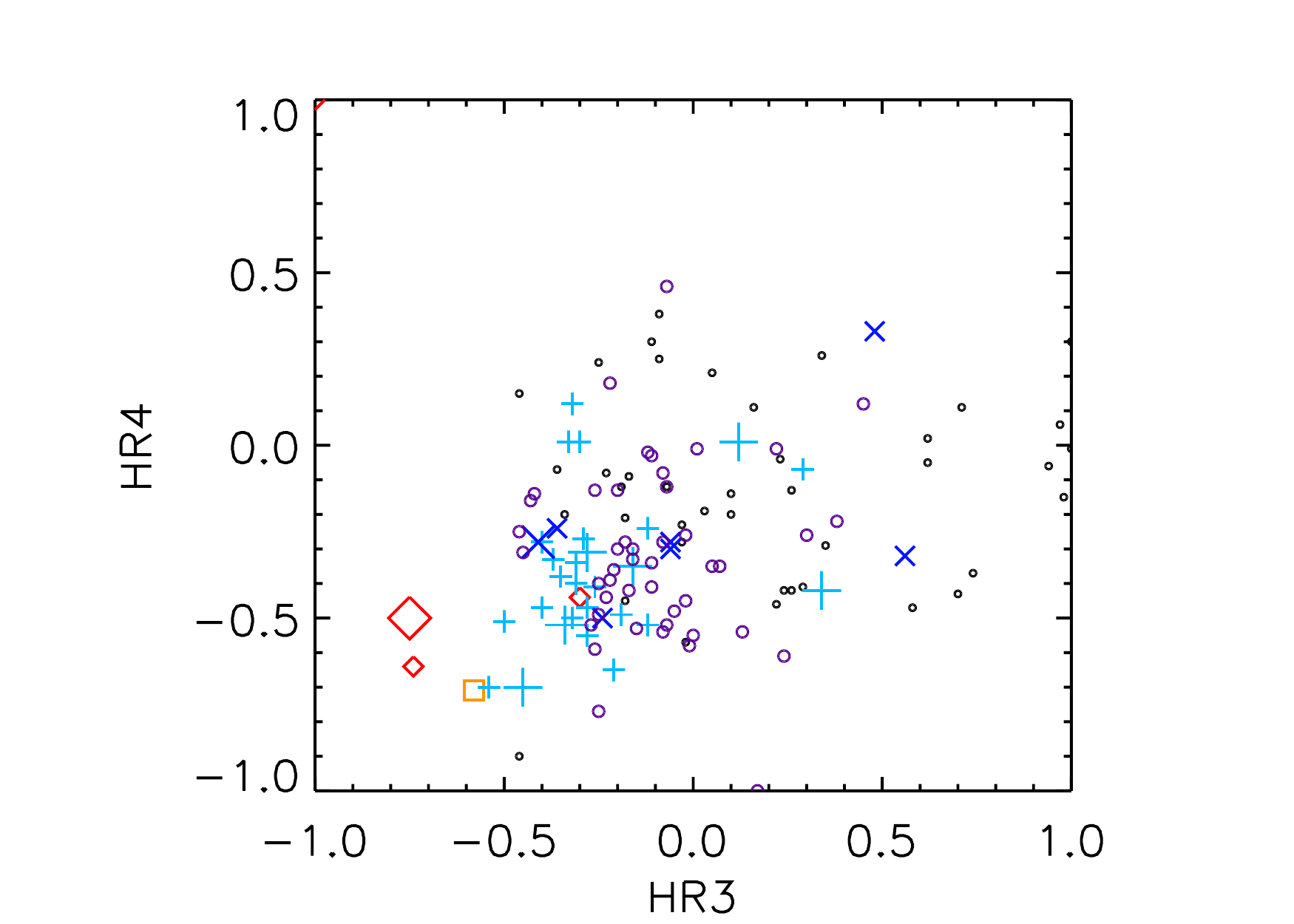}
\caption{
Hardness ratio diagrams for sources detected in the four new \xmm\ 
observations of the northern disc of \mth.
Only sources with hardness ratio errors $EHR_i < 0.3$ are shown.
Different symbols are used for different source classes:
red diamonds for foreground stars, orange squares for SNRs, 
cyan triangles for SSSs, light-blue crosses for X-ray binaries, and 
dark-blue Xs for background sources. 
Hard sources, which either can be AGNs or X-ray binaries, are marked with 
purple circles. Small symbols indicate candidates for each class,
confirmed classifications are marked with large symbols.
Black dots are used for unclassified sources.
In $HR_2$-$HR_3$ diagram lines indicate the predicted position
for sources with a power-law spectrum (dash-dotted lines) for 
$\Gamma$ = 1 (black), 2 (red), 3 (blue) and a disc black-body spectrum 
(dashed-dot-dotted lines) for $kT$ = 0.5 keV (black), 1.0 keV (light blue), 
2.0 keV (green) for different absorbing 
foreground \nh\ (thin dotted for $10^{20}$ cm$^{-2}$, dashed 
for $10^{21}$ cm$^{-2}$, and thick solid for $5 \times 10^{21}$ cm$^{-2}$).
}
\label{hrplots}
\end{figure*}

Since we detect sources in different energy bands, we can obtain information
about their spectral properties by calculating their hardness ratios, which
are defined as:
\begin{equation}
HR_\mathrm{i}=\frac{B_\mathrm{i+1}-B_\mathrm{i}}{B_\mathrm{i+1}+B_\mathrm{i}}\,;\qquad
EHR_\mathrm{i}=2\frac{\sqrt{(B_\mathrm{i+1}EB_\mathrm{i})^2+(B_\mathrm{i}EB_\mathrm{i+1})^2}} {(B_\mathrm{i+1}+B_\mathrm{i})^2}\,,
\end{equation}
for i=1,...4; $B_\mathrm{i}$ is the count rate and 
$EB_\mathrm{i}$ is the corresponding error in each energy band. 
The count rates and errors are determined by the source detection routine 
based on the maximum-likelihood algorithm, which is applied to
particle-background filtered, vignetting and exposure corrected images
in each band for each EPIC for each observation
\citep[see, e.g., SPH11 or][for details]{2013A&A...558A...3S}.
All detections fulfill the requirement that the detection maximum likelihood
$ML_\mathrm{det} = - ln(P) > 6$, with $P$ being the detection probability 
for Poissonian background fluctuations.
We assume that the count statistics is high enough in each band to
calculate the hardness ratios and to propagate the errors.
However, this is of course not the case
for all sources: e.g., for a super-soft source (SSS), the hardness ratios
$HR_3$ and $HR_4$, and most likely also $HR_2$ will not yield meaningful
values. Therefore, for further discussion, we only consider hardness ratios
$HR_\mathrm{i}$ with errors $<$0.3.
The hardness ratio diagrams are shown in Fig.\,\ref{hrplots}.

We have also calculated the hardness ratios for different types of spectral
models in order to compare the distribution of sources in the hardness ratio 
diagram to expected spectral properties of the sources. 
In particular, we would like to understand if we see a difference between 
the XRBs in \mth\ and the background sources.
For AGNs we assume a power-law spectrum with $\Gamma$ = 1 -- 3
\citep{2011A&A...530A..42C}.
The spectrum of an XRB in the hard state is also dominated by a similar
power-law model. In the soft state, however, a disc black-body model will 
better describe the source 
\citep[see, e.g.,][and references therein]{2005ApJ...630..465B,2006ARA&A..44...49R}.

There is a clear separation in $HR_2$ between soft (foreground stars, 
super-soft sources [SSSs], and SNRs) and hard (XRBs and AGNs) sources as
can be seen in the upper panels of Fig.\,\ref{hrplots}. 
In addition, there seems to be a separating trend between the sources in the 
background and in the \mth, which are all found in the upper-right quadrant 
of the 
$HR_1$-$HR_2$ diagram (Fig.\,\ref{hrplots}, top). 

\begin{figure}
\centering
\includegraphics[width=0.5\textwidth,trim=45 0 55 20,clip=]{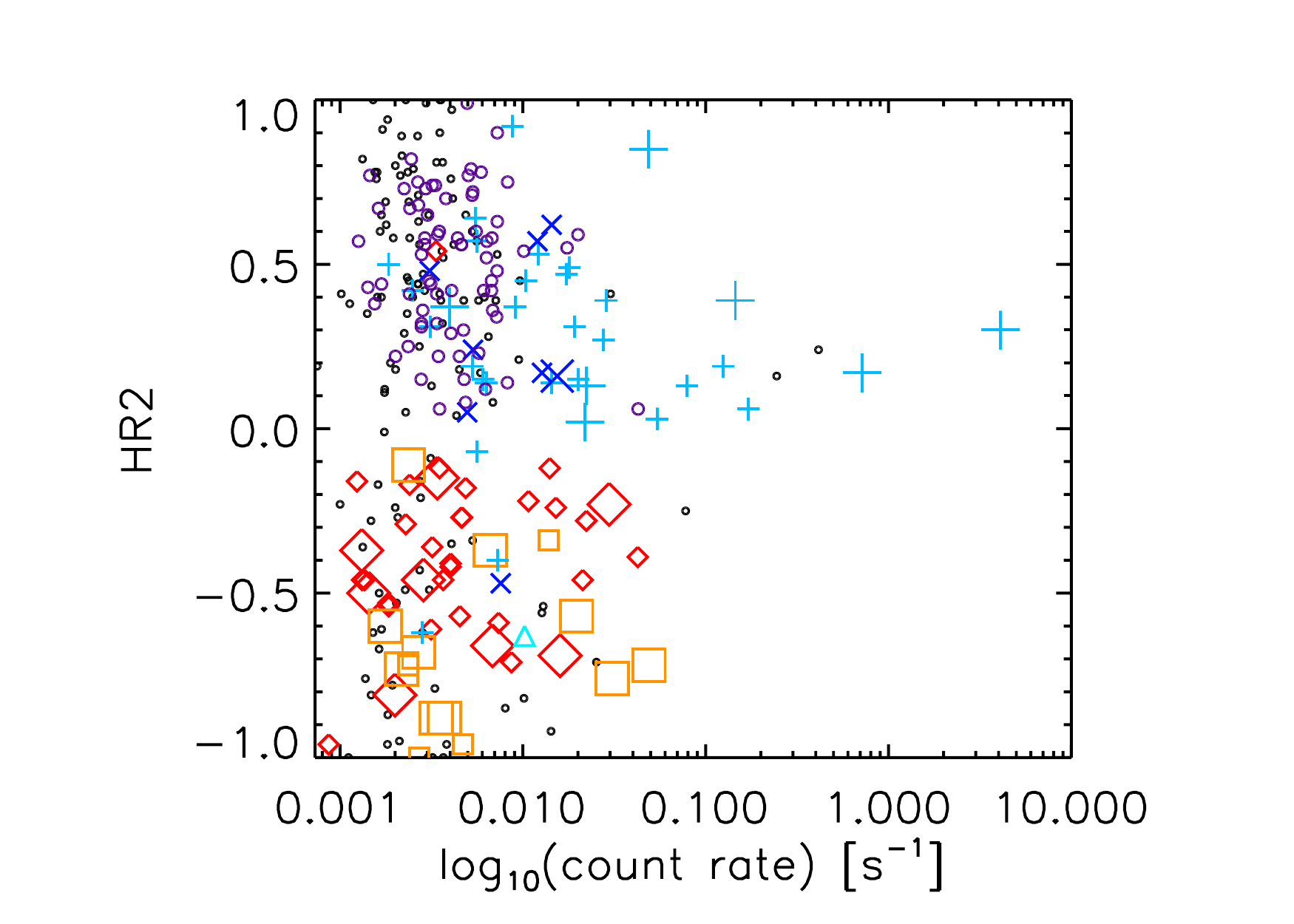}
\caption{
Hardness ratio $HR_2$ over logarithm of count rate (s$^{-1}$). 
Same symbols as in Fig.\,\ref{hrplots} are used.
}
\label{hrrate}
\end{figure}

In Fig.\,\ref{hrrate} we show the count rate of each source vs. 
hardness ratio $HR_2$. It is obvious that the brightest sources are XRBs.
In addition, the diagram reveals that the brightest XRBs 
are neither soft nor hard ($HR_2 = 0.0 - 0.5$).

\subsection{Spectra}
\label{spectra}

\begin{figure}[h]
\includegraphics[width=0.45\textwidth,trim=60 0 60 0,clip]{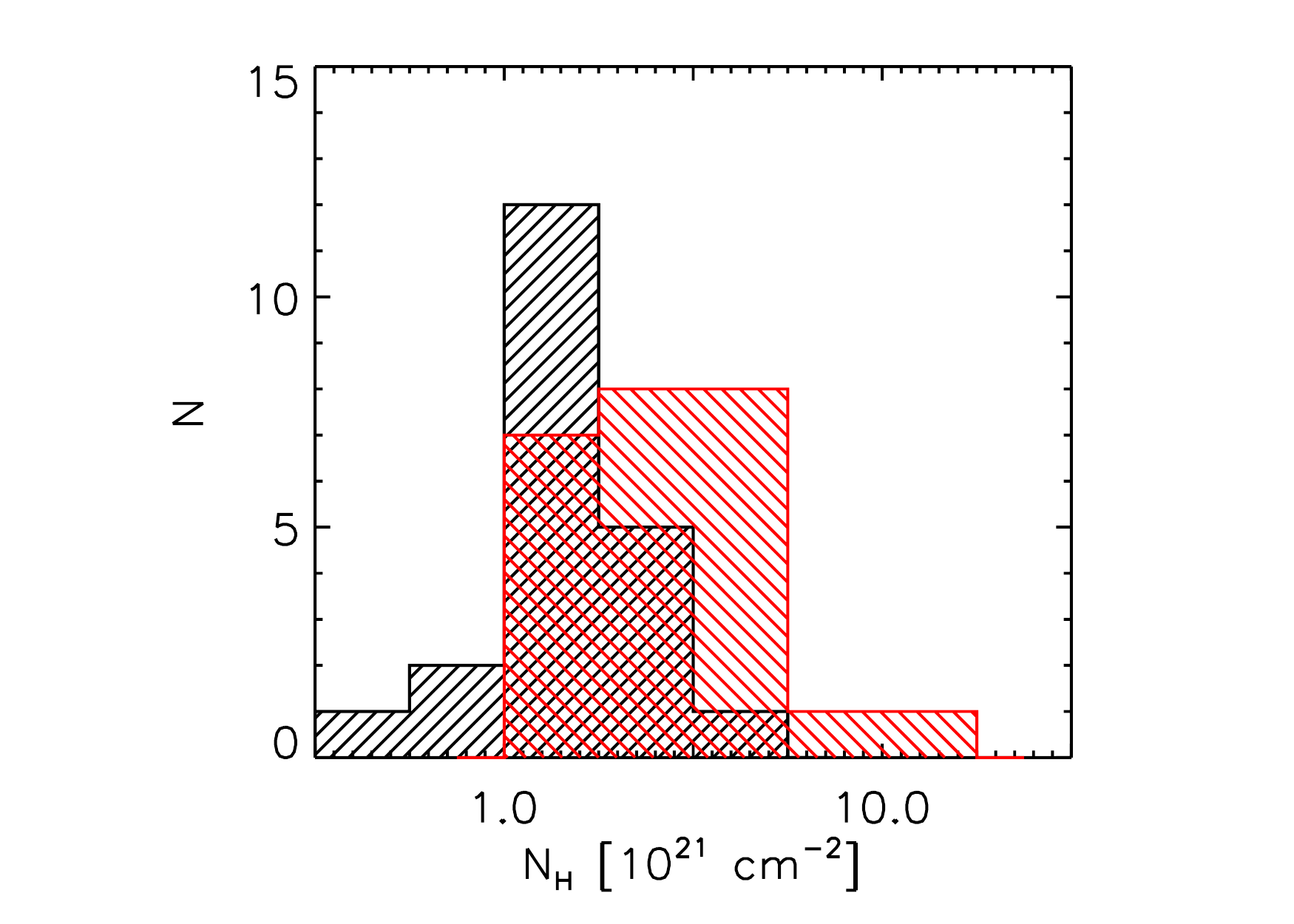}\\
\includegraphics[width=0.45\textwidth,trim=60 0 60 0,clip]{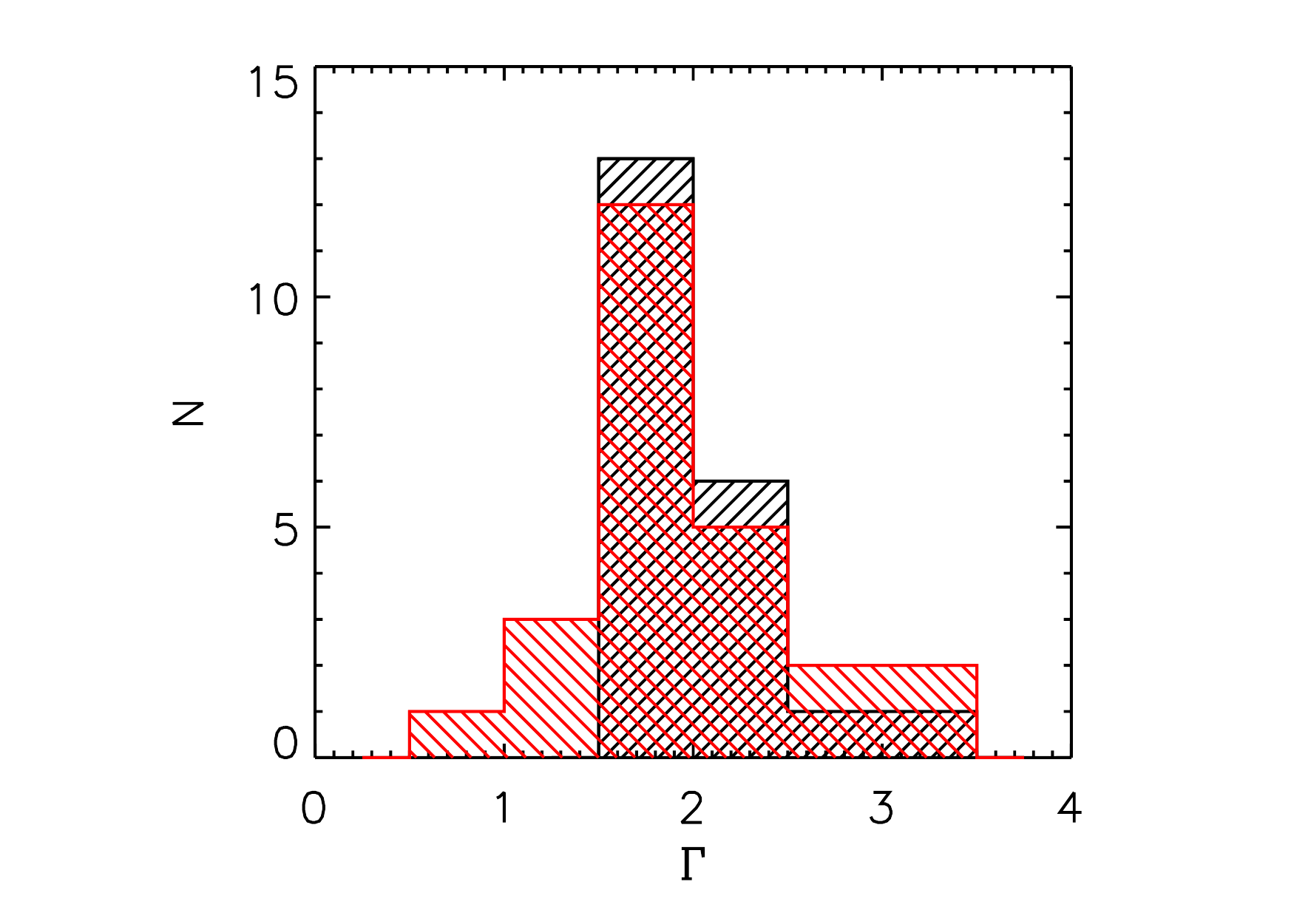}
\caption{Histograms showing the distribution of spectral fit
parameters \nh\ (Galactic and additional \nh\ combined) 
and $\Gamma$ for \xmm\ sources suggested to be 
located in \mth\ (black) and to be background sources (red) based 
on PHAT data.}
\label{histo}
\end{figure}

We extracted spectra for all sources with number of counts $\ga100$.
The spectra of these sources were fitted with  a 
power-law model.
The aim of the analysis was to determine if the hard X-ray sources
are members of \mth, background galaxies, or AGNs. 
If the column density is comparable to the Galactic foreground column density 
$N_\mathrm{H, MW}$ in the direction of \mth\
\citep{1992ApJS...79...77S},
the source is most likely located in \mth. 
If the column density is higher than the Galactic  $N_\mathrm{H, MW}$
the source either has an additional significantly high intrinsic \nh\ or
is located far behind \mth. 
The optical counterpart of the source will allow us to verify whether the 
source is a member of \mth\ or is a background source. Therefore, the column 
density of a yet unclassified X-ray source will help us to determine its
nature.

The results of the spectral analysis are given in Tables~\ref{spec-result-m31}
and \ref{spec-result-bkg}. Table~\ref{spec-result-m31} lists sources, 
for which the optical PHAT counterparts indicate that they are members of \mth, 
while sources in Table~\ref{spec-result-bkg} have PHAT counterparts that are
likely background sources.
Both tables list the source ID, coordinates, the  Galactic 
foreground \nh, additional column density, photon index of the power-law
spectrum, reduced $\chi^2$ with degrees of freedom, and the name of the 
observation and the EPIC instruments, 
which have been used for the spectral analysis.
An additional $N_\mathrm{H}$ was applied for sources, for which the column 
density was higher than the Galactic column density 
when we first fitted the spectrum with one
absorbing \nh. In such a case, we froze the first $N_\mathrm{H}$ to the 
Galactic $N_\mathrm{H}$ and included a second $N_\mathrm{H}$ which
accounts for absorption in \mth\ or for intrinsic absorption of the source.

If the source was detected in more than one EPIC, a simultaneous fit was 
performed for the source using the spectral data of all available 
EPIC detectors.
For brighter sources that have been detected in two observations, the 
result of the fit of the data with higher statistics is given in 
Tables~\ref{spec-result-m31}
and \ref{spec-result-bkg}. If a source was too faint for a spectral
analysis in one observation, but was detected in two 
observations and the fluxes of the source were similar in the
two observations, 
we used the SAS task \texttt{epicspeccombine} to combine the source 
spectra, background spectra, and ancillary response and response matrix 
files of each EPIC of two observations.
We thus could improve the statistics of the spectra of these sources.

\begin{figure*}[t!]
\includegraphics[width=.49\textwidth,trim=50 365 115 70,clip]{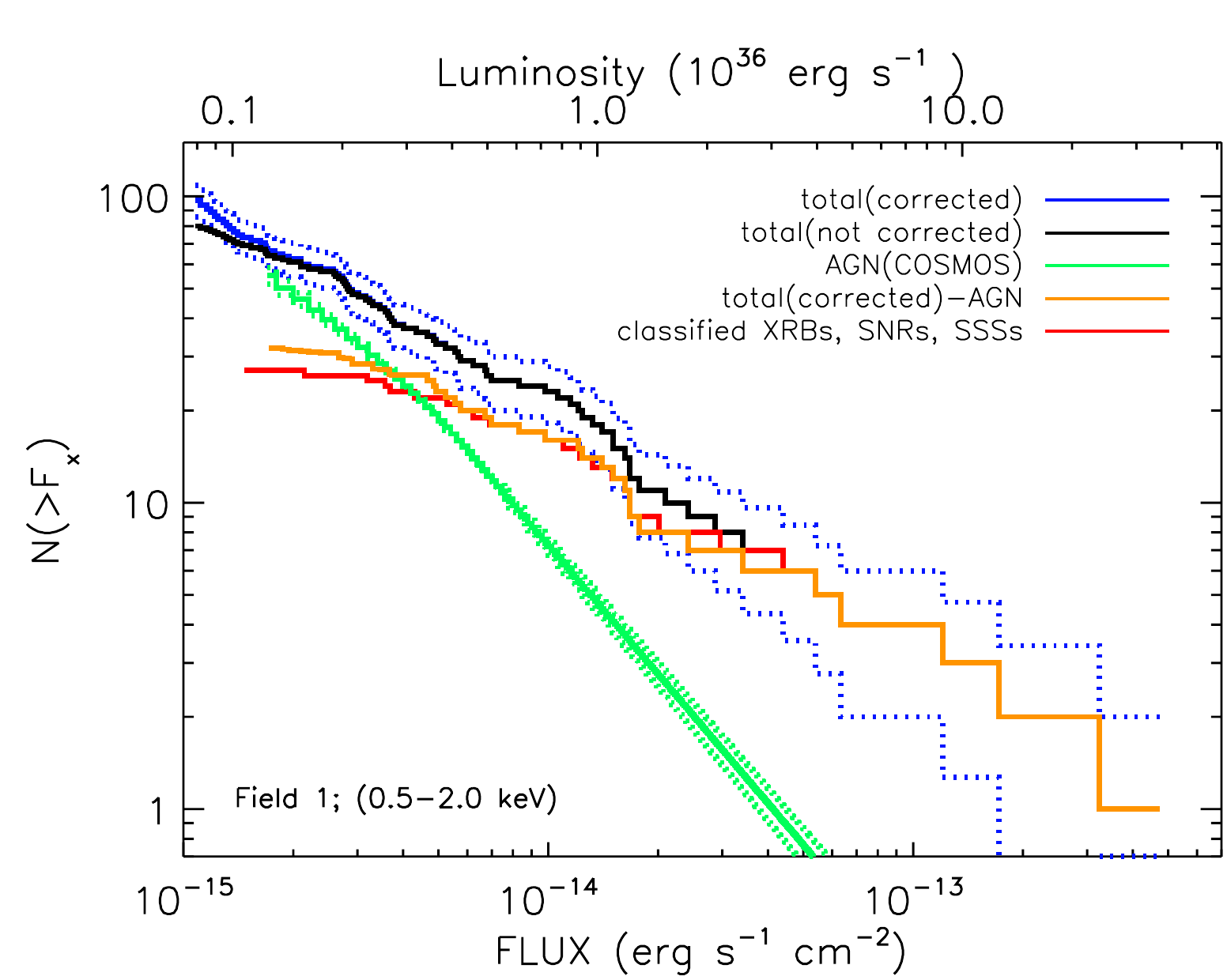}
\includegraphics[width=.49\textwidth,trim=50 365 115 70,clip]{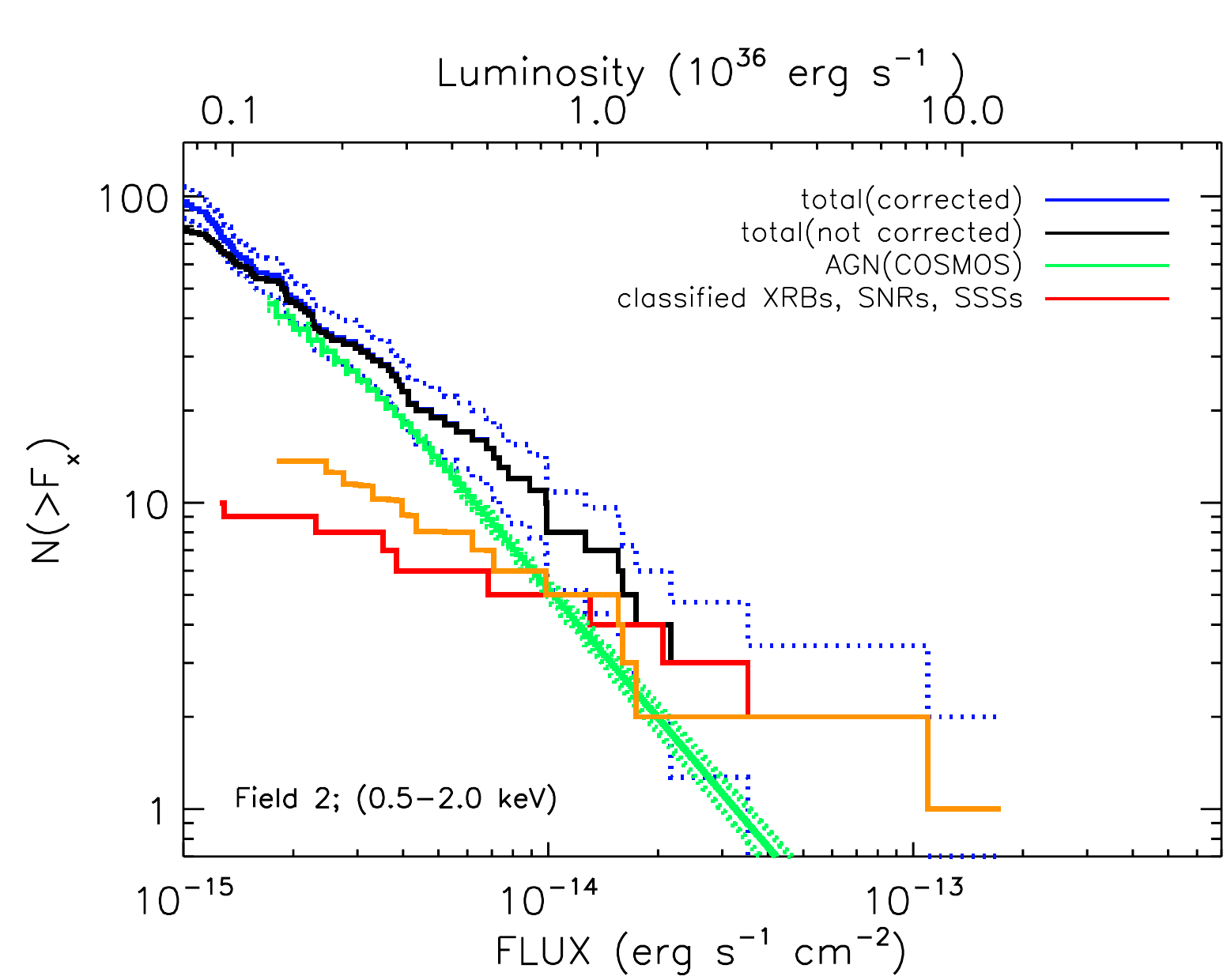}\\
\includegraphics[width=.49\textwidth,trim=50 365 115 70,clip]{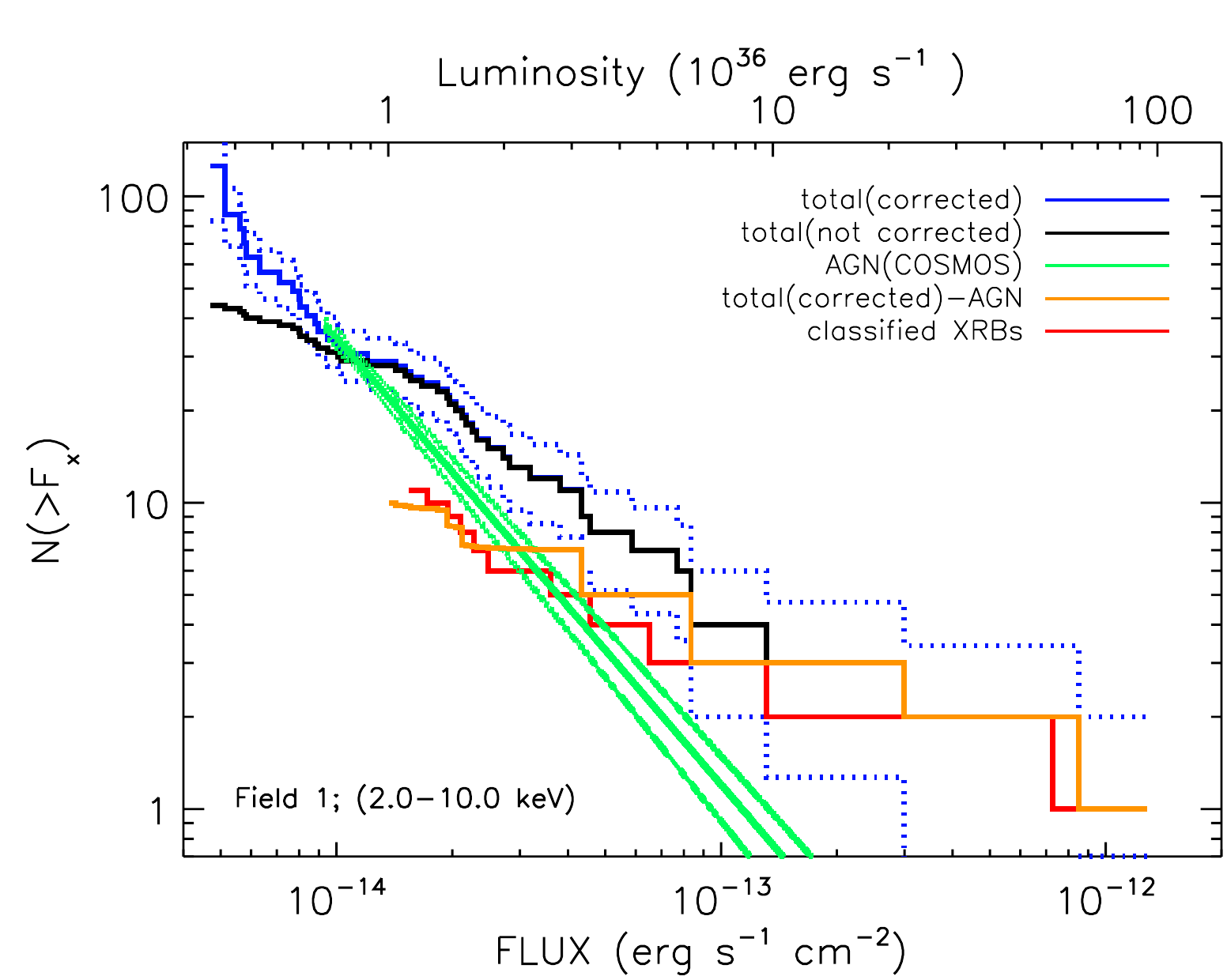}
\includegraphics[width=.49\textwidth,trim=50 365 115 70,clip]{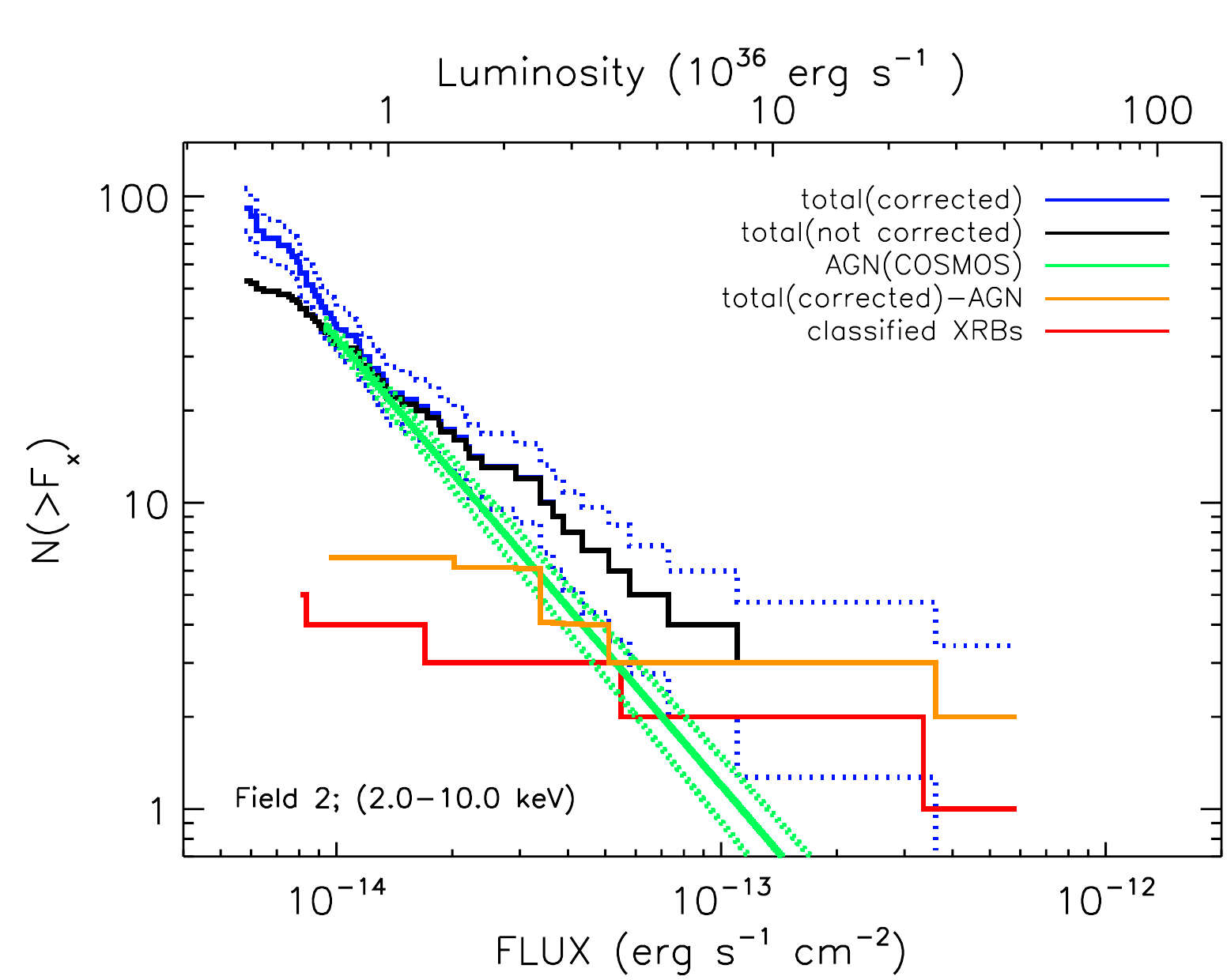}
\caption{
Cumulative X-ray luminosity functions for Fields 1 (left panel) and 2 
(right panel) in the bands of 
0.5 -- 2.0 keV (upper diagrams) and 2.0 -- 10.0 keV
(lower diagrams).
The cumulative XLF of all observed sources without foreground stars 
is shown as black solid line (error ranges with dotted lines),
and those that have been incompleteness-corrected as blue lines.
The green lines show the estimated contribution of background AGNs in the
field based on the study of \citet{2009A&A...497..635C}. The XLF after
subtracting the background AGN contribution is shown in orange. 
The XLF of classified \mth\ sources is shown in red.
\label{xlfplot}}
\end{figure*}

In Fig.\,\ref{histo} we show the distribution of the obtained
\nh\ and photon index $\Gamma$ for the two samples. The \nh\ is
the sum of the Galactic foreground \nh\ and the additional \nh\
obtained by the spectral fit. The
distribution of $\Gamma$ seems to be similar for \mth\ and
background sources. The distribution of \nh, by contrast, 
has a peak around $\nh = 1.5 \times 10^{21}$ cm$^{-2}$ 
corresponding to the average  \nh\  in the direction of 
\mth\ (Milky Way and \mth\ combined), and an additional distribution
at higher  $\nh$ for the sources suggested to be
background galaxies based on PHAT data. We thus confirm the
classification of these sources to be in background of
\mth\ through their X-ray spectra.
However, this result also shows that there are background sources
with $\nh$ values that are not significantly higher than the 
Galactic foreground. Therefore, the absorbing column density $\nh$ alone
is not sufficient to separate \mth\ sources from background sources.

\section{X-ray luminosity function}\label{xlf}

\begin{figure}[t!]
\includegraphics[trim=60 370 90 110,width=0.5\textwidth]{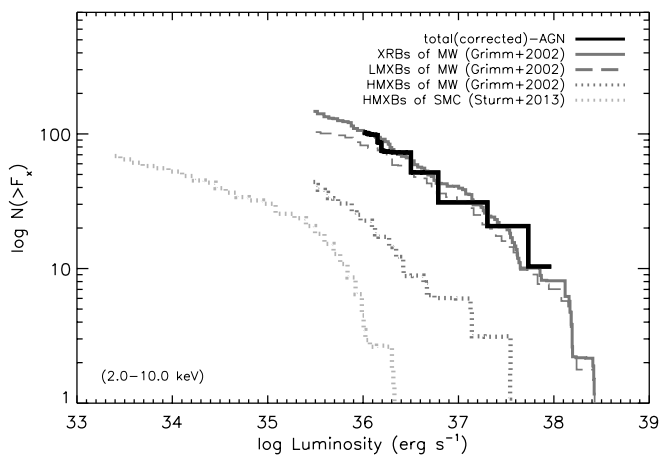}
\caption{XLFs of sources in \mth\ (thick black line), 
X-ray binaries in the Milky Way (solid grey line), 
low-mass X-ray binaries (LMXBs) in the Milky Way (dashed grey line),
high-mass X-ray binaries (HMXBs) in the Milky Way (dotted grey line),
and HMXBs in the Small Magellanic Cloud (SMC, dotted light-grey line).
The data for the Milky Way are taken from \citet{2002A&A...391..923G}
and those for the SMC from \citet{2013A&A...558A...3S}.
The XLF of \mth\ is rescaled to the size of the entire
luminous disk (see Sect.\,\ref{xlf}).
\label{XLFcomp}
}
\end{figure}

Using the sources from the final source list, we calculated the 
X-ray luminosity functions (XLFs) for the energy ranges of  0.5 -- 2.0 keV and 
2.0 -- 10.0 keV.
To estimate the contribution of background sources, we modelled
the background based on the AGN XLF of the COSMOS survey 
\citep[][shown in green in Fig.\,\ref{xlfplot}]{2009A&A...497..635C}.
The flux limits of the COSMOS survey were
$1.7 \times 10^{-15}$ erg cm$^{-2}$ s$^{-1}$ and
$9.3 \times 10^{-15}$ erg cm$^{-2}$ s$^{-1}$ 
for the 0.5 -- 2 keV and  2 -- 10 keV bands, respectively. 
We calculated the numbers of background sources 
by taking into account that they are more absorbed
\citep[\nh = $2.3 \times 10^{21}$ cm$^{-2}$ for Field 1 and 
$3.0 \times 10^{21}$ cm$^{-2}$ for Field 2,][]
{2005A&A...440..775K} 
than sources in the foreground
of \mth\ without intrinsic absorption 
(\nh = $7 \times 10^{20}$ cm$^{-2}$) and extrapolated to lower fluxes.
The flux was calculated assuming 
a power-law spectrum with a photon index of $\Gamma = 2.0$ for
the softer band (0.2 -- 5.0 keV) and $\Gamma = 1.7$ for
2.0 -- 10.0 keV \citep{2009A&A...497..635C}.

In Fig.\,\ref{xlfplot} we show the cumulative X-ray luminosity
functions of Fields 1 and 2 in the bands of 0.5 -- 2.0 keV and 2.0 -- 10.0 keV,
calculated from the sources detected in observations OBS1 and OBS3.
Sources identified as foreground stars are not included.
The black lines show the cumulative XLF calculated from the detected sources,
while the blue lines show the same XLFs after correcting for incompleteness.
The completeness correction is based on sky-coverage function of each
observation. The details can be found, e.g., in 
\citet{2013A&A...553A...7D,2016A&A...586A..64S}.
The green line is the calculated XLF of background AGNs 
\citep{2009A&A...497..635C}, 
corrected for the absorption through \mth\ and scaled for the
observed field size. If we subtract this contribution from the corrected
XLF, we should obtain the XLF of sources in \mth\ (orange lines). This
is compared to the XLFs of sources, which have been classified as SSSs,
SNRs, or XRBs in the softer band and as XRBs in the harder band (red lines).

We have also calculated the cumulative XLFs for the merged data of
OBS1 and OBS2 for Field 1 and OBS3 and OBS4 for Field 2. However, 
since OBS2 and OBS4 are shorter than OBS1 and OBS3, merging the results
only introduces more artefacts. While merging the data will
help to detect more fainter sources, it will not be possible to clearly 
separate artefacts from real sources. In addition, both fields have been
observed with large offsets of a few arcminutes between the two pointings, 
which increases the systematic errors due to inconsistent coverage of the 
fields. Therefore, we decided to use only the longer exposures.

If we compare the XLFs of Fields 1 and 2 it is obvious that more sources are 
detected in Field 1, which is closer to the nuclear region of 
\mth. The majority of these sources is probably low-mass X-ray binaries. 
We also compare the XLF to those of the XRBs in 
the Milky Way and the Small Magellanic
Cloud (SMC) in Fig.\,\ref{XLFcomp}.
While the data of the Milky Way and the SMC cover the entire galaxy, our \mth\ 
data only includes the northern disk covering an area of  0.13 deg$^2$. 
For the comparison with the Milky Way and the SMC, we scaled the XLF of
\mth\ to the apparent size of the disk of 1.34 deg$^2$, based on the
parameters of the luminous disk obtained by
\citet{2011ApJ...739...20C}.
The XLF of \mth\ is consistent with that of the entire XRB population
in the Milky Way.

\section{Variability studies}\label{variability}

\subsection{Flux change between two epochs}

\begin{figure}[t!]
\includegraphics[width=.5\textwidth]{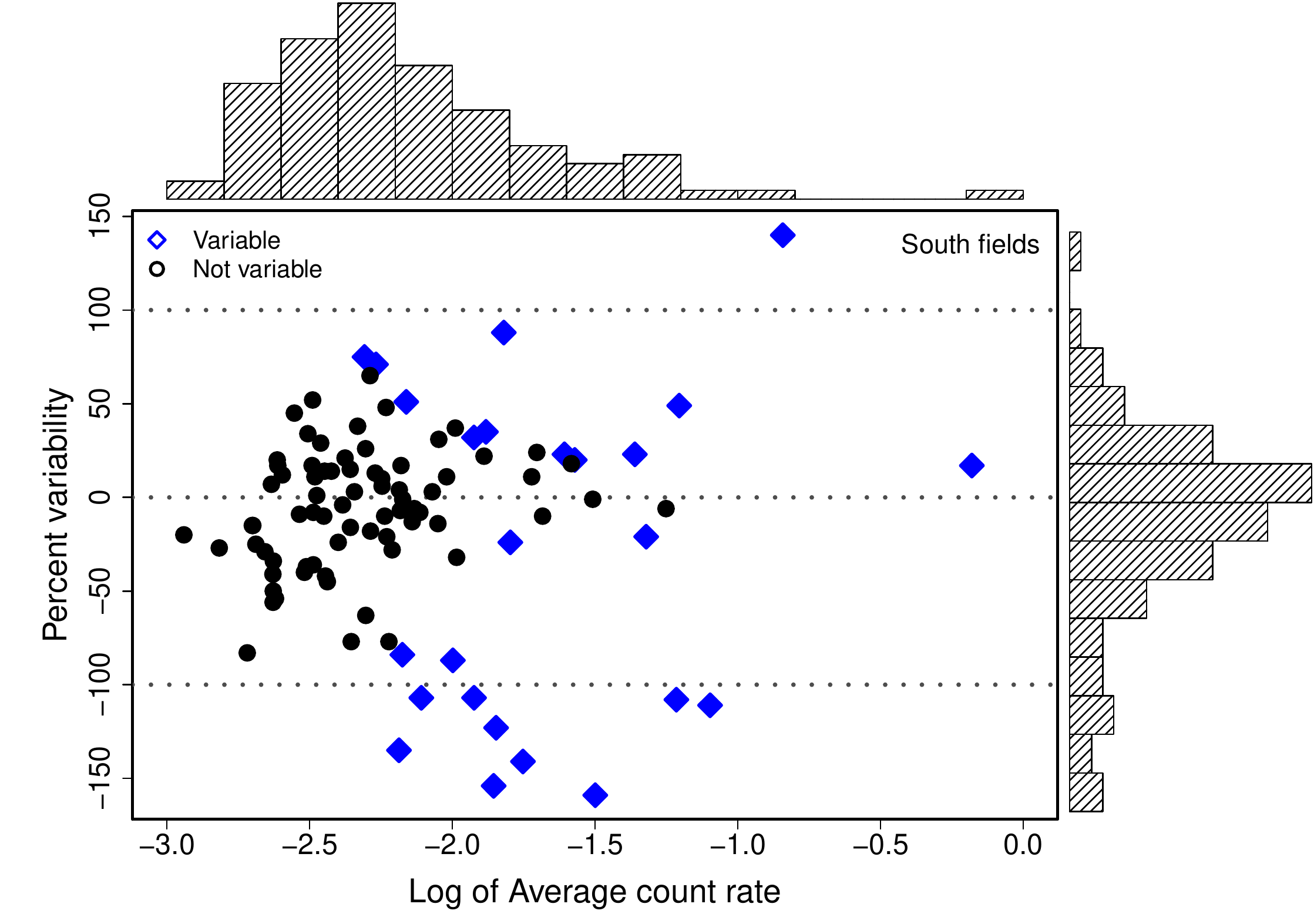}\\
\includegraphics[width=.5\textwidth]{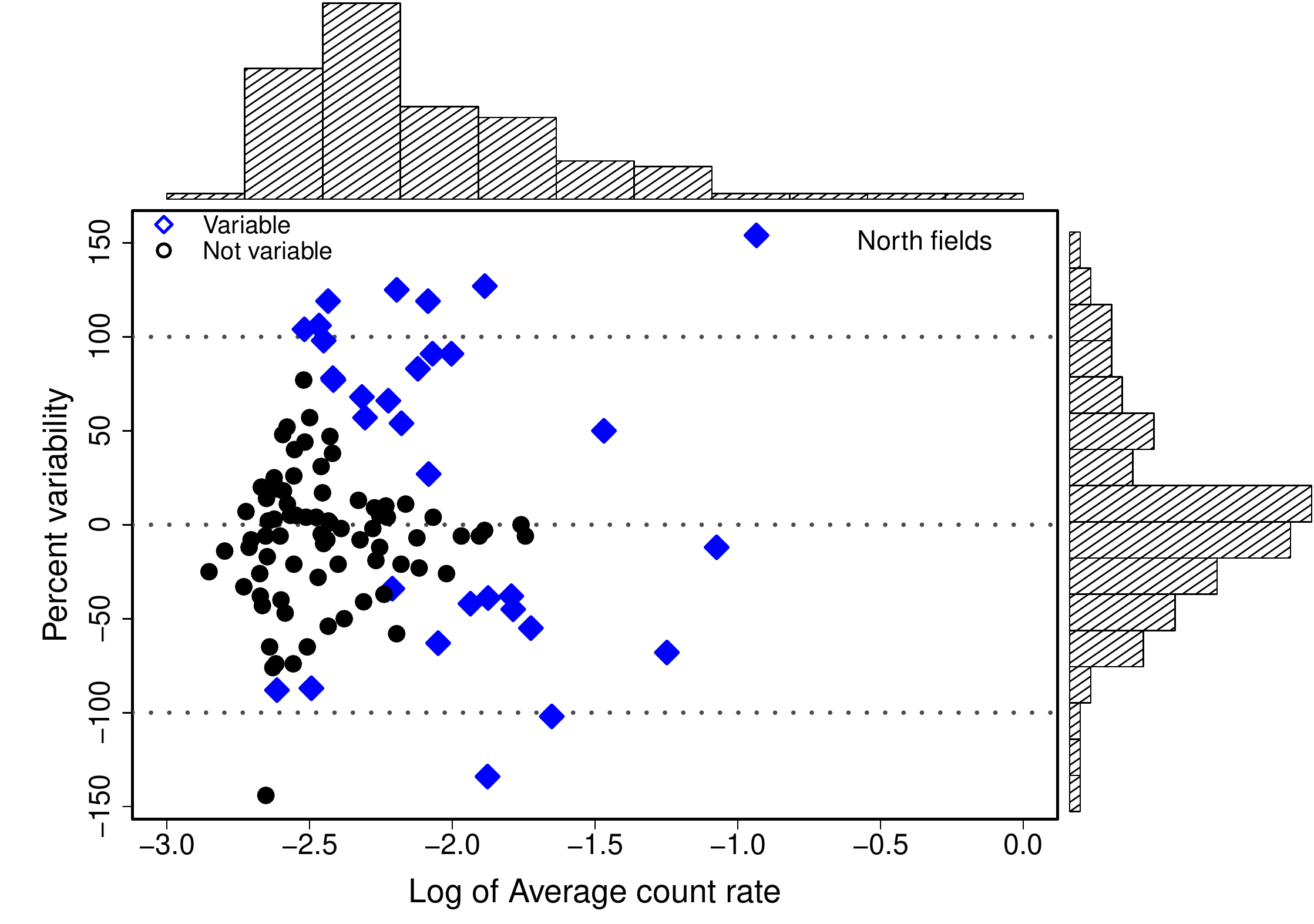}
\caption{Fractional normalised variability versus average count rate for all sources detected in each of the two epochs of the south (Field 1, upper panel) and north (Field 2, lower panel) \xmm\ EPIC fields. These diagrams show the change of flux between two observations, which are about half a year apart. Sources with significant variability are marked with filled blue diamonds. Non-variable sources, where the real difference in their count rates is smaller than their (quadratically added) uncertainty, are marked with filled black circles. Dashed grey lines mark the zero and 100\% variability levels. Marginal histograms show the distributions of average count rates and fractional normalised variability.}
\label{fig:var}
\end{figure}

We analysed the long-term variability of the objects in the cleaned source 
catalogue based on their EPIC count rates. 
Since the number of counts of all detections used for the variability 
study is high enough ($>$ 20), one can assume Gaussian statistics.
The search for change in flux between the two observations at a specific 
position
is based on 
cross-matching of the coordinates in each epoch of the south and north fields
(Fields 1 and 2, respectively). 
The matching radius was chosen to be 6\arcsec\ to account for 
systematic offsets and the EPIC point spread function (PSF) especially for 
faint objects or those close to chip gaps. The result is shown in 
Fig.~\ref{fig:var}. Variable sources are marked with blue diamonds, 
whereas those that show no changes are indicated as black circles.

The corresponding count rates and source numbers are listed in 
Tables~\ref{tab:var_south} and \ref{tab:var_north} for Fields 1 and 2,
respectively.
A source was classified as variable (blue sources in 
Fig.~\ref{fig:var}) if its difference in count rate between OBS1 and OBS2 
or OBS3 and OBS4 (column "Difference") was
larger than three times 
the square root of the quadratically added uncertainties 
(columns "Error1" and "Error2", or "Error3" and "Error4"). 
In the column "Significance", the ratio between the "Difference" and
$\sqrt{\mathrm{Error1}^2 + \mathrm{Error2}^2}$ or
$\sqrt{\mathrm{Error3}^2 + \mathrm{Error4}^2}$ 
is given.
The fractional variability in 
Fig.~\ref{fig:var}, listed as "Variability" in percent in 
Tables~\ref{tab:var_south} and \ref{tab:var_north}, was computed as the 
"Difference" divided by the "Mean Rate". 
In total, 55 sources show significant variability between the two epochs
in Fields 1 and 2.
A total of 154 sources could not be matched and were detected only in 
one observation, even though they were observed at least twice.
They are listed in Table~\ref{unmatched} and
can be transient sources.

\subsection{Light curves}

\subsubsection{Light curve construction and inspection}

\begin{figure}
\centering
\includegraphics[width=.5\textwidth,trim=30 150 70 150,clip]{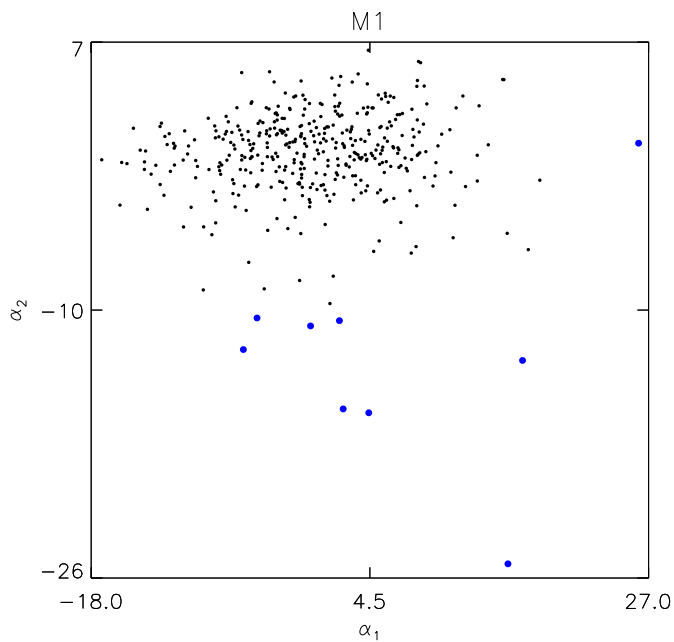}
\caption{
Distribution of the admixture coefficients $\alpha_1$ and
$\alpha_2$ of the principal component analysis of MOS1 light curves.
This diagram shows the variability on time scales of 1 ks.
The sources marked with blue dots show significant variability consistently
in all data. In this plot, source 146 appears twice since it was detected in 
both fields.
}
\label{admixplot}
\end{figure}

For each source in the cleaned catalogue we extracted a source and background 
light curve using the XMMSAS tools \texttt{evselect} and \texttt{epiclccorr}. 
The source region sizes were scaled based on the EPIC count rate. For pn, the 
background regions were determined using the new \texttt{ebkgreg} tool. For 
the MOS data we assumed a generic background annulus for a first 
approximation. The light curves were binned to 1-ks resolution for the few 
bright objects or 10-ks resolution for the many fainter objects. 
A standard energy band of 0.2 -- 5.0~keV was used.
Light curves of variable sources will be shown and discussed 
in Sect.\,\ref{varsources}.

We made an initial classification of the variability in each 
background-subtracted light curve based on its variance and linear trend. 
In addition, we computed a smoothing fit (moving average) to identify 
non-linear patterns like short-term flares. We examined each light curve by 
eye to validate our method.

Light curves that show significant change in flux were studied in 
more detail in a second step. Here, we chose a tailored time binning and 
energy range to characterize the light curve shape in more quantitative detail. 
In addition, we defined source and background regions by hand to optimise the 
event extraction. 
All errors are $1\sigma$ and all upper limits are $3\sigma$ unless specifically 
stated otherwise. The majority of the statistical analysis was carried out 
within the \texttt{R} software environment \citep{R_manual}.
We thus found a total of six sources that are clearly variable, 
two of which show flares during an observation
(Sect.~\ref{varsources}). 

\subsubsection{Variability search with the PCA}

We also followed a different approach for variability detection in our sample. This approach is inspired by variability search techniques
used in the optical band where accurate estimates of brightness measurement errors are often not available. We characterize each light
curve with the logarithm of its average count rate and a set of 18 variability indices quantifying the light curve smoothness and scatter
\citep[for the definition of these indices see][]{2017MNRAS.464..274S}.
All the indices are designed to highlight variability, but we do not know a priori which indices will work best for our data set and
variability patterns that are present in it. We therefore combine the variability indices using the
Principal Component Analysis \citep[PCA,][]{1901......2....559P}
and identify sources that are more variable (according to the PCA combination of indices) than the majority of sources in the field.
The Principal Component Analysis
is an unsupervised, non-parametric, linear decomposition of data into new coordinates (admixture coefficients $\alpha$) of an optimal set of uncorrelated axes (the eigenvectors of the variance-covariance matrix of the input data set, the Principal Components, hereafter, PC).  The principal components ($PC_1$ - $PC_{18}$) can be thought of as variability indices, each being a linear combination of the input variability indices, while the admixture coefficients ($\alpha_1$ - $\alpha_{18}$) are the values of these indices. The first few PC are expected to be optimal indices. The first principal component $PC_1$ accounts for the highest data variance possible (more widespread information), $PC_2$ (uncorrelated to $PC_1$) encodes most of the remaining variance, etc. High values of the first two admixture coefficients ($\alpha_1$ and $\alpha_2$) have been found to generally indicate variable sources 
\citep{2017MNRAS.464..274S,2018MNRAS.477.2664M}. 
For a general
discussion of the impact of measurement errors on the PCA analysis see
\cite{doi:10.1111/sjos.12083}.

The method was applied to the light curves of detections in the two M 31 fields, separately for the three 
EPIC detectors
MOS1/2 and pn. Sources which were detected in both fields are first treated as 
two sources (due to largely different off-axis angles). 
Any source with $\alpha_1$ or  $\alpha_2$ values that differ by at least three standard deviations (3$\sigma$) from the corresponding median values was considered to be a variable candidate (Fig.\,\ref{admixplot}). 
Thus, the variability of the selected candidates is significant at
$>3\sigma$ level.
A total of 8, 12, and 8 sources were identified as candidate variables in MOS1, MOS2 and pn respectively, while these numbers have not been screened yet for sources which have been detected by many cameras or in two observations. Table \ref{pcatab} lists the final candidates for variable sources identified with this method and their classification.
Sources 46, 146, and 171, which are classified as XRBs,
were found to be variable.
In addition, the PCA indicates that source 141, which is most likely a flaring 
star in the foreground (Sect.\,\ref{sec:disc_flare}), and 
eight additional sources are probably variable
(see Table \ref{pcatab}).
Of particular interest are 
sources 290 and 333, which have been classified to be XRB candidates 
as a star was found in the PHAT data inside the error circle 
of the \chandra\ counterpart in each case.
Source 300 is a bright SNR with a spectrum that is consistent with
thermal emission from shocked plasma, with no significant
hard emission detected at $\ga$ 2 keV (Sect.\,\ref{spec300}). It is
rather unlikely that this source shows variability on timescales of ks.
This indicates that using only statistical indices 
is not sufficient to clearly identify variable 
sources. Sources with clearly visible variability in their light curves are 
further discussed in Sect.\,\ref{varsources}.

\begin{table}
\caption{Sources found to be variable based on PCA}
\label{pcatab}
\centering
\begin{tabular}{ll}
\hline\hline
ID & Classification$^\dagger$ \\
\hline
40 & 	$<$fgstar$>$ \\
46 &	XRB 2E161 \\
141 & 	fgstar \\
146  &	XRB \\
171  &	XRB \\
290 &	$<$XRB$>$ \\
295  &	-- \\
300  &	SNR \\
321  &	$<$hard$>$ \\
325  &	$<$hard$>$ \\
333  &	$<$XRB$>$ \\
382  &	galaxy \\
\hline\hline
\multicolumn{2}{l}{$^\dagger$ Candidates are in $< >$.}
\end{tabular}
\end{table}

\subsection{Pulsations}

We also searched for pulsations using the 
barycentre-corrected events for each source.
We applied different methods of timing analyses. First, 
we performed a $Z_{n}^{2}$ analysis 
\citep{1983A&A...128..245B,1988A&A...201..194B} using the first and the second 
harmonic values of $n$ =1, 2 for 
sources that were detected on EPIC-pn. We searched for periodic signals in the 
period range of the Nyquist limit of 0.146 s (corresponding to twice the time 
resolution of EPIC-pn in full frame mode) to a period corresponding to the 
length of the observation. For the bright sources ($>$300 counts in one EPIC), 
we also used the Lomb-Scargle technique for unevenly sampled time series 
\citep{1982ApJ...263..835S}. We calculated the Lomb-Scargle periodogram for the 
light curves of all observations together.  
However, no significant pulsation was detected for any of the sources.

\subsection{Variable sources}
\label{varsources}

\begin{figure}
\raisebox{5.5cm}{\hspace{1.3cm}Source 46}\hspace{-2.6cm}\includegraphics[width=.475\textwidth]{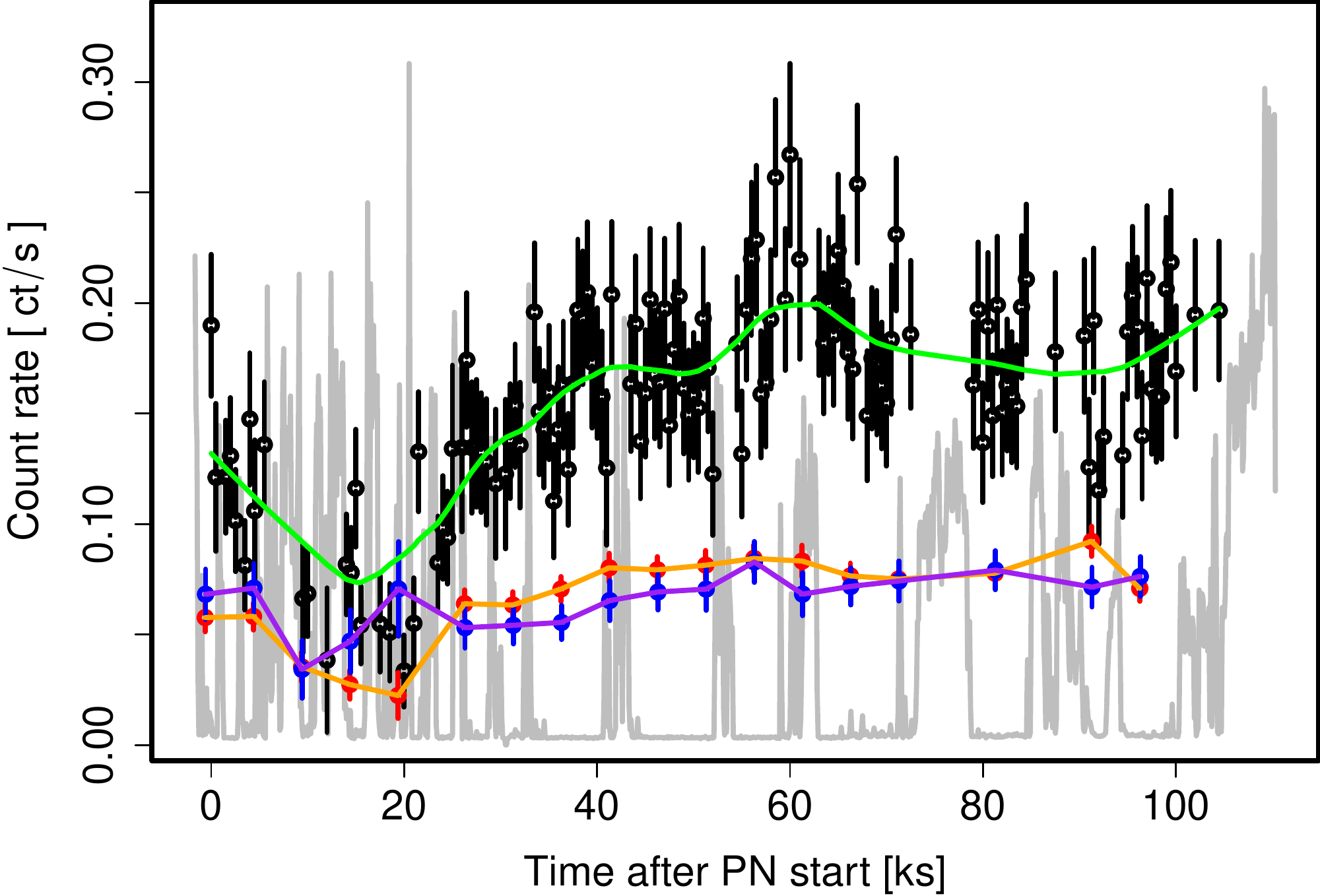}\\[2mm]
\raisebox{6.8cm}{\hspace{1.3cm}Source 255}\hspace{-2.9cm}\includegraphics[width=.48\textwidth]{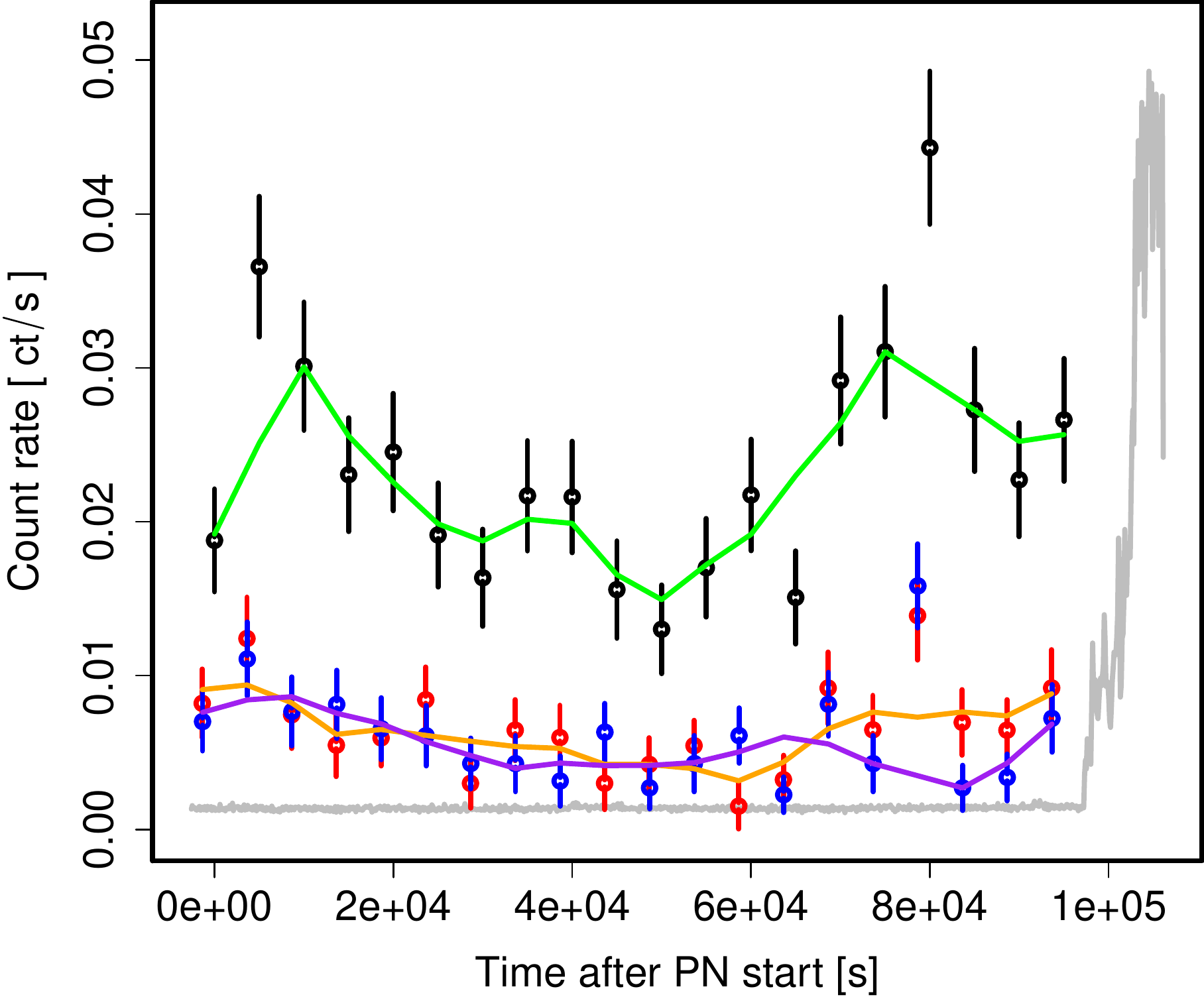}
\caption{\xmm\ EPIC light curves of sources 46 and 255 (0.2 -- 5.0 keV), which show variability on short timescales: pn (black), MOS1 (red), and MOS2 (blue) data points are plotted with the corresponding uncertainties. The green (pn), orange (MOS1), and purple (MOS2) curves are smoothing fits. Time resolution was optimised to the source count rate. The gray light curve shows the (scaled pn) background.
\label{lightcurves1}}
\end{figure}

\begin{figure*}[h!]
\centering
\raisebox{6.6cm}{\hspace{1.5cm}Source 141}\hspace{-3.0cm}\includegraphics[width=.95\textwidth]{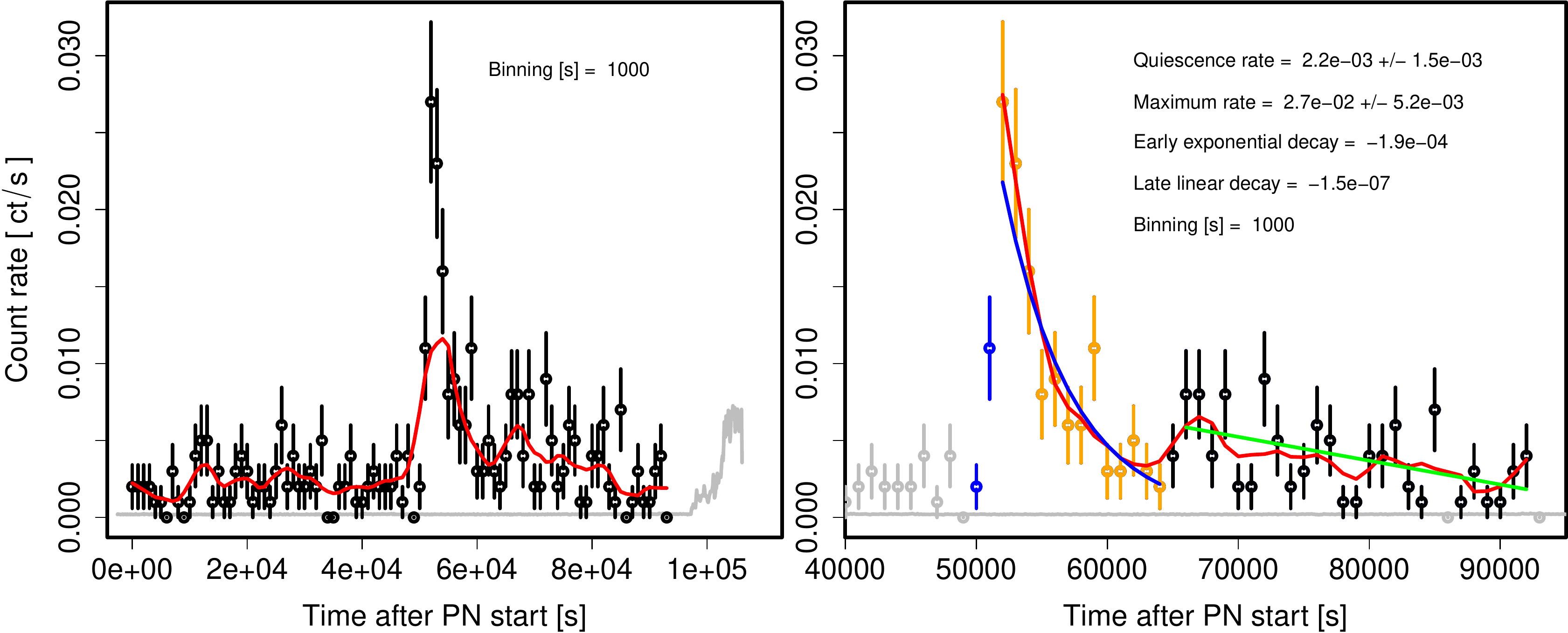}\\[3mm]
\raisebox{6.6cm}{\hspace{1.5cm}Source 153}\hspace{-3.0cm}\includegraphics[width=.95\textwidth]{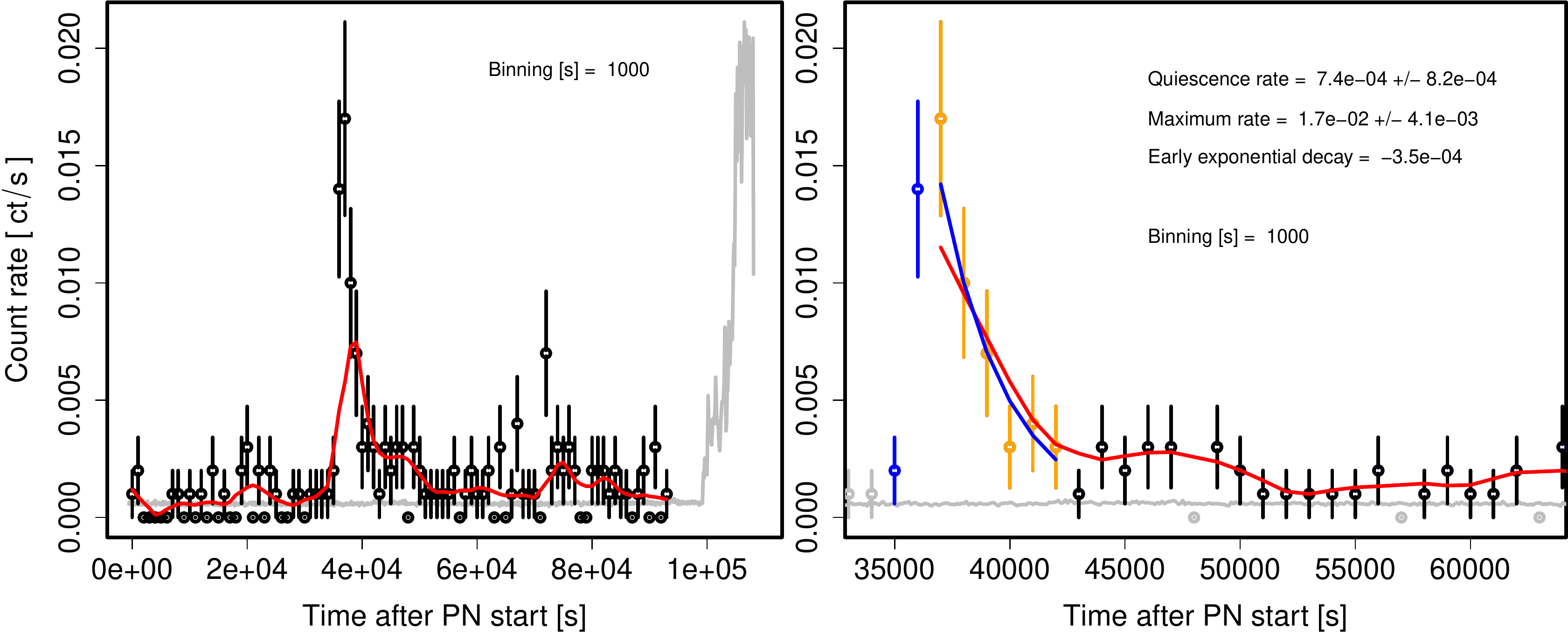}
\caption{\textit{Top:}
Light curves of source 141 (upper panels) and source 153 (lower panels). \textit{Left panels:} entire pn light curve (black, 0.2 -- 5.0 keV) with a smoothing fit overlaid in red. The background light curve is shown in grey. MOS light curves are not shown since the photon statistics are too low. \textit{Right panels:} analysis of the individual stages of the flare in a time window around it. Blue data points show the rise to maximum. The early decline (orange) in both sources can be reasonably well modelled with an exponential decline (blue fit curve; compare the red smoothing fit). The decline rates are given in units of ct/s/s. Before the flare there are no indications for variability in either source. The binned count rates are entirely consistent with a Poisson process.
\label{fig:diag_flare}}
\end{figure*}

\begin{figure}[ht]
\raisebox{6.1cm}{\hspace{1.5cm}Source 168}\hspace{-3.1cm}\includegraphics[width=.47\textwidth]{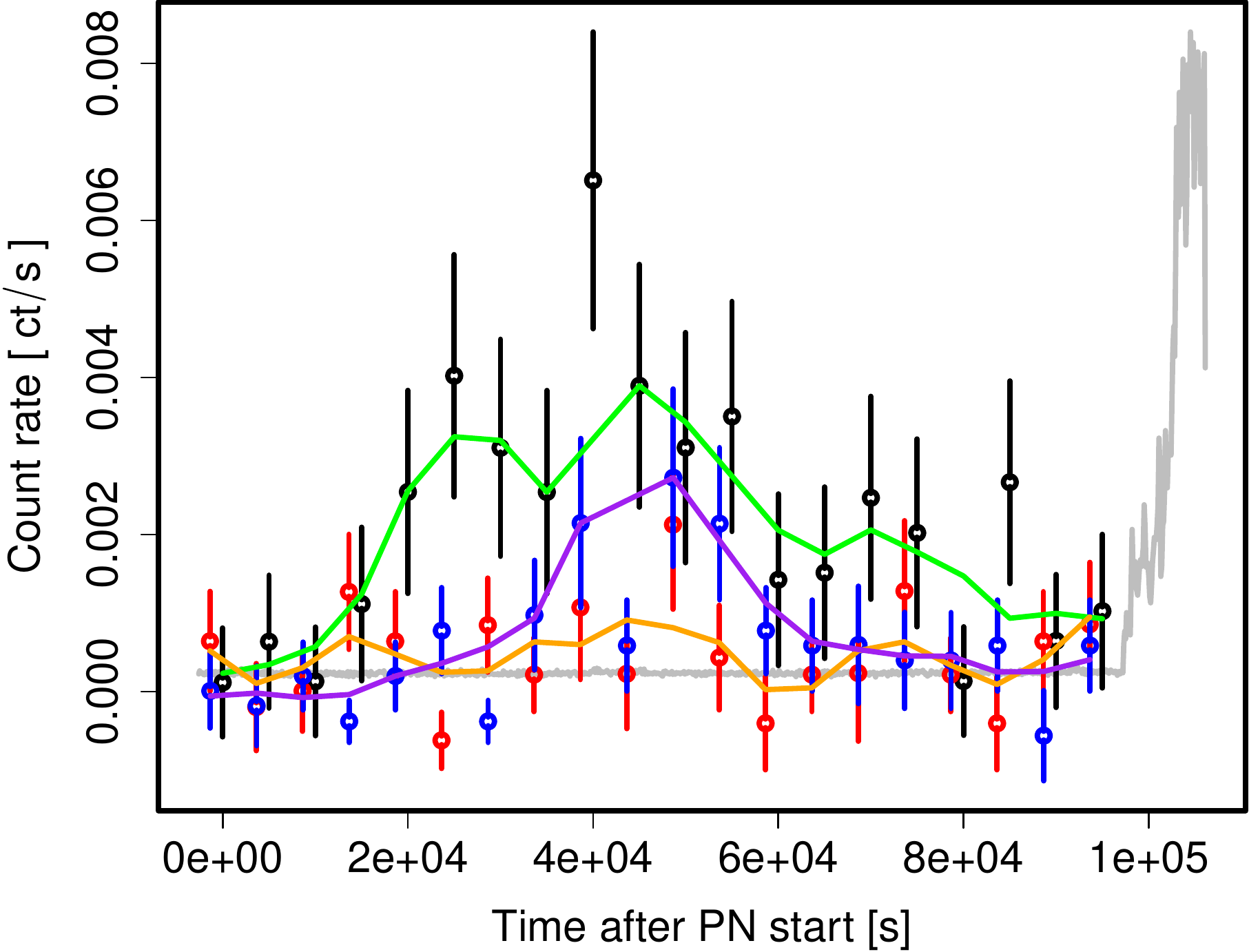}\\
\includegraphics[width=.47\textwidth]{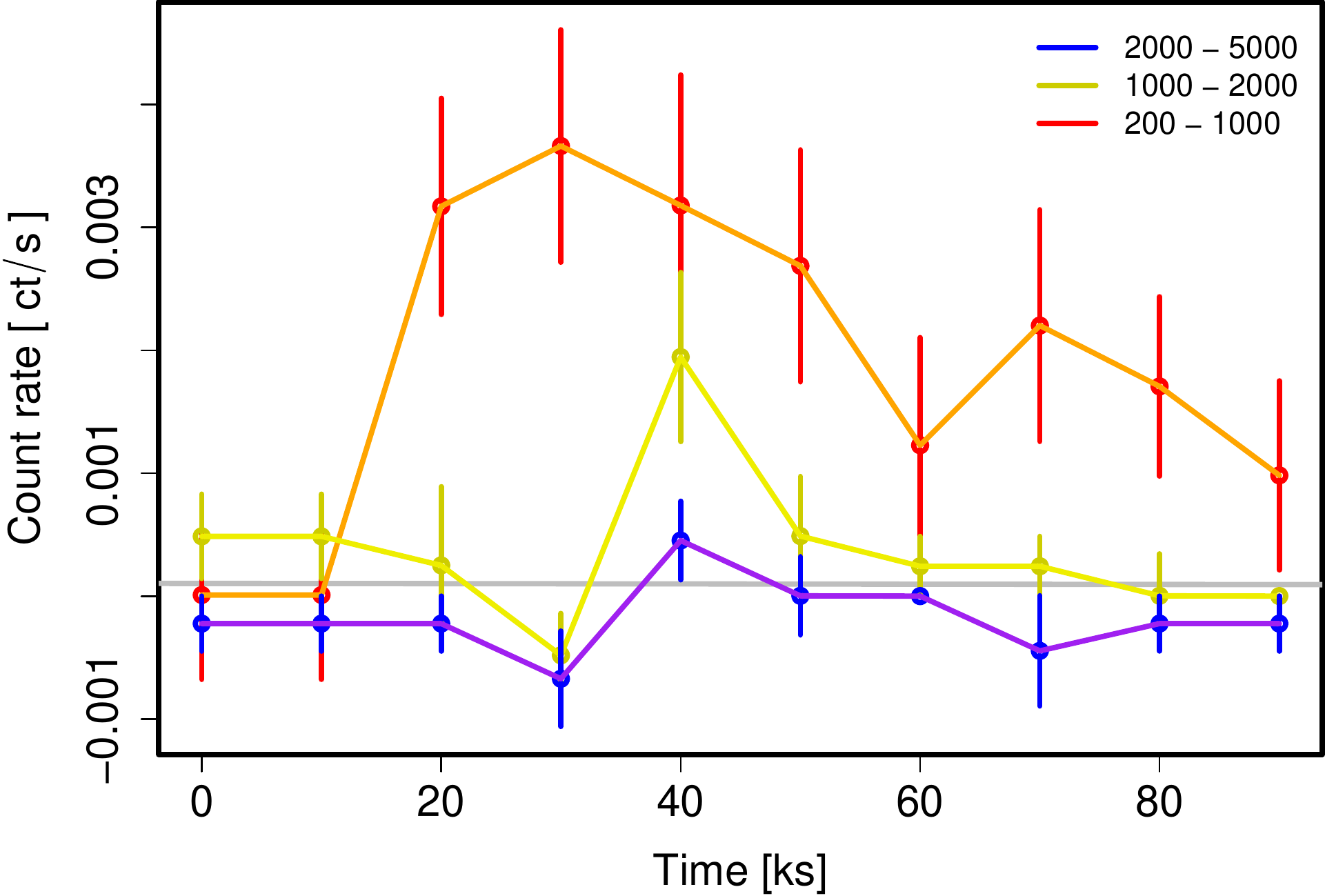}

\vspace{-1mm}
\caption{{\it Upper:} \xmm\ EPIC light curves (0.2 -- 5.0 keV) 
of possibly variable source 168 (pn in black, MOS1 in red, and MOS2 in blue).
The colours are the same as in  Fig.\,\ref{lightcurves1}.
{\it Lower:} The light curves of the source 168 in three energy bands: 
0.2 -- 1.0 keV (red), 1.0 -- 2.0 keV (yellow), and 2.0 -- 5.0 keV (blue).
The variability is significant in the softest band.
\label{lightcurves3}}
\end{figure}

\subsubsection{Sources with variability on short timescales}
\label{sec:res_ind_var}

\paragraph{Source 46 (J004344.5+412410)}
This is the recurrent transient 2E~161, which was first detected with the \einstein\ Observatory \citep{1991ApJ...382...82T}. Its detection in our new data was first reported by \citet[][]{2016ATel.8827....1H}. The source was neither detected by \pcat nor by \scat in their \mth\ mosaics. It was classified as a black hole candidate by \citet{2014ApJ...780...83B}. A long-term X-ray light curve was published by \citet{2013A&A...555A..65H} showing indications of flare-like variability. 

The source shows noticeable variability on time scales of hours in our survey data. The most prominent feature is an apparent dip at the beginning of observation 2, followed by a gradual increase in flux with potential plateaus and a continuing variability on smaller scales (Fig.~\ref{lightcurves1}). An analysis of different energy bands indicates that the variability happened primarily at energies below 2~keV but extended to higher energies as well.
Its variability was also confirmed by the PCA.

\paragraph{Source 255 ([SPH11] 1551, J004456.3+415937)}
This is a bright, long-known ROSAT source. \pcat and \scat classified it as 
a foreground star candidate. In the 3XMM-DR4
\citep[the third XMM-Newton serendipitous source catalogue,][]{2016A&A...590A...1R}, it was classified as not variable,
even though some EPIC light curves suggest variability. 
The pn light curve in Fig.~\ref{lightcurves1} shows a long, slow dip of about 70~ks duration.

\subsubsection{Flare sources}
\label{sec:disc_flare}

Two sources (141 and 153) showed clear indications for a flaring event; both during observation 3 (which had low background up until the final 10~ks). Here we study their flux variability in detail. The light curves of the entire observation as well as a time range focused on the flare are shown in Fig.~\ref{fig:diag_flare}.

\paragraph{Source 141 (J004417.7+415033)}
This source showed a clear and prominent flare detected in the pn and MOS data (Fig.~\ref{fig:diag_flare}, upper panels), 
which was also confirmed by PCA.
The flare appears to be a primarily soft event, based on the light curves in different energy bands. 

The flux rises from quiescence by an order of magnitude to peak within about 1.5 -- 2 ks (i.e. $\sim1$ hour; blue data points in Fig.~\ref{fig:diag_flare}, upper right). This time-scale was verified using shorter bins. Following the initial (exponential) decline, there seemed to be a rebrightening with a subsequent slow decline. This decline is consistent with a linear (green line) or exponential trend; there is no strong reason to favour either model. 
Source 141 is present in the catalogues of \scat (number 1426) and \pcat (number 546) and was classified as a foreground star. The suggested counterpart is the object 2MASS~00441774+4150327, which has a NIR colour of 
$J - K_{s}$ = 0.878$\pm$0.058. This colour is consistent with the source
being a M-type dwarf. Overall, the event is likely to have been a stellar flare.

\paragraph{Source 153 (J004423.0+415536)}
This source is also a flare object. It was located outside of MOS1 and on a chip gap of MOS2, resulting in a pn-only light curve (Fig.~\ref{fig:diag_flare}). 
Similar to source 141, the flare is observed in the softer band.

As can be seen in Fig.~\ref{fig:diag_flare} (lower panels), the flux of source 153 rises from quiescence by an order of magnitude to peak within about 1-1.5 ks (i.e. $\sim20$ mins; blue data points). Again the rise time was verified using shorter bins. The light curve suggests that immediately after the initial (exponential) decline (orange data points) the count rate remains somewhat elevated for about 6~ks until dropping back to a lower level. 

During this elevated stage (about 52~ks after start) the light curve count rate appears to be Poissonian but its mean is about a factor of 2 higher than before the flare. A statistical rank-sum test \citep[U-test or "Mann-Whitney test";][]{utest} indicates that the respective distributions are significantly different at the 95\% level (p-value of 0.003). This test is a non-parametric equivalent of the popular "Student's" t-test and is used to check for shifts in the mean location between two (cumulative) distributions.

The significant finding indicates that the count rate remains elevated for a certain time after the flare. Source 153 is not present in any X-ray catalogue so far. We note that there is a 2MASS source, 2MASS~00442305+4155351 with
$J - K_s$ = 0.878$\pm$0.058, which is a late K- or early M-type 
star, only 1.1\arcsec\ away from our X-ray position. The 2MASS object is a likely counterpart for our X-ray source in which case the source can be 
classified as a foreground flaring star.

\begin{figure}
\centering
\includegraphics[height=0.45\textwidth,trim=30 0 30 0,clip,angle=270]{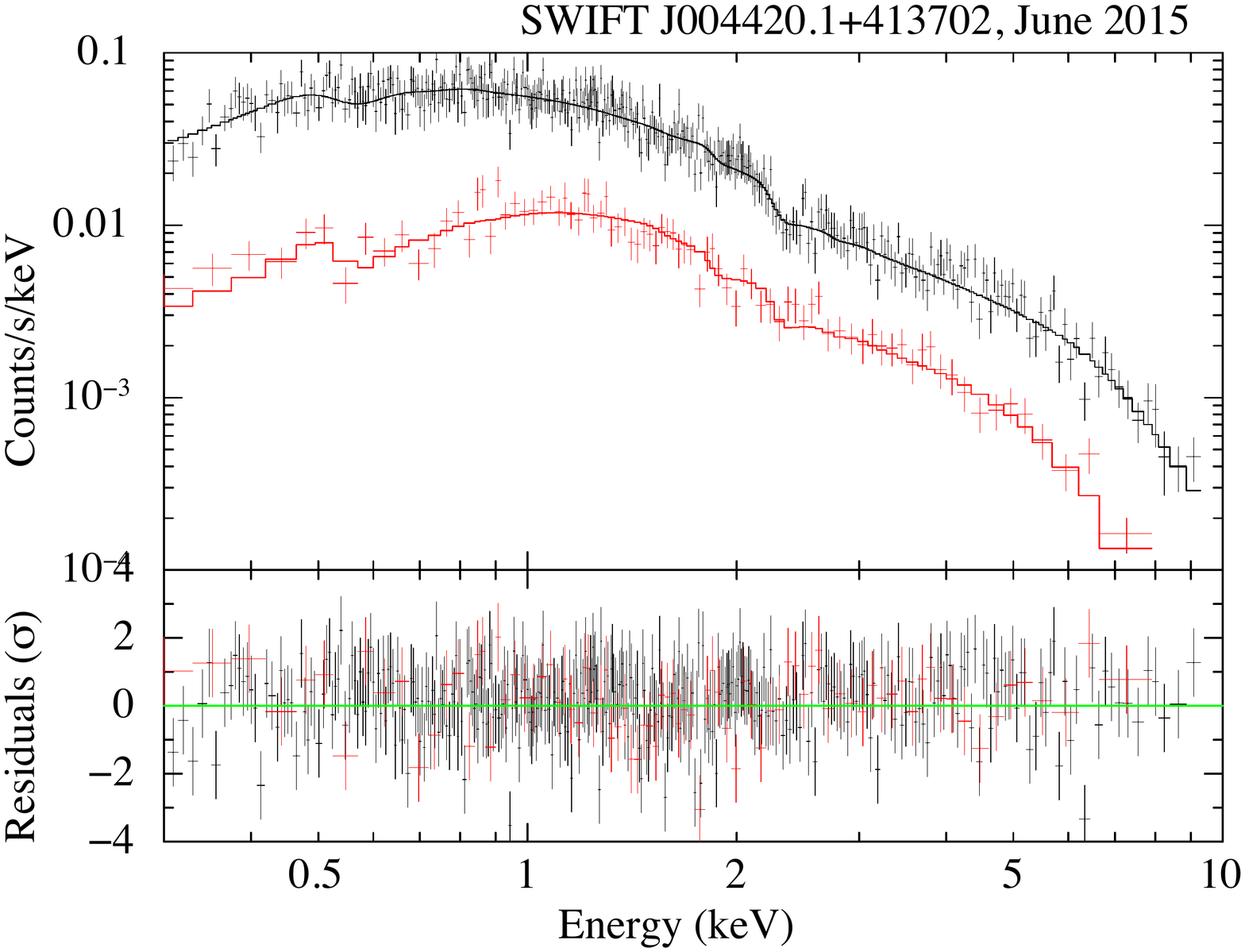}\\
\includegraphics[height=0.45\textwidth,trim=30 0 30 0,clip,angle=270]{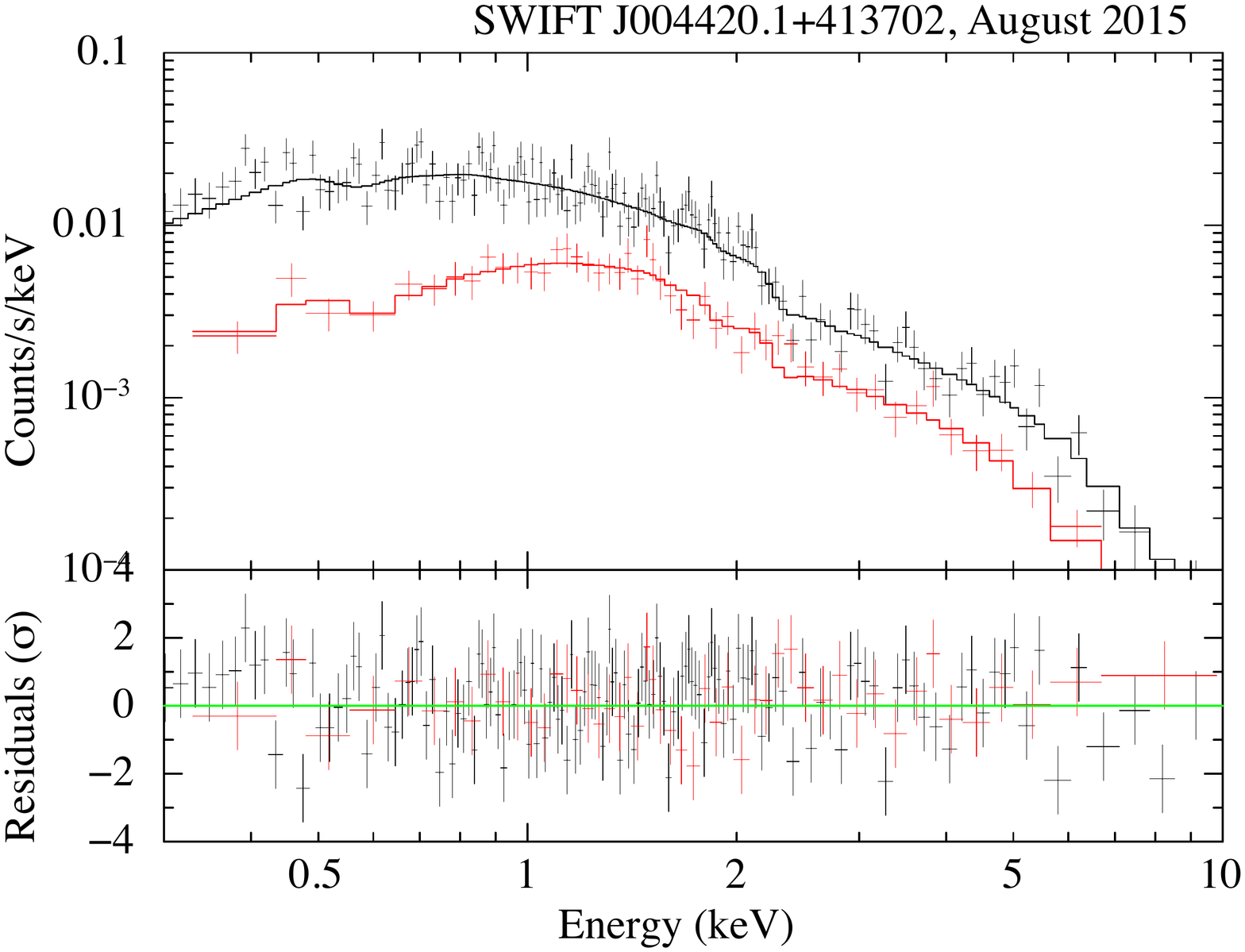}\\
\includegraphics[height=0.45\textwidth,trim=30 0 30 0,clip,angle=270]{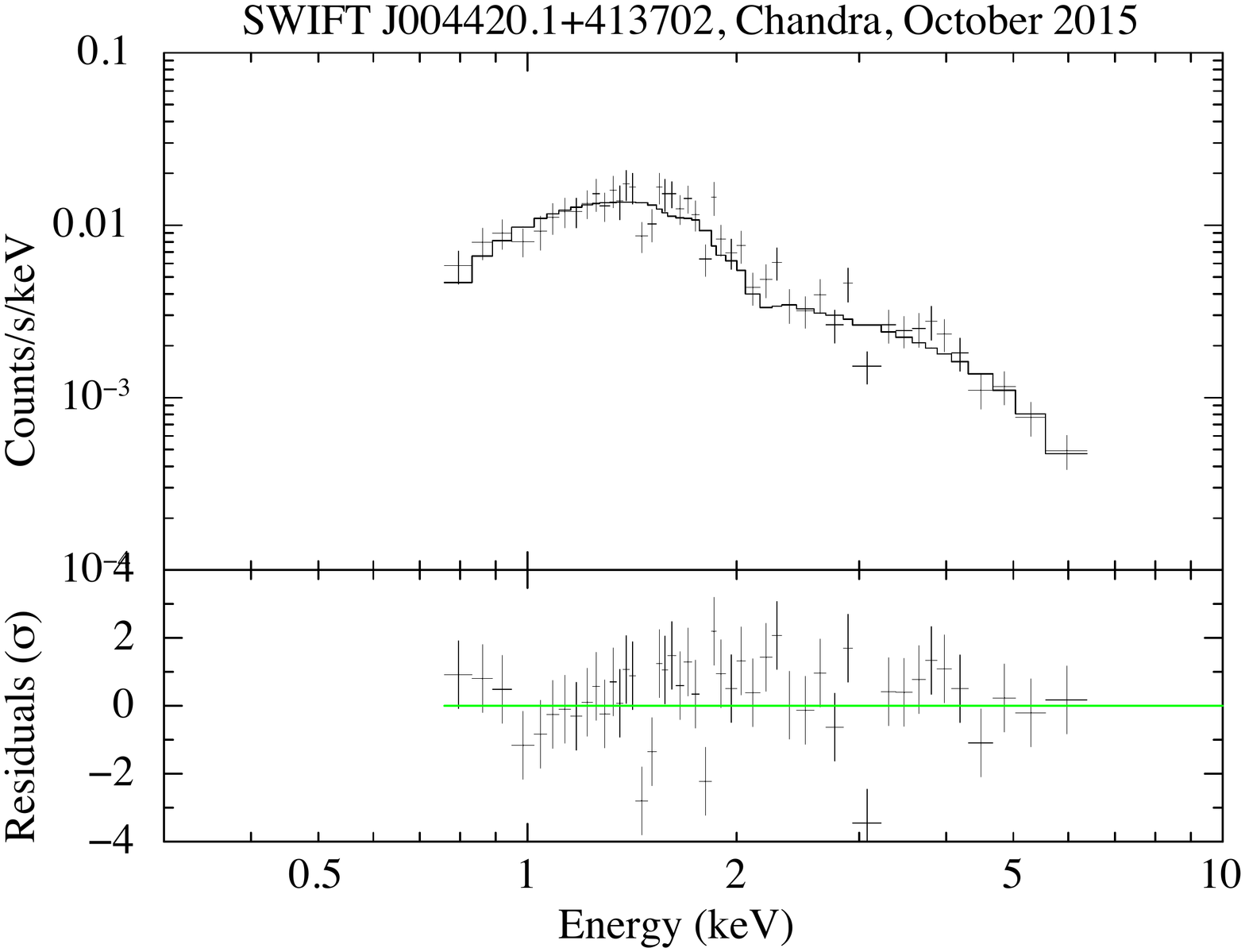}\\
\includegraphics[height=0.45\textwidth,trim=30 0 30 0,clip,angle=270]{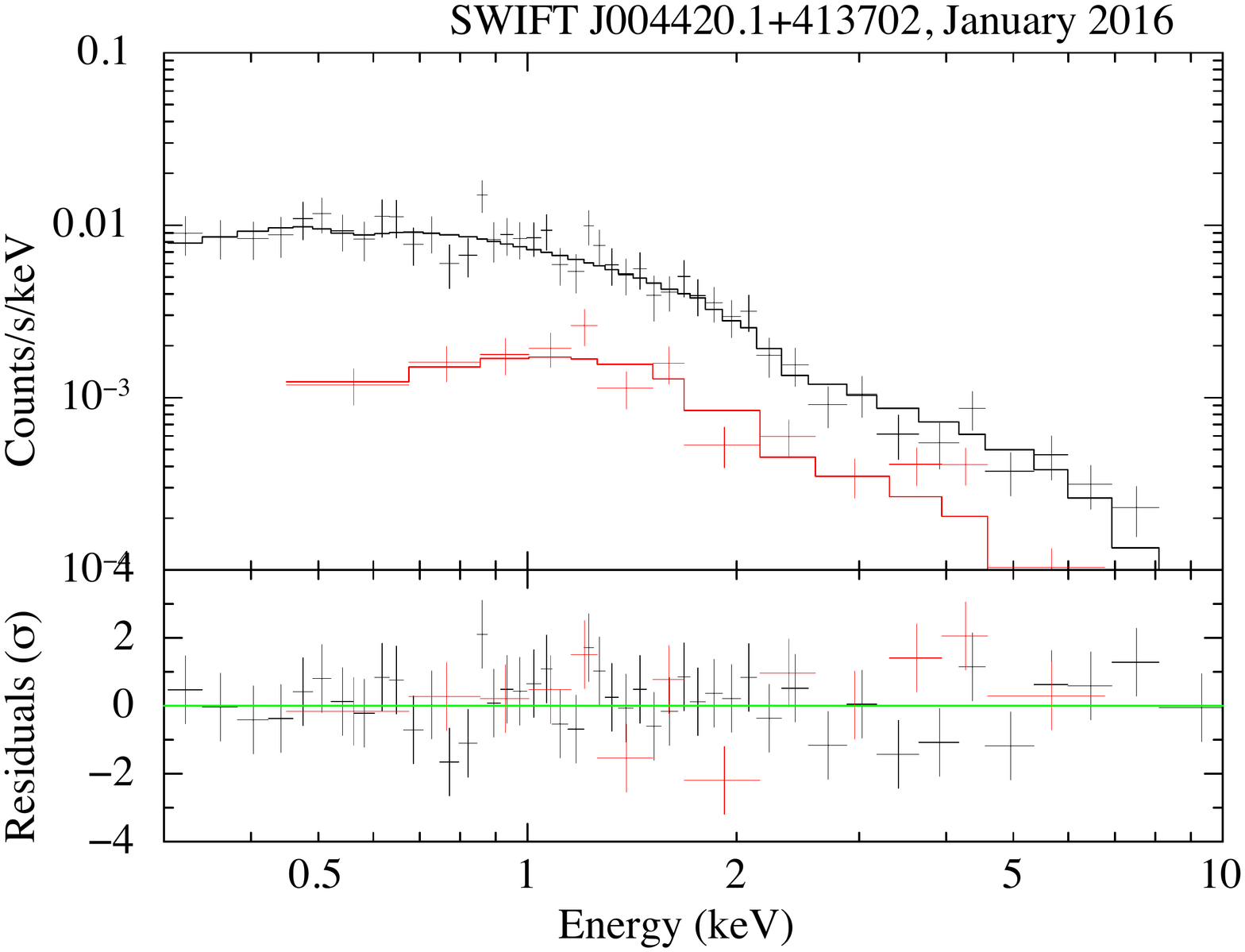}
\caption{
Spectra of SWIFT J004420.1+413702 with an absorbed power-law model (\nh\ and $\Gamma$ free). For June and August 2015, and January 2016, the pn spectrum
is shown in black and the MOS (1 or 2, see text) spectrum in red.
}
\label{swift0044}
\end{figure}

\begin{table*}
\caption{Spectral fit results for SWIFT J004420.1+413702}
\label{lp146_spectab}
\centering
\begin{tabular}{llrrrrr}
\hline\hline
Time & Instrument & \multicolumn{1}{l}{\nh} & \multicolumn{1}{l}{$\Gamma$} & \multicolumn{1}{l}{norm} & \multicolumn{1}{l}{red. $\chi^2$} & \multicolumn{1}{l}{DOF} \\
&  & \multicolumn{1}{l}{[$10^{21}$ cm$^{-2}$]} &  & \multicolumn{1}{l}{[ph~keV$^{-1}$~cm$^{-2}$~s$^{-1}$} & &  \\
 &  &  &  & \multicolumn{1}{l}{at 1 keV]} & &  \\
\hline
\multicolumn{7}{c}{\nh\ and $\Gamma$ free}\\
\hline
June 2015 & EPIC pn/MOS1 & 1.3$\pm$0.1 & 1.8$\pm$0.1 & (1.1$\pm$0.1)e--4 & 0.97 & 447 \\
August 2015 & EPIC pn/MOS1 & 1.1$\pm$0.2 & 1.7$\pm$0.1 & (9.6$\pm$0.1)e--5 &1.1  & 191 \\
October 2015 & ACIS-I & 1.8$\pm$1.2 & 1.9$\pm$0.2 & (8.7$\pm$2.2)e--5 & 1.5 & 44 \\
January 2016 & EPIC pn/MOS2 & 0.5$\pm$0.3 & 1.6$\pm$0.2 & (1.3$\pm$0.2)e--5 & 0.92 & 57\\
\hline
\multicolumn{7}{c}{\nh\ fixed to Galactic foreground \nh}\\
\hline
June 2015 & EPIC pn/MOS1 & 0.7 & 1.6$\pm$0.1 & (8.5$\pm$0.2)e--5 & 1.22 & 448 \\
August 2015 & EPIC pn/MOS1 & 0.7 & 1.5$\pm$0.1 & (8.1$\pm$0.3)e--5 &1.2  & 192 \\
October 2015 & ACIS-I & 0.7 & 1.7$\pm$0.2 & (6.7$\pm$0.5)e--5 & 1.5 & 45 \\
January 2016 & EPIC pn/MOS2 & 0.7 & 1.7$\pm$0.1 & (1.4$\pm$0.1)e--5 & 0.93 & 58\\
\hline
\multicolumn{7}{c}{Photon index frozen to $\Gamma = 1.7$}\\
\hline
June 2015 & EPIC pn/MOS1 & 1.0$\pm$0.1 & 1.7 & (0.9$\pm$0.1)e--4 & 1.03 & 448 \\
August 2015 & EPIC pn/MOS1 & 1.1$\pm$0.1 & 1.7 & (1.0$\pm$0.1)e--4 &1.1  & 192 \\
October 2015 & ACIS-I & 0.8$\pm$0.6 & 1.7 & (6.5$\pm$0.6)e--5 & 1.5 & 45 \\
January 2016 & EPIC pn/MOS2 & 0.6$\pm$0.2 & 1.7 & (1.4$\pm$0.2)e--5 & 0.92 & 58\\
\hline\hline
\end{tabular}
\end{table*}

\subsubsection{Sources with potential variability}
\label{sec:res_ind_pot}

\paragraph{Source 168 ([SPH11]  1461, J004428.7+414948)}
This source shows possible variability on time scales of hours (Fig.~\ref{lightcurves3}). Here, the signal/noise ratio and the flux are rather low overall but the behaviour in the three EPIC detectors looks consistent. The variability appears to be strongest in the 0.2 -- 1~keV band.

\scat classified this sources as a possible foreground star. We have found a 
2MASS source (2MASS~00442880+4149476) only 0.4\arcsec\ away from the \xmm\ position. The \txmm catalogue contains two faint potential past detections (ML = 6 and 12; 3XMM J004428.8+414947) with no sign of variability. The NIR counterpart
is a likely M-Type star with $J - K_s$ = 0.876$\pm$0.048.

\subsubsection{Source with spectral variability: Source 146 (SWIFT J004420.1+413702)}\label{swifttransient}

The source SWIFT J004420.1+413702 was first detected as a transient source
by \citet{2008ATel.1671....1P} based on observations with the
Swift satellite in combination with earlier ROSAT data.
The unabsorbed flux at the time of the Swift detection in 2008 was
$(4.6 \pm 0.7) \times 10^{-13}$ erg s$^{-1}$ cm$^{-2}$
(0.5 -- 5.0 keV) assuming an absorbed 
power-law model with an absorption of \nh\ = $6.6 \times 10^{20}$ cm$^{-2}$ and
a photon index of $\Gamma = 1.7$.

The PCA has shown that it is significantly variable.
We have identified a spectral change of SWIFT J004420.1+413702,
for which we have data from 4 epochs: 
June 2015 (\xmm),
August 2015 (\xmm),
October 2015 (\chandra), and 
January 2016 (\xmm).
Figure \ref{swift0044} shows the \xmm\ EPIC as well as \chandra\ ACIS-I spectra of the source
SWIFT J004420.1+413702, sorted chronologically 
(June 2015 [pn, MOS1],
August 2015 [pn, MOS1],
October 2015 [Chandra ACIS-I],
January 2016 [pn, MOS2]).
We fit all spectra with an absorbed power-law model.
The fit results are summarised in Table \ref{lp146_spectab}
for the three cases: 1. both \nh\ and $\Gamma$ are free,
2. \nh\ is fixed to the Galactic foreground value of 
\nh\ = $7 \times 10^{20}$ cm$^{-2}$, and
3. photon index is frozen to $\Gamma$ = 1.7.

If we fix the \nh\ to the Galactic foreground value, the photon index does
not vary significantly (case 2), however, the fit statistics get worse for the
spectra taken in June and August 2015, when the source was brighter.

If we assume that the shape of the intrinsic spectrum does not change
($\Gamma = 1.7 = const$, case 3) the \nh\ values change significantly from 
June 2015 to January 2016, decreasing from 
$\nh = 1.1 \pm\ 0.1 \times 10^{21}$ cm$^{-2}$ in August 2015 when
the source was bright 
(absorbed flux of $3.6 \pm 0.4 \times 10^{-13}$ erg s$^{-1}$ cm$^{-2}$ and
unabsorbed flux of $4.2 \pm 0.5 \times 10^{-13}$ erg s$^{-1}$ cm$^{-2}$
in the energy range of 0.5 -- 5.0 keV) 
to $0.6 \pm\ 0.2 \times 10^{21}$ cm$^{-2}$
about 5 months later when the source was about 7 times fainter
(absorbed flux of $5.4 \pm 0.8 \times 10^{-14}$ erg s$^{-1}$ cm$^{-2}$ and
unabsorbed flux of $5.9 \pm 0.8 \times 10^{-14}$ erg s$^{-1}$ cm$^{-2}$).

Therefore, the source seems to show higher absorption when it is brighter
and to be less absorbed when it becomes fainter without changing its intrinsic
spectrum.
The photon index of $\Gamma\ = 1.7$ is rather inconsistent with 
a high-mass X-ray binary which typically have a harder spectrum.
At the X-ray position, a low-mass star was found in the Chandra/PHAT
survey 
(Williams et al. \citeyear{2018arXiv180810487W}, accepted),
also suggesting the classification of SWIFT J004420.1+413702 as a
low-mass X-ray binary.

\begin{table*}
\caption{
\label{xsnrfl}
List of SNR detected in the new \xmm\ LP data and their X-ray flux.}
\centering
\begin{tabular}{rrrrlrrrr}
\hline\hline
\multicolumn{1}{l}{ID} & \multicolumn{1}{c}{RA J2000.0} & \multicolumn{1}{c}{Dec J2000.0}        & SPH11 ID& Class$^\star$& LL14 ID &$F_\mathrm{abs}$ & $F_\mathrm{abs}$ & $L_\mathrm{X,abs}$ \\
    &  \multicolumn{1}{c}{(hms)} &   \multicolumn{1}{c}{(dms)}           &      &           &     &(0.35 -- 2 keV) & (0.3 -- 10 keV)  & (0.3 -- 10 keV) \\
    &  &      &      &           &     &[erg s$^{-1}$ cm$^{-2}$] & [erg s$^{-1}$ cm$^{-2}$]  & [erg s$^{-1}$] \\
\hline
  24& 0:43:27.887& +41:18:29.33& 1234 & SNR       & --  &6.1e--14 & 6.5e--14 & 4.8e+36 \\
  38& 0:43:39.157& +41:26:53.44& 1275 & SNR       & --  &2.5e--14 & 2.6e--14 & 1.8e+36 \\
  39& 0:43:41.219& +41:34:07.07& 1282 & $<$SNR$>$ & --  &1.6e--15 & 1.6e--15 & 1.2e+35 \\
  68& 0:43:53.846& +41:20:43.79& 1332 & $<$SNR$>$ & --  &4.0e--15 & 4.0e--15 & 2.9e+35 \\
 102& 0:44:04.278& +41:48:45.11& 1372 & SNR       & 83  &1.9e--15 & 1.9e--15 & 1.4e+35 \\
 104& 0:44:04.480& +41:58:04.41& 1370 & SNR       & --  &3.3e--15 & 3.3e--15 & 2.4e+35 \\
 118& 0:44:09.530& +41:33:22.00& 1386 & SNR       & --  &1.5e--15 & 1.5e--15 & 1.1e+35 \\
 134& 0:44:13.358& +41:19:52.70& 1410 & SNR       & --  &7.7e--15 & 7.8e--15 & 5.7e+35 \\
 192& 0:44:34.963& +41:25:13.67& 1481 & SNR       & --  &2.5e--15 & 2.5e--15 & 1.8e+35 \\
 213& 0:44:43.461& +41:26:56.10&   -- & SNR       & 102 &4.2e--16 & 4.2e--16 & 3.1e+34 \\
 222& 0:44:45.823& +41:52:59.21&   -- & SNR       & 105 &9.9e--16 & 9.9e--16 & 7.3e+34 \\
 226& 0:44:47.132& +41:29:20.11&  522 & SNR       & --  &2.2e--15 & 2.2e--15 & 1.6e+35 \\
 238& 0:44:49.860& +41:53:07.79&   -- & SNR       & 107 &8.4e--16 & 8.5e--16 & 6.2e+34 \\
 240& 0:44:50.586& +41:54:21.45&   -- & SNR       & 109 &1.6e--15 & 1.6e--15 & 1.2e+35 \\
 242& 0:44:51.053& +41:29:06.26& 1535 & $<$SNR$>$ & --  &1.0e--14 & 1.6e--14 & 1.2e+36 \\
 246& 0:44:52.823& +41:55:00.67& 1539 & SNR       & --  &1.5e--15 & 1.7e--15 & 1.3e+35 \\
 253& 0:44:55.775& +41:56:56.95& 1548 & SNR       & --  &1.6e--15 & 1.6e--15 & 1.2e+35 \\
 300& 0:45:13.866& +41:36:15.54& 1599 & SNR       & --  &1.6e--14 & 1.6e--14 & 1.2e+36 \\
\hline\hline
\multicolumn{8}{l}{$^\star$Based on SPH11, updated by SPH12, and this work. $<$SNR$>$ indicates that the source is a SNR candidate.}
\end{tabular}
\end{table*}

\begin{table*}
\caption{
\label{osnrul}
List of optical SNR detected by LL14 in the field of view of the new \xmm\ LP observations and the X-ray flux upper limits (EPIC-pn).}
\centering
\begin{tabular}{rrrrrrr}
\cmidrule{1-3}\cmidrule{5-7}\\[-6mm]
\cmidrule{1-3}\cmidrule{5-7}
\multicolumn{1}{l}{LL14 ID} & $F_\mathrm{X}$ upper limit & $L_\mathrm{X}$ upper limit  & & \multicolumn{1}{l}{LL14 ID} & $F_\mathrm{X}$ upper limit & $L_\mathrm{X}$ upper limit  \\
 & (0.3 -- 10 keV) & (0.3 -- 10 keV) &  &  & (0.3 -- 10 keV) & (0.3 -- 10 keV) \\
& [erg s$^{-1}$ cm$^{-2}$]  & [erg s$^{-1}$ cm$^{-2}$] & & & [erg s$^{-1}$ cm$^{-2}$] & [erg s$^{-1}$ cm$^{-2}$]  \\
\cmidrule{1-3}\cmidrule{5-7}
   54 &  1.5e--13 & 1.1e+37 & &  113 &  1.7e--14 & 1.3e+36 \\
   61 &  4.9e--14 & 3.6e+36 & &  114 &  7.8e--15 & 5.7e+35 \\
   67 &  6.1e--14 & 4.5e+36 & &  115 &  7.2e--15 & 5.3e+35 \\
   69 &  4.2e--14 & 3.1e+36 & &  116 &  2.2e--14 & 1.6e+36 \\
   72 &  4.7e--14 & 3.5e+36 & &  117 &  7.4e--14 & 5.4e+36 \\
   73 &  1.6e--14 & 1.2e+36 & &  118 &  1.6e--14 & 1.2e+36 \\
   74 &  3.6e--14 & 2.6e+36 & &  119 &  2.7e--14 & 2.0e+36 \\
   77 &  5.1e--14 & 3.7e+36 & &  120 &  1.5e--14 & 1.1e+36 \\
   78 &  2.1e--14 & 1.5e+36 & &  121 &  4.1e--14 & 3.0e+36 \\
   79 &  1.3e--13 & 9.5e+36 & &  122 &  4.0e--14 & 2.9e+36 \\
   81 &  5.5e--14 & 4.0e+36 & &  123 &  1.6e--14 & 1.2e+36 \\
   82 &  5.0e--15 & 3.7e+35 & &  125 &  6.8e--14 & 5.0e+36 \\
   84 &  2.9e--14 & 2.1e+36 & &  126 &  7.5e--14 & 5.5e+36 \\
   85 &  5.9e--14 & 4.3e+36 & &  127 &  7.0e--14 & 5.1e+36 \\
   86 &  1.8e--14 & 1.3e+36 & &  128 &  5.0e--14 & 3.7e+36 \\
   88 &  1.7e--14 & 1.3e+36 & &  129 &  7.2e--15 & 5.3e+35 \\
   89 &  6.8e--14 & 5.0e+36 & &  130 &  3.2e--14 & 2.4e+36 \\
   90 &  4.5e--15 & 3.3e+35 & &  131 &  2.4e--13 & 1.8e+37 \\
   91 &  2.8e--15 & 2.1e+35 & &  132 &  5.7e--15 & 4.2e+35 \\
   92 &  7.3e--14 & 5.4e+36 & &  133 &  7.5e--15 & 5.5e+35 \\
   93 &  3.2e--14 & 2.4e+36 & &  134 &  1.9e--13 & 1.4e+37 \\
   94 &  3.0e--14 & 2.2e+36 & &  135 &  5.1e--14 & 3.7e+36 \\
   95 &  2.5e--14 & 1.8e+36 & &  136 &  1.5e--14 & 1.1e+36 \\
   96 &  4.0e--14 & 2.9e+36 & &  137 &  4.6e--14 & 3.4e+36 \\
   97 &  2.4e--14 & 1.8e+36 & &  138 &  2.9e--14 & 2.1e+36 \\
   98 &  5.6e--14 & 4.1e+36 & &  139 &  3.9e--14 & 2.9e+36 \\
   99 &  5.5e--15 & 4.0e+35 & &  140 &  1.1e--13 & 8.0e+36 \\
  100 &  2.5e--14 & 1.8e+36 & &  141 &  2.9e--14 & 2.1e+36 \\
  101 &  2.9e--14 & 2.1e+36 & &  142 &  3.5e--14 & 2.6e+36 \\
  103 &  1.2e--13 & 8.8e+36 & &  143 &  2.6e--14 & 1.9e+36 \\
  104 &  2.7e--14 & 2.0e+36 & &  144 &  1.9e--13 & 1.4e+37 \\
  108 &  1.5e--14 & 1.1e+36 & &  146 &  3.1e--13 & 2.3e+37 \\            
\cmidrule{5-7}\\[-6mm]
\cmidrule{5-7}\\[-5mm]
  110 &  7.8e--14 & 5.7e+36 \\
\cmidrule{1-3}\\[-6mm]
\cmidrule{1-3}
\end{tabular}
\end{table*}

\section{Supernova remnants}

One of the major objectives of the \xmm\ LP observations of the northern
disc of \mth\ was the study of the population of SNRs.
Detailed studies of several SNRs in \mth\ using \chandra\ were performed
by, e.g., \citet{2002ApJ...580L.125K,2003ApJ...590L..21K,2004ApJ...615..720W}.
We had studied the SNRs detected in the \xmm\ survey 
of the entire \mth\ galaxy using the \xmm\ data in X-rays and the LGS data in 
the optical (SPH12). 
We cross-correlated the new source list of the northern disc
with the list of SNRs and SNR candidates of SPH12.

Lee \& Lee (\citeyear{2014ApJ...786..130L}, LL14 hereafter) 
created a list of optical SNRs 
using the LGS data and classified different types of SNRs based on the
optical morphology and possible progenitor. The cross-correlation of the
list of X-ray SNRs with the list of LL14 has shown that not all optical SNRs
are bright enough in X-rays to be detected and vice versa
\citep[see also, e.g.,][]{2017ApJS..230....2B}.
All SNRs detected in the \xmm\ LP data 
have $HR_1 > 0.0$ and $HR_2 < 0.0$, with the majority having
$HR_2 < -0.5$. X-ray sources which have an optical SNR as a counterpart
and fulfill these hardness ratio criteria were newly classified as SNRs
(sources 102, 213, 222, 238, and 240).
In Table \ref{xsnrfl} we list the SNRs detected in the \xmm\ LP of the
northern disc. For all the SNRs and candidates in this list we extracted
the spectrum. 
The faintest detected SNR has a flux of $4.2 \times 10^{-16}$ 
erg s$^{-1}$ cm$^{-2}$ (0.3 -- 10.0 keV), corresponding to a luminosity of
$3.1 \times 10^{34}$ erg s$^{-1}$ (0.3 -- 10.0 keV).

\begin{figure}
\centering
\includegraphics[width=0.46\textwidth,trim=0 40 0 35,clip=]{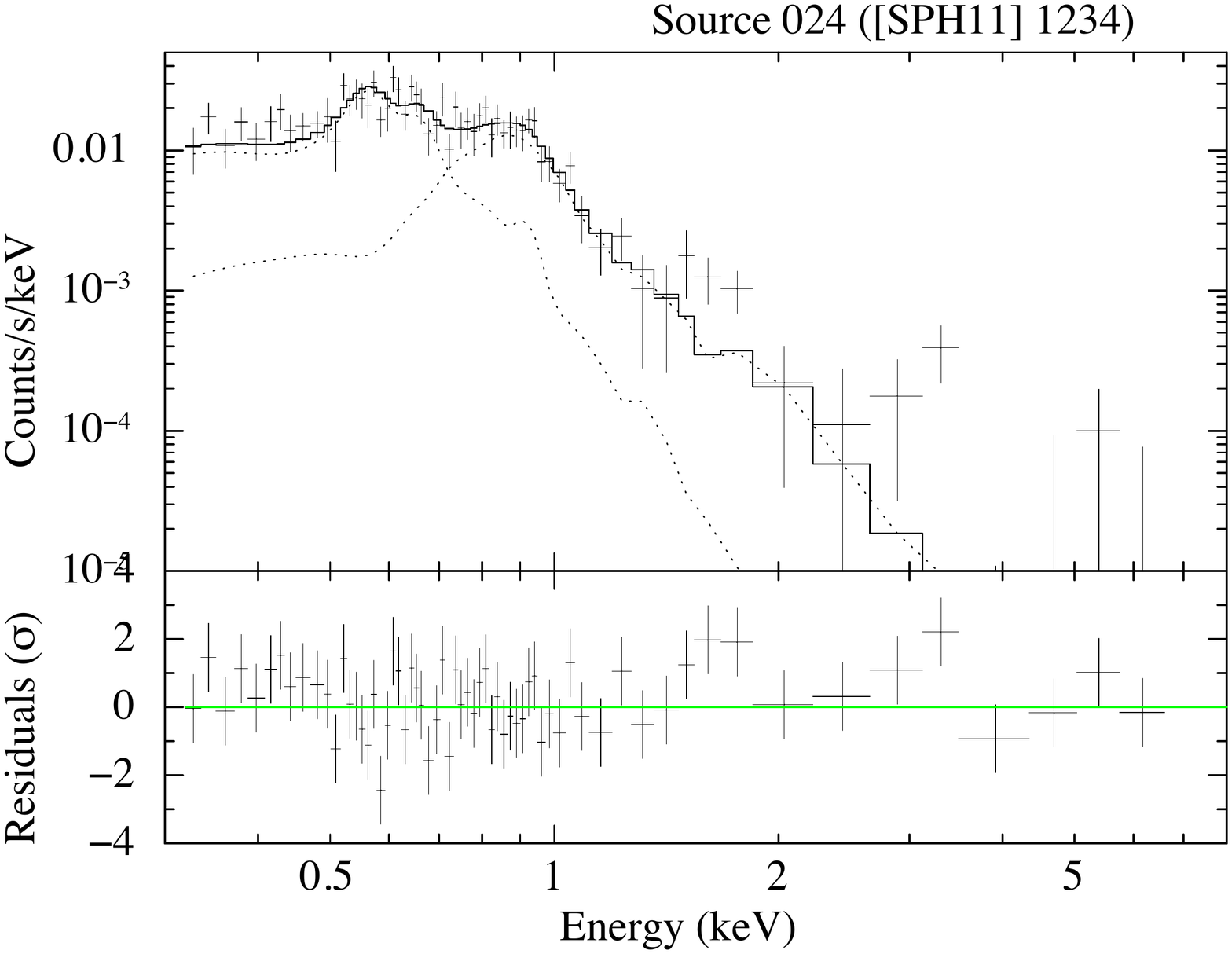}\\
\includegraphics[width=0.46\textwidth,trim=0 40 0 35,clip=]{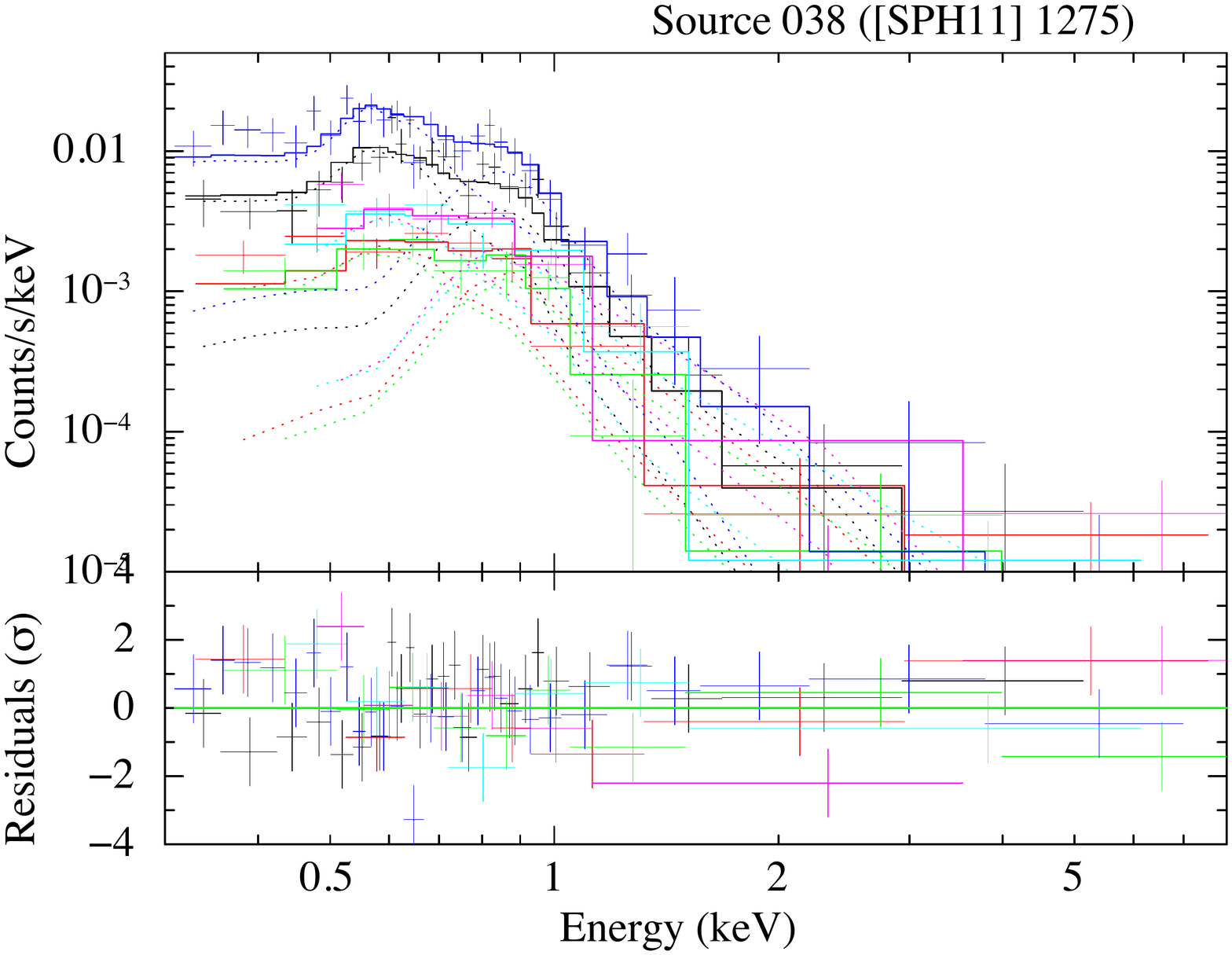}\\
\includegraphics[width=0.46\textwidth,trim=0 40 0 35,clip=]{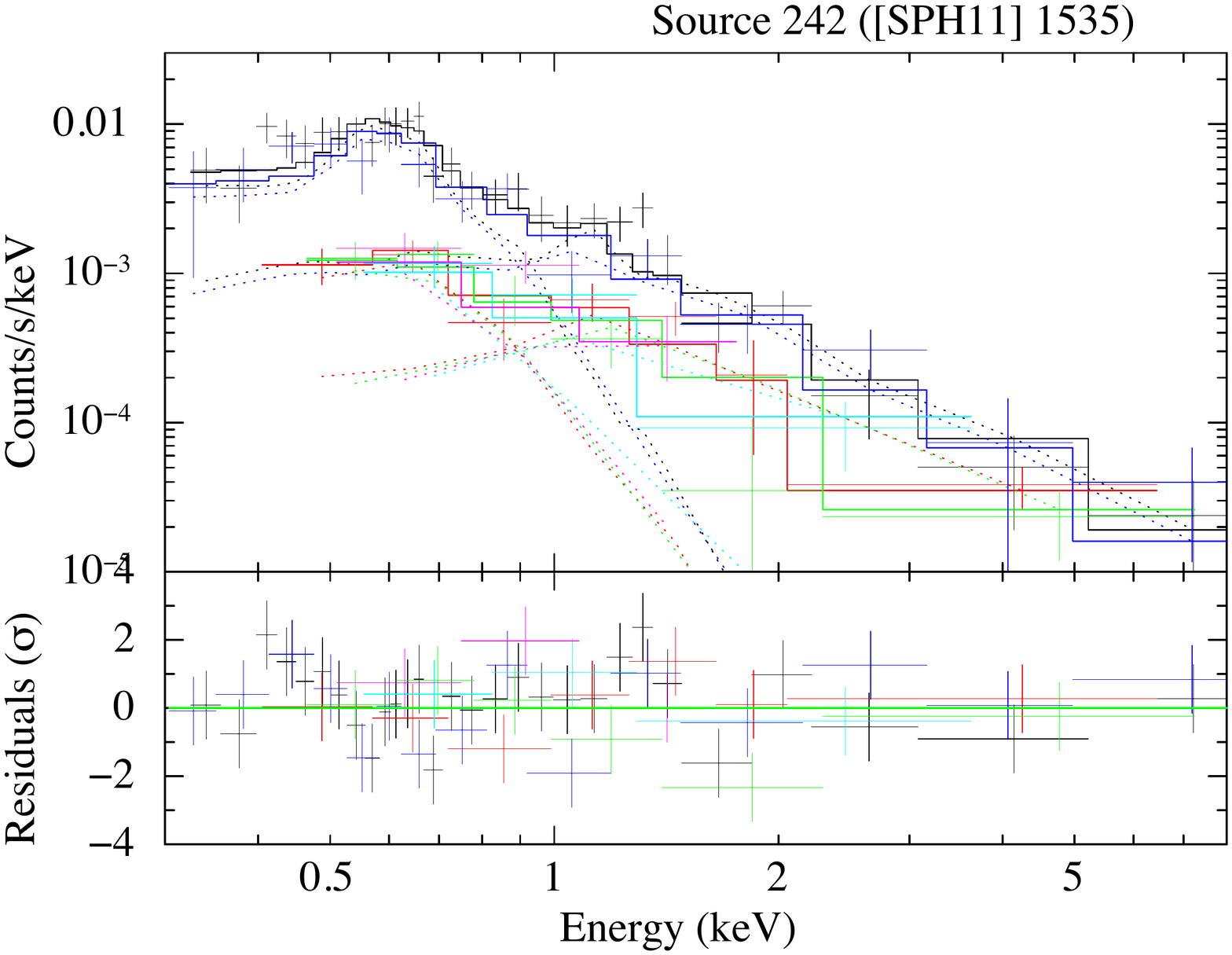}\\
\includegraphics[width=0.46\textwidth,trim=0 40 0 35,clip=]{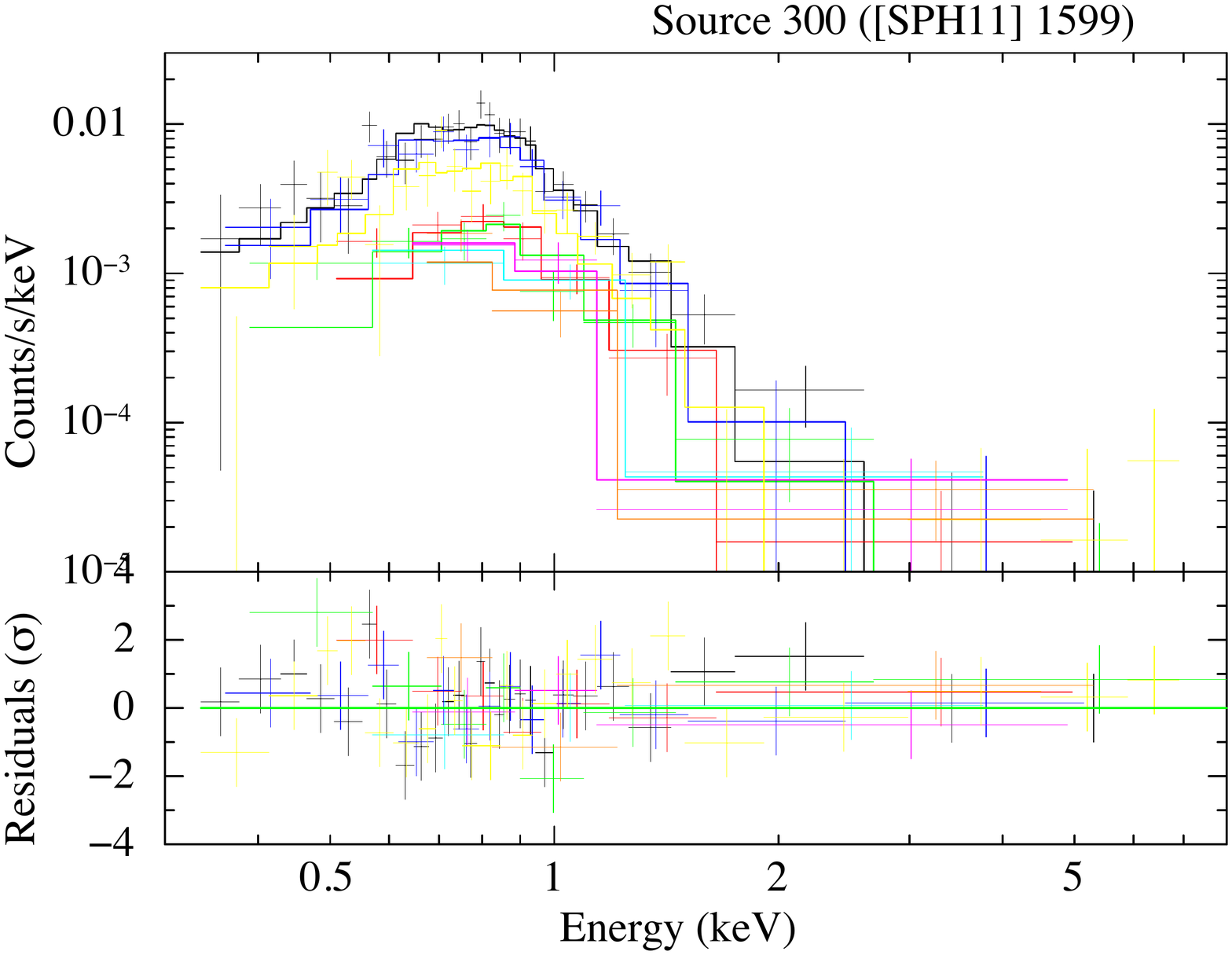}
\caption{
Spectra and best fit models of the SNRs [SPH11] 1234, 1275, 1535 ({\tt APEC+APEC}),
and 1599 ({\tt APEC}). Different colours are used to show spectra taken from 
different 
EPIC detectors
and observations.
}
\label{snrspectra}
\end{figure}

There are four sources with enough counts to make spectral fitting worthwhile
(pn source counts $\ga$500, MOS source counts  $\ga$150):
[SPH11] 1234, 1275, 1535, and 1599. 
The fit results are summarised in Table \ref{snrfitlst} and the spectra 
are shown in Fig.\,\ref{snrspectra}.
For the rest of the
X-ray detected SNRs, the flux was derived from the extracted spectra
by assuming a thermal model ({\tt APEC}).
We also derived X-ray upper limits for the optical SNRs of LL14, which
are listed in Table \ref{osnrul}.
For the flux calculation based on count rates, we assumed the same thermal
spectrum as used by SPH12, i.e., with $kT$ = 0.2 keV, absorbed by a 
column density of \nh\ = $7 \times 10^{20}$ cm$^{-2}$ for the Milky Way
and an additional column density of \nh\ = $1 \times 10^{21}$ cm$^{-2}$.

\begin{table*}
\caption{
\label{snrfitlst}
Spectral fit parameters for the best fit models for the three brightest SNRs in the FOVs of the \xmm\ LP.
}
\centering

\begin{tabular}{rrccccc}
\hline\hline
ID & SPH11 ID & $N_{\rm H}$ ({\tt PHABS2}) & $kT_1$ ({\tt APEC}) & $kT_2$ ({\tt APEC}) & Red.\ $\chi^2$ & DOF \\
 &  & [$10^{21}$~cm$^{-2}$] & [keV] & [keV] & & \\
\hline
  24& 1234 & 0.0 (0.0 -- 0.5) & 0.18 (0.16 -- 0.21) & 0.80 (0.72 -- 0.91) & 1.0 & 60 \\
  38& 1275 & 0.0 (0.0 -- 0.1) & 0.19 (0.17 -- 0.22) & 0.75 (0.64 -- 0.87) & 1.2 & 76 \\
 242& 1535 & 0.0 (0.0 -- 0.6) & 0.19 (0.18 -- 0.21) & 3.5 (2.6 -- 5.1) & 1.1 & 56 \\
 300& 1599 & 4.1 (3.4 -- 4.8) & 0.24 (0.22 -- 0.27) & & 1.1 & 83\\
\hline\hline
\end{tabular}
\end{table*}

\subsection{Source 24 ([SPH11] 1234, J004327.8+411829)}

This is the brightest X-ray SNR in \mth\ and is also confirmed
in the optical and radio.
The X-ray SNR was spatially resolved with \chandra\
\citep{2002ApJ...580L.125K}. [SPH11] 1234 was located at the edge
of the FOV of observation 1 and was also only observed with pn.
The new spectrum can be fitted well with two {\tt APEC} components
(see Table \ref{snrfitlst}).
The two components have temperatures of 0.18 keV and 0.8 keV, indicating
a mix of different plasma components. Neither a fit with a 
single non-equilibrium ionisation (NEI) model nor with a thermal model 
combined with a power law yielded a reasonable result. 
The two temperature spectrum is consistent with a young SNR,
with the lower-temperature component corresponding to ISM emission
and the higher-temperature component to that of ejecta.
Its unabsorbed flux is 
$9.6 \times 10^{-14}$ erg s$^{-1}$ cm$^{-2}$ (0.35 -- 2.0 keV) or
$1.1 \times 10^{-13}$ erg s$^{-1}$ cm$^{-2}$ (0.3 -- 10.0 keV), 
corresponding to an intrinsic luminosity of 
$7.2 \times 10^{36}$ erg s$^{-1}$ cm$^{-2}$ (0.35 -- 2.0 keV).
This source is therefore significantly fainter than the brightest 
SNRs in the Magellanic Clouds
\citep{2004A&A...421.1031V,2008A&A...485...63F,2016A&A...585A.162M}.

\subsection{Source 38 ([SPH11] 1275, J004339.1+412653)}

The SNR [SPH11] 1275 was also detected in the optical and radio. 
The optical narrow band 
images of LGS and the [\sii]/\ha\ image show an extended, but rather compact
source (SPH12). The spectrum of [SPH11] 1275
can be fit with two thermal components ({\tt APEC}s). 
A power law component instead of the harder thermal component does not improve
the quality of the fit.
This source also seems to show both ISM and ejecta components.
Its unabsorbed flux is 
$2.5 \times 10^{-14}$ erg s$^{-1}$ cm$^{-2}$ (0.35 -- 2.0 keV) or
$2.6 \times 10^{-14}$ erg s$^{-1}$ cm$^{-2}$ (0.3 -- 10.0 keV), 
corresponding to an intrinsic luminosity of 
$1.9 \times 10^{36}$ erg s$^{-1}$ cm$^{-2}$ (0.35 -- 2.0 keV).

\subsection{Source 242 ([SPH11] 1535, J004451.0+412906)}

SPH12 have reported about the detection of two 
stars at the position of this SNR, one detected in $V$ and one in $I$. 
Therefore, depending on the nature of these two optical point sources, 
the X-ray emission might be contaminated.
In the LGS narrow band images, one can see a circular shell source. The
[\sii]/\ha\ image also suggests the optical shell to be an SNR.
The X-ray spectrum of [SPH11] 1535 is well fit with a thermal model
consisting of two {\tt APEC} components (Table \ref{snrfitlst}).
The temperatures are $kT$ = 0.2 keV and 3.5 keV. 
The harder component can also be well modelled with a  power-law with a
photon index of $\Gamma = 2.4 (-0.4,+0.3)$, while the absorption and the
softer component are the same as in the fit with a hot thermal component.
The lower temperature
is consistent to what has been derived for [SPH11] 1234 and 1599.
Most likely this component corresponds to shocked ISM.
The additional harder component either corresponds to ejecta emission
with a relatively high temperature of 3.5 keV or 
it might indicate the existence of a pulsar wind nebula.
The unabsorbed flux of this source is 
$1.6 \times 10^{-14}$ erg s$^{-1}$ cm$^{-2}$ (0.35 -- 2.0 keV) or
$2.3 \times 10^{-13}$ erg s$^{-1}$ cm$^{-2}$ (0.3 -- 10.0 keV).
At the distance of \mth, this corresponds to an unabsorbed luminosity of
$1.2 \times 10^{36}$ erg s$^{-1}$ cm$^{-2}$  in the energy range of
0.35 -- 2.0 keV.

\subsection{Source 300 ([SPH11] 1599, J004513.8+413615)}\label{spec300}

The spectrum of the SNR [SPH11] 1599 is well fit with one {\tt APEC} component.
The temperature is $kT$ = 0.24 keV, similar to the lower temperatures of
the two components for [SPH11] 1234 and 1535. Therefore,
its emission seems to be dominated by that of the shocked ISM. The
intrinsic absorption is high (\nh\ = $4.1 \times 10^{21}$ cm$^{-2}$) 
and indicates that the SNR is embedded in higher density gas.
The unabsorbed flux of this source is 
$2.1 \times 10^{-13}$ erg s$^{-1}$ cm$^{-2}$ (0.35 -- 2.0 keV) or
$2.4 \times 10^{-13}$ erg s$^{-1}$ cm$^{-2}$ (0.3 -- 10.0 keV), 
which corresponds to an intrinsic luminosity of
$1.6 \times 10^{37}$ erg s$^{-1}$ in the energy range of 0.35 -- 2.0 keV,
making it similarly bright as the brightest SNR in the Magellanic Clouds or
M\,33 \citep{2004A&A...421.1031V,2008A&A...485...63F,2010ApJS..187..495L,2016A&A...585A.162M,
2017MNRAS.472..308G}.

\section{Summary}

We carried out deep \xmm\ observations of two fields in the northern disc
of \mth. We obtained a list of 389 sources through a combined analysis of
all four observations. The new catalogue of \xmm\ sources in the northern 
disc of \mth\ contains 43 foreground stars and candidates, 50 background 
sources, 34 XRBs and candidates, and 18 SNRs and candidates.
\begin{itemize}
\item Based on the \xmm\ spectra of sources for which an optical counterpart 
was found in the PHAT data in the Chandra/PHAT survey, 
we find higher \nh\ for sources which are 
likely background galaxies. Optical photometry combined with X-ray spectral
analysis allows us to distinguish between
\mth\ and background sources.
\item 55 sources show variability between the two epochs. 
154 sources, including bright objects, are only detected in one epoch each and can be transient sources. However, there are also many faint sources with low detection likelihood among these "single-detection" objects.
We have identified sources with interesting variability in their light curves: flares (sources 141 and 153), likely flares (source 168), and dips (sources 255 and 46).
Based on Principal Component Analysis, we identify 
12 sources as variable.
\item We have detected the transient source 
SWIFT J004420.1+413702 and show that its flux changes
significantly from June 2015 to January 2016. We analysed its spectrum
at different epochs and show that the change in flux is correlated with
a change in absorption. This source is classified as a low-mass X-ray binary,
since a low-mass star was detected in the PHAT data.
\item The cumulative XLFs calculated for the two fields in two energy bands
(0.5 -- 2.0 keV and 2.0 -- 10.0 keV) show that a higher number of 
sources are detected in the field closer to the nuclear region. In this
field also more luminous sources are found.
We have detected \mth\ sources down to luminosities of 
$\sim7\times10^{34}$ erg s$^{-1}$ in the softer band (0.5 -- 2.0 keV) and
to $\sim4\times10^{35}$ erg s$^{-1}$ in the harder band (2.0 -- 10.0 keV).
\item We performed spectral analysis of the four brightest SNRs. All four
sources indicates a shocked ISM component with a temperature of
$kT \approx 0.2$ keV. Two SNRs have an additional thermal component
with a temperature of $kT \approx 0.8$ keV, which might correspond to
ejecta emission.
\item Source 242 ([SPH11] 1535, J004451.0+412906) has a significantly harder
X-ray spectrum with emission in addition to the shocked ISM
component, which can either be modelled as that of a thermal plasma
with a temperature of $kT = 3.5$ keV or with a power-law with a
photon index of $\Gamma = 2.4$. This source might harbour a pulsar wind
nebula and requires further investigation.
\item We identified five sources as new X-ray SNRs based on their hardness ratios and optical counterpart. Many of the detected SNR are very faint in X-rays 
with fluxes of $<10^{-15}$ erg s$^{-1}$ cm$^{-2}$, corresponding to luminosities of 
$<10^{34}$ erg s$^{-1}$ in the energy band of 0.35 -- 2.0 keV. 
We thus have a sample of SNRs in the northern disc with $\sim$ 2 times the 
sensitivity of the SNR population obtained from the XMM M31 survey (SPH12). 
A deep coverage of the entire \mth\ galaxy like in our
new study of the northern disc will allow us to study the complete sample
of faint X-ray SNRs, giving information about environments in which 
SNRs originate, their interaction with the ambient ISM, and  how long SNRs 
survive in the ISM.
A detailed analysis of the ISM using the new \xmm\ data will be presented in
Kavanagh et al. (in prep.).
\end{itemize}

\begin{acknowledgements}
This research was funded by the German Bundesministerium f\"ur Wirtschaft und 
Technologie and the Deutsches Zentrum f\"ur Luft- und Raumfahrt (BMWi/DLR) 
through the grant FKZ 50 OR 1510.
M.S.\ acknowledges support by the Deutsche Forschungsgemeinschaft (DFG)
through the Heisenberg fellowship SA 2131/3-1, the Heisenberg research
grant SA 2131/4-1, and the Heisenberg professor grant SA 2131/5-1.
M.S., B.W., and P.P.\ acknowledge partial support for this research through 
the Chandra Research Visitors Program.
M.H.\ acknowledges the support of the Spanish Ministry of Economy and Competitiveness (MINECO) under the grant FDPI-2013-16933 as well as the support of the Generalitat de Catalunya/CERCA programme.
Support for B.F.W.\ was provided by Chandra Award Number GO5-16085X issued by the Chandra X-ray Observatory Center, which is operated by the Smithsonian Astrophysical Observatory for and on behalf of the National Aeronautics and Space Administration under contract NAS8-03060.
D.H., A.K., and K.S.\ are supported by the European Space Agency (ESA) under the ``Hubble Catalog of Variables'' programme, contract No. 4000112940.
D.B.\ acknowledges partial support from the DFG through project ISM-SPP 1573.
This research has made use of the the SIMBAD database, and the VizieR catalogue 
access tool operated at CDS, Strasbourg, France.
\end{acknowledgements}

\bibliographystyle{aa} 
\bibliography{refs}

\begin{appendix}
\setcounter{section}{1}

\onecolumn

\include{sources}

\onecolumn

\include{spec_m31}

\include{spec_bkg}

\twocolumn

\label{sec:tab}

\include{var12}

\include{var34}

\include{onlyonce}

\end{appendix}
\end{document}

%% file: sources.tex
\onecolumn

\begin{landscape}
{\tiny
\begin{longtable}{rllrlrrrrrrrll}
\caption{Sources detected in the \xmm\ LP of the northern disc of M\,31}
\label{soutab}
\\
\hline\hline
ID & RA (J2000.0) & Dec (J2000.0) & PosErr & Chandra$^{\dagger}$ & DetML & Cts & Rate [cts/s] & HR1 & HR2 & HR3 & HR4 & [SPH11] ID & Class \\
\hline
\endfirsthead

\multicolumn{14}{c}%
{\tablename\ \thetable{} -- continued from previous page} \\
\hline
ID & RA (J2000.0) & Dec (J2000.0) & PosErr & Chandra$^{\dagger}$ & DetML & Cts & Rate [cts/s] & HR1 & HR2 & HR3 & HR4 & [SPH11] ID & Class \\
\hline
\endhead

\hline 
\multicolumn{14}{l}{{\tiny $^{\dagger}$ Chandra source (Williams et al., 2018, accepted).}} \\ 
\multicolumn{14}{l}{{\tiny $^{\star}$ Based on Chandra/PHAT (Williams et al., 2018, accepted).}} \\ 
\multicolumn{14}{l}{{\tiny Continued on next page}} \\ 
\endfoot

\hline 
\multicolumn{14}{l}{{\tiny $^{\dagger}$ Chandra source (Williams et al., 2018, accepted).}} \\ 
\multicolumn{14}{l}{{\tiny $^{\star}$ Based on Chandra/PHAT (Williams et al., 2018, accepted).}} \\ 
\endlastfoot

  1 & 00:42:59.031 & +41:31:29.28 & 3.65\arcsec &   &      6.73 &    65.93$\pm$18.99 &0.003131 & 0.61$\pm$0.33 &-0.09$\pm$0.30 &-1.00$\pm$0.49 & 1.00$\pm$0.28 &  -- &   \\
  2 & 00:42:59.045 & +41:34:21.28 & 2.72\arcsec &   &     11.77 &    68.09$\pm$16.69 &0.002908 & 1.00$\pm$0.59 & 0.58$\pm$0.29 & 0.08$\pm$0.23 &-0.41$\pm$0.57 &1098 & $<$hard$>$    \\
  3 & 00:43:01.335 & +41:30:16.08 & 0.58\arcsec &   &   4529.49 &  3209.88$\pm$71.59 &0.145200 & 0.43$\pm$0.05 & 0.39$\pm$0.03 & 0.12$\pm$0.03 & 0.01$\pm$0.03 &1109 & $<$XRB$>$ GlC        \\
  4 & 00:43:08.287 & +41:24:58.19 & 1.71\arcsec &   &      8.82 &    81.45$\pm$19.57 &0.004602 & 0.65$\pm$0.43 &-0.63$\pm$0.51 & 0.62$\pm$0.52 & 0.68$\pm$0.17 &  -- &   \\
  5 & 00:43:08.592 & +41:30:22.97 & 1.54\arcsec &   &     19.87 &   124.01$\pm$20.67 &0.004857 & 0.40$\pm$0.25 & 0.08$\pm$0.22 &-0.22$\pm$0.25 & 0.18$\pm$0.29 &1141 & $<$hard$>$    \\
  6 & 00:43:14.552 & +41:25:13.06 & 1.43\arcsec &   &     27.43 &   105.81$\pm$18.61 &0.003972 & 0.51$\pm$0.31 & 0.37$\pm$0.17 &-0.50$\pm$0.20 &-1.00$\pm$1.06 &1173 & $<$XRB$>$ GlC        \\
  7 & 00:43:14.967 & +41:20:15.03 & 2.36\arcsec &   &      9.62 &    89.87$\pm$21.81 &0.004282 & 0.09$\pm$0.44 &-0.26$\pm$0.52 & 0.62$\pm$0.30 &-0.07$\pm$0.37 &  -- &   \\
  8 & 00:43:17.760 & +41:39:14.70 & 1.50\arcsec &   &     17.90 &    88.52$\pm$18.78 &0.003451 & 0.84$\pm$0.32 & 0.22$\pm$0.26 & 0.15$\pm$0.21 &-0.32$\pm$0.50 &1185 & $<$hard$>$    \\
\hline\hline
\end{longtable}
}
\end{landscape}

\twocolumn

%% file: spec_m31.tex
\begin{table}
{\footnotesize
\caption{Spectral fit results of \mth\ \xmm\ LP sources which are most likely located in \mth
\label{spec-result-m31}}
\begin{tabular}{rrllllcc}
\hline\hline
Source No & \xmm\  & $N_\mathrm{H, Gal}^{*}$ &Intrinsic $N_\mathrm{H, \mth}^{*}$& Photon-Index &  $\chi^2$ (d.o.f) & OBS & Instrument\\
 & coordinates &$10^{21}$ cm$^{-2}$&$10^{21}$ cm$^{-2}$&&&&\\
\hline 
33  & J004336.0+413319 &0.7 frozen  &$4.96^{+4.2}_{-2.8}$&$2.25^{+0.94}_{-0.66}$ &1.94\,(19) & OBS1, OBS2& pn$^{\dagger}$, MOS2$^{\dagger}$ $^{\bigstar}$    \\
47  & J004345.4+413657 &$0.7^{+0.6}_{-0.5}$ && $2.10^{+0.39}_{-0.29}$& 1.12\,(72)& OBS1, OBS2& pn$^{\dagger}$  \\
60  & J004350.8+412116 &0.7 frozen& $1.3^{+0.9}_{-0.7}$ & $1.88^{+0.30}_{-0.30}$ &0.99\,(71) & OBS1, OBS2 & pn$^{\dagger}$  \\
74  & J004356.4+412202 & $0.2^{+0.6}_{-0.2}$ && $1.78^{+0.32}_{-0.21}$ &1.17\,(39)& OBS1 & pn, MOS1, MOS2$^{\bigstar}$   \\
75  & J004356.6+413410 &0.7 frozen &$4.1^{+15.6}_{-3.7}$ & $2.93^{+2.92}_{-1.52}$ &1.40\,(20) & OBS1, OBS2 &pn$^{\dagger}$    \\
83  & J004357.4+413056 &0.7 frozen &$1.2^{+3.2}_{-2.7}$ & $1.94^{+0.32}_{-0.27}$ &0.76\,(43) &OBS1, OBS2&   pn$^{\dagger}$, MOS1$^{\dagger}$, MOS2$^{\dagger}$ $^{\bigstar}$  \\
\hline
\hline
\end{tabular}

{\footnotesize $^*$: The spectra were first fit with one absorption component. If the fitted value was significantly higher than the Galactic foreground \nh, an additional absorption component was introduced. In this case, the first component was fixed to the Galactic foreground \nh, and the second component was used to model the additional \nh.\\
$^\dagger$: EPIC data of two observations were combined.\\
$^{\star}$: Simultaneous fit was performed using all EPIC data.\\}
}
\end{table}

%% file: spec_bkg.tex
\begin{table}
{\footnotesize
\caption{Spectral fit results of \mth\ \xmm\ LP sources which are most likely background sources
\label{spec-result-bkg} }
\begin{tabular}{rrllllcc}
\hline\hline
Source No & \xmm\ &  $N_\mathrm{H, Gal}^{*}$ &$N_\mathrm{H, \mth}^{*}$& Photon-Index &  $\chi^2$ (d.o.f) & OBS & Instrument\\
 & coordinates &$10^{21}$ cm$^{-2}$&$10^{21}$ cm$^{-2}$&&&&\\
\hline 
16   & J004323.5+413145 &$1.2^{+0.9}_{-0.7}$ && $1.69^{+0.36}_{-0.30}$& 1.31\,(33)& OBS2& pn, MOS, MOS2$^{\bigstar}$\\
43   & J004342.9+412850 &0.7 frozen  &$1.8^{+1.0}_{-0.9}$&$1.67^{+0.25}_{-0.22}$ &0.71\,(55) &OBS1, OBS2& pn$^{\dagger}$, MOS1$^{\dagger}$, MOS2$^{\dagger}$ $^{\bigstar}$  \\
58   & J004350.3+413247 &0.7 frozen & $7.3^{10.0}_{-5.7}$ & $2.37^{+2.16}_{-1.10}$ & 0.69\,(16)& OBS1, OBS2 & pn$^{\dagger}$ \\
96   & J004402.6+413926 &0.7 frozen&$4.1^{+1.6}_{-1.3}$&2.75$^{+0.52}_{-0.42}$ &1.05\,(46) &OBS1, OBS2&   pn$^{\dagger}$, MOS2$^{\dagger}$ $^{\bigstar}$ \\
100  & J004403.6+412415 &0.7 frozen & $12.4^{+26.9}_{-10.9}$ &1.95$^{+2.25}_{-1.36}$ &1.17\,(5) & OBS1, OBS2 & pn, MOS2$^{\bigstar}$   \\
105  & J004404.7+412126 &$1.6^{+3.1}_{-1.2}$ & & $0.78^{+0.39}_{-0.34}$  & 0.83\,(32) & OBS1, OBS2 & pn$^{\dagger}$, MOS2$^{\dagger}$ $^{\bigstar}$    \\
138  & J004415.9+413057 &0.7 frozen&$14.0^{+2.0}_{-2.2}$ & $1.66^{+0.15}_{-0.14}$&1.24\,(108)&OBS1 &pn, MOS2$^{\bigstar}$     \\
144  & J004418.9+413212 &0.7 frozen& $18.9^{+11.3}_{-12.9}$& 3.36$^{+1.92}_{-3.01}$& 1.39\,(8) & OBS1 & pn, MOS1$^{\bigstar}$    \\
148  & J004421.1+414657 &$<$0.8 &&$ 1.40^{+0.29}_{-0.27}$ &1.69\,(18) & OBS3 & pn, MOS1$^{\bigstar}$   \\
176  & J004430.2+412307 & 0.7 frozen & $11.9^{+7.7}_{-5.4}$ & $1.89^{+0.63}_{-0.52}$ &0.61\,(30)& OBS1, OBS2 & pn$^{\dagger}$    \\
185  & J004432.1+415505 & $<38.$ && $0.49^{+2.57}_{-0.97}$& 1.12\,(3)& OBS3 & pn    \\
193  & J004435.1+414733 & $1.21^{+2.7}_{-1.2}$ && $1.44^{+0.58}_{-0.45}$ & 1.26\,(13)& OBS3,OBS4& pn, MOS1, MOS2$^{\bigstar}$ $^{\dagger}$    \\
197  & J004437.0+411951 &0.7 frozen &$16.3^{+18.4}_{-10.5}$ & $1.84^{+1.48}_{-1.00}$&  1.28\,(22)& OBS1, OBS2&  pn$^{\dagger}$   \\
210  & J004442.6+415340 & 0.7 frozen &$186.2^{+51.6}_{-42.0}$ &$2.37^{+0.64}_{-0.55}$ & 0.91\,(71)&  OBS3, OBS4 & pn$^{\dagger}$   \\
219  & J004444.9+415154 & 0.7 frozen & $3.2^{+2.1}_{-1.6}$& $2.10^{+0.43}_{-0.36}$ &1.065\,(54) & OBS3, OBS4 & pn$^{\dagger}$, MOS1$^{\dagger}$, MOS2$^{\dagger}$ $^{\bigstar}$ \\
232  & J004448.7+415720 & 0.7 frozen&$8.8^{+13.5}_{-6.4}$&$1.98^{+1.85}_{-1.07}$ &0.78\,(6) &OBS3&  pn  \\
237  & J004449.6+415139 & 0.7 frozen&$19.6^{+21.0}_{-12.4}$&$2.58^{+1.51}_{-1.01}$ &0.33\,(10) &OBS3&   pn  \\
254  & J004456.0+414830 &<66.&&<4.21 &0.39\,(5) &OBS4&   pn, MOS1$^{\bigstar}$  \\
260  & J004458.1+414625 &0.7 frozen &$2.04^{+1.0}_{-0.9}$ &$1.99^{+0.28}_{-0.25}$&1.13\,(84)&OBS3, OBS4 & pn$^{\dagger}$, MOS1$^{\dagger}$, MOS2$^{\dagger}$ $^{\bigstar}$   \\
270  & J004501.2+415609 &0.7 frozen&$36.64^{+57.9}_{-23.6}$&$3.04^{+3.37}_{-1.67}$ &093\,(7) &OBS3&   pn  \\
272  & J004502.3+414943 &<37.&&$0.98^{+1.97}_{-0.54}$ &1.66\,(11) &OBS3, OBS4&   pn$^{\dagger}$ \\
307  & J004516.1+412302 &<6.8&&$0.85^{+0.80}_{-0.51}$ &1.38\,(13) &OBS1&   pn  \\
311  & J004518.6+413936 & 0.7 frozen &$9.9^{+9.4}_{-6.9}$ &$1.68^{+0.80}_{-0.69}$ & 0.65\,(28) & OBS3,OBS4 & pn$^{\dagger}$    \\
328  & J004526.8+413217 & 0.7 frozen& $2.0^{+1.5}_{-1.2}$ & $2.02^{+0.61}_{-0.46}$& 0.45\,(10) & OBS2& pn      \\
329  & J004527.3+413253 &0.7 frozen& $1.2^{+0.7}_{-0.6}$&$1.37^{+0.20}_{-0.18}$ &0.61\,(39) & OBS1, OBS2 & pn$^{\dagger}$   \\
334  & J004528.8+415015 &0.7 frozen&$7.9^{+10.5}_{-5.1}$&$1.51^{+0.82}_{-0.61}$ & 0.73\,(9) &OBS3&   pn  \\
339  & J004530.5+413600 &0.7 frozen &$4.8^{+7.9}_{-3.7}$ &$1.93^{+1.89}_{-1.00}$ &1.63\,(11) & OBS2& pn    \\
343  & J004532.1+414526 & 0.7 frozen& $5.4^{+7.2}_{-4.5}$ &$2.27^{+1.41}_{-0.90}$ & 1.04\,(18) &OBS3,OBS4& pn, MOS2 $^{\bigstar}$   \\
358  & J004537.6+415122 & <6.9 && $1.77^{+1.39}_{-0.88}$ & 1.33\,(31) & OBS3, OBS4 & pn$^{\dagger}$   \\
360  & J004538.0+414857 & <511. && <3.03 & 0.97\, (1) & OBS3& pn   \\
371  & J004544.8+415859 & <2.4 && $1.38^{+0.73}_{-0.30}$ &0.75\,(4) & OBS3&pn     \\
379  & J004555.1+415647 &<11.&&$1.08^{+1.54}_{-0.49}$ &1.88\,(6) &OBS4& pn  \\
380  & J004555.4+415212 & 0.7 frozen & $17.09^{+38.47}_{-12.84}$ & $1.73^{+1.96}_{-1.09}$ & 0.83\,(11) & OBS3, OBS4& pn$^{\dagger}$  \\
382  & J004557.0+414831 &0.7 frozen&$3.98^{+0.82}_{-0.73}$&$1.43^{+0.14}_{-0.13}$ &1.00\,(157) &OBS3, OBS4&   pn$^{\dagger}$, MOS2$^{\dagger}$ $^{\bigstar}$ \\
\hline
\hline
\end{tabular}

{\footnotesize $^*$: The spectra were first fit with one absorption component. If the fitted value was significantly higher than the Galactic foreground \nh, an additional absorption component was introduced. In this case, the first component was fixed to the Galactic foreground \nh, and the second component was used to model the additional \nh.\\
$^\dagger$: EPIC data of two observations were combined.\\
$^{\star}$: Simultaneous fit was performed using all EPIC data.\\}
}
\end{table}

%% file: var12.tex
\onecolumn

\captionsetup{width=\textwidth}

\begin{longtable}{rrrrrrrrrr}
\caption{Variability for Field 1 (OBS1 vs.\ OBS2).
We list the source ID, EPIC count rates and corresponding $1\sigma$ errors in each observation (OBS1 \& OBS2).
The table also includes the mean count rate, the difference in count rate between OBS1 and OBS2, 
the variability in percent of the count rate difference over the mean count rate, and the significance of
the variability.}
\label{tab:var_south}
\\
\hline\hline
ID & Rate1 & Rate2 & Mean Rate & Difference & Error1 & Error2 & Variability & Signicance & Significantly\\
& \multicolumn{2}{c}{[ct s$^{-1}$]} & [ct s$^{-1}$] & [ct s$^{-1}$] & \multicolumn{2}{c}{[ct s$^{-1}$]} & [\%] & & variable? \\ \hline \noalign{\smallskip}
\endfirsthead

\multicolumn{10}{c}%
{\tablename\ \thetable{} -- continued from previous page} \\
\hline
ID & Rate1 & Rate2 & Mean Rate & Difference & Error1 & Error2 & Variability & Signicance & Significantly\\
& \multicolumn{2}{c}{[ct s$^{-1}$]} & [ct s$^{-1}$] & [ct s$^{-1}$] & \multicolumn{2}{c}{[ct s$^{-1}$]} & [\%] & & variable? \\ \hline \noalign{\smallskip}
\hline
\endhead

\hline \multicolumn{10}{l}{{\footnotesize Continued on next page}} \\
\endfoot

\hline
\endlastfoot

    6 &   0.00397 &   0.00294 &   0.00346 &    0.00103 &   0.00076 &   0.00067 &     29 &    1.0 &     no \\
   13 &   0.00241 &   0.00224 &   0.00232 &    0.00016 &   0.00058 &   0.00055 &      7 &    0.2 &     no \\
   16 &   0.00320 &   0.02470 &   0.01390 &   --0.02150 &   0.00058 &   0.00139 &   --154 &   14.3 &    yes \\
   17 &   0.00404 &   0.00421 &   0.00413 &   --0.00017 &   0.00066 &   0.00090 &     --4 &    0.1 &     no \\
   18 &   0.00632 &   0.00682 &   0.00657 &   --0.00050 &   0.00076 &   0.00102 &     --7 &    0.4 &     no \\
   20 &   0.00350 &   0.00446 &   0.00398 &   --0.00096 &   0.00076 &   0.00089 &    --24 &    0.8 &     no \\
   23 &   0.00283 &   0.00435 &   0.00359 &   --0.00152 &   0.00055 &   0.00077 &    --42 &    1.6 &     no \\
   25 &   0.00462 &   0.00446 &   0.00454 &    0.00017 &   0.00067 &   0.00086 &      3 &    0.2 &     no \\
   27 &   0.00574 &   0.00500 &   0.00537 &    0.00074 &   0.00084 &   0.00112 &     13 &    0.5 &     no \\
   31 &   0.00250 &   0.00367 &   0.00308 &   --0.00117 &   0.00054 &   0.00072 &    --37 &    1.3 &     no \\
   33 &   0.00550 &   0.00609 &   0.00579 &   --0.00059 &   0.00059 &   0.00078 &    --10 &    0.6 &     no \\
   36 &   0.00350 &   0.00295 &   0.00323 &    0.00055 &   0.00054 &   0.00073 &     17 &    0.6 &     no \\
   37 &   0.00518 &   0.03020 &   0.01770 &   --0.02500 &   0.00082 &   0.00176 &   --141 &   12.9 &    yes \\
   38 &   0.03080 &   0.03120 &   0.03100 &   --0.00041 &   0.00108 &   0.00126 &     --1 &    0.2 &     no \\
   39 &   0.00271 &   0.00614 &   0.00442 &   --0.00343 &   0.00044 &   0.00152 &    --77 &    2.2 &     no \\
   41 &   0.00187 &   0.00284 &   0.00235 &   --0.00097 &   0.00039 &   0.00075 &    --41 &    1.1 &     no \\
   42 &   0.00586 &   0.00548 &   0.00567 &    0.00038 &   0.00069 &   0.00080 &      6 &    0.4 &     no \\
   43 &   0.01400 &   0.01790 &   0.01600 &   --0.00394 &   0.00077 &   0.00100 &    --24 &    3.1 &    yes \\
   47 &   0.02190 &   0.00841 &   0.01510 &    0.01350 &   0.00084 &   0.00087 &     88 &   11.2 &    yes \\
   48 &   0.00349 &   0.00311 &   0.00330 &    0.00038 &   0.00050 &   0.00058 &     11 &    0.5 &     no \\
   50 &   0.00728 &   0.00443 &   0.00586 &    0.00285 &   0.00056 &   0.00186 &     48 &    1.5 &     no \\
   53 &   0.07760 &   0.04700 &   0.06230 &    0.03060 &   0.00138 &   0.00131 &     49 &   16.1 &    yes \\
   57 &   0.00682 &   0.00347 &   0.00515 &    0.00335 &   0.00086 &   0.00126 &     65 &    2.2 &     no \\
   58 &   0.00404 &   0.00350 &   0.00377 &    0.00054 &   0.00063 &   0.00064 &     14 &    0.6 &     no \\
   60 &   0.02770 &   0.02170 &   0.02470 &    0.00592 &   0.00133 &   0.00143 &     23 &    3.0 &    yes \\
   63 &   0.00734 &   0.00347 &   0.00540 &    0.00387 &   0.00067 &   0.00068 &     71 &    4.0 &    yes \\
   68 &   0.00470 &   0.00564 &   0.00517 &   --0.00094 &   0.00064 &   0.00160 &    --18 &    0.5 &     no \\
   72 &   0.00278 &   0.00305 &   0.00291 &   --0.00027 &   0.00034 &   0.00047 &     --9 &    0.5 &     no \\
   74 &   0.02220 &   0.01730 &   0.01980 &    0.00484 &   0.00103 &   0.00135 &     24 &    2.9 &     no \\
   75 &   0.00184 &   0.00214 &   0.00199 &   --0.00030 &   0.00031 &   0.00043 &    --15 &    0.6 &     no \\
   76 &   0.00386 &   0.00950 &   0.00668 &   --0.00564 &   0.00083 &   0.00071 &    --84 &    5.2 &    yes \\
   83 &   0.00565 &   0.01440 &   0.01000 &   --0.00877 &   0.00042 &   0.00081 &    --87 &    9.6 &    yes \\
   87 &   0.00404 &   0.00475 &   0.00439 &   --0.00071 &   0.00097 &   0.00116 &    --16 &    0.5 &     no \\
   89 &   0.00718 &   0.00604 &   0.00661 &    0.00114 &   0.00049 &   0.00058 &     17 &    1.5 &     no \\
   93 &   0.00528 &   0.00702 &   0.00615 &   --0.00174 &   0.00050 &   0.00091 &    --28 &    1.7 &     no \\
   96 &   0.00866 &   0.01200 &   0.01040 &   --0.00339 &   0.00059 &   0.00104 &    --32 &    2.8 &     no \\
  100 &   0.00408 &   0.00240 &   0.00324 &    0.00169 &   0.00049 &   0.00046 &     52 &    2.5 &     no \\
  103 &   0.00168 &   0.00302 &   0.00235 &   --0.00134 &   0.00028 &   0.00056 &    --56 &    2.1 &     no \\
  105 &   0.00866 &   0.00834 &   0.00850 &    0.00032 &   0.00088 &   0.00091 &      3 &    0.3 &     no \\
  113 &   0.00281 &   0.00448 &   0.00365 &   --0.00167 &   0.00039 &   0.00072 &    --45 &    2.0 &     no \\
  114 &   0.00111 &   0.00272 &   0.00191 &   --0.00161 &   0.00025 &   0.00051 &    --83 &    2.8 &     no \\
  118 &   0.00269 &   0.00238 &   0.00254 &    0.00031 &   0.00033 &   0.00040 &     12 &    0.6 &     no \\
  119 &   0.00865 &   0.00513 &   0.00689 &    0.00352 &   0.00071 &   0.00073 &     51 &    3.5 &    yes \\
  120 &   0.00184 &   0.00216 &   0.00200 &   --0.00031 &   0.00029 &   0.00043 &    --15 &    0.6 &     no \\
  127 &   0.02010 &   0.01780 &   0.01900 &    0.00223 &   0.00064 &   0.00094 &     11 &    2.0 &     no \\
  128 &   0.00241 &   0.00365 &   0.00303 &   --0.00123 &   0.00047 &   0.00059 &    --40 &    1.6 &     no \\
  131 &   0.00312 &   0.00340 &   0.00326 &   --0.00027 &   0.00034 &   0.00045 &     --8 &    0.5 &     no \\
  134 &   0.00665 &   0.00639 &   0.00652 &    0.00026 &   0.00065 &   0.00084 &      4 &    0.2 &     no \\
  138 &   0.04870 &   0.03850 &   0.04360 &    0.01020 &   0.00093 &   0.00110 &     23 &    7.1 &    yes \\
  140 &   0.00736 &   0.00803 &   0.00769 &   --0.00066 &   0.00045 &   0.00060 &     --8 &    0.9 &     no \\
  144 &   0.00188 &   0.00253 &   0.00220 &   --0.00064 &   0.00026 &   0.00056 &    --29 &    1.0 &     no \\
  145 &   0.00562 &   0.00432 &   0.00497 &    0.00130 &   0.00042 &   0.00052 &     26 &    1.9 &     no \\
  146 &   0.24500 &   0.04290 &   0.14400 &    0.20200 &   0.00217 &   0.00145 &    140 &   77.4 &    yes \\
  147 &   0.00338 &   0.00332 &   0.00335 &    0.00006 &   0.00034 &   0.00064 &      1 &    0.1 &     no \\
  149 &   0.00266 &   0.00224 &   0.00245 &    0.00042 &   0.00032 &   0.00039 &     17 &    0.8 &     no \\
  151 &   0.00649 &   0.05680 &   0.03170 &   --0.05040 &   0.00065 &   0.00632 &   --159 &    7.9 &    yes \\
  152 &   0.00175 &   0.00305 &   0.00240 &   --0.00130 &   0.00026 &   0.00045 &    --54 &    2.5 &     no \\
  157 &   0.02850 &   0.02370 &   0.02610 &    0.00476 &   0.00069 &   0.00181 &     18 &    2.5 &     no \\
  160 &   0.02960 &   0.02410 &   0.02680 &    0.00554 &   0.00082 &   0.00105 &     20 &    4.2 &    yes \\
  163 &   0.00711 &   0.00760 &   0.00736 &   --0.00049 &   0.00071 &   0.00079 &     --6 &    0.5 &     no \\
  166 &   0.00179 &   0.00231 &   0.00205 &   --0.00052 &   0.00040 &   0.00057 &    --25 &    0.7 &     no \\
  171 &   0.71700 &   0.60300 &   0.66000 &    0.11500 &   0.00466 &   0.00593 &     17 &   15.2 &    yes \\
  176 &   0.00526 &   0.00653 &   0.00589 &   --0.00127 &   0.00055 &   0.00078 &    --21 &    1.3 &     no \\
  177 &   0.00382 &   0.00332 &   0.00357 &    0.00050 &   0.00045 &   0.00062 &     14 &    0.7 &     no \\
  182 &   0.00594 &   0.00537 &   0.00566 &    0.00057 &   0.00101 &   0.00052 &     10 &    0.5 &     no \\
  184 &   0.00131 &   0.00174 &   0.00153 &   --0.00043 &   0.00033 &   0.00039 &    --27 &    0.8 &     no \\
  187 &   0.00103 &   0.00126 &   0.00115 &   --0.00023 &   0.00021 &   0.00033 &    --20 &    0.6 &     no \\
  197 &   0.00666 &   0.00673 &   0.00670 &   --0.00007 &   0.00076 &   0.00097 &     --1 &    0.1 &     no \\
  199 &   0.00209 &   0.01090 &   0.00650 &   --0.00882 &   0.00034 &   0.00186 &   --135 &    4.7 &    yes \\
  203 &   0.00678 &   0.00307 &   0.00492 &    0.00371 &   0.00040 &   0.00049 &     75 &    5.8 &    yes \\
  204 &   0.00473 &   0.00404 &   0.00439 &    0.00069 &   0.00040 &   0.00057 &     15 &    1.0 &     no \\
  206 &   0.00555 &   0.00377 &   0.00466 &    0.00178 &   0.00064 &   0.00066 &     38 &    1.9 &     no \\
  211 &   0.00343 &   0.00217 &   0.00280 &    0.00126 &   0.00033 &   0.00049 &     45 &    2.1 &     no \\
  223 &   0.00196 &   0.00277 &   0.00236 &   --0.00081 &   0.00034 &   0.00044 &    --34 &    1.5 &     no \\
  226 &   0.00336 &   0.00372 &   0.00354 &   --0.00036 &   0.00036 &   0.00048 &    --10 &    0.6 &     no \\
  227 &   0.00176 &   0.00295 &   0.00236 &   --0.00119 &   0.00052 &   0.00072 &    --50 &    1.3 &     no \\
  229 &   0.01210 &   0.00834 &   0.01020 &    0.00381 &   0.00116 &   0.00092 &     37 &    2.6 &     no \\
  242 &   0.01380 &   0.00999 &   0.01190 &    0.00381 &   0.00058 &   0.00084 &     32 &    3.7 &    yes \\
  243 &   0.00366 &   0.00257 &   0.00312 &    0.00109 &   0.00053 &   0.00064 &     34 &    1.3 &     no \\
  244 &   0.04270 &   0.05280 &   0.04770 &   --0.01000 &   0.00133 &   0.00156 &    --21 &    4.9 &    yes \\
  245 &   0.00269 &   0.00218 &   0.00244 &    0.00051 &   0.00038 &   0.00049 &     20 &    0.8 &     no \\
  249 &   0.00907 &   0.00211 &   0.00559 &    0.00696 &   0.00118 &   0.00048 &    125 &    5.5 &    yes \\
  250 &   0.01010 &   0.00898 &   0.00955 &    0.00113 &   0.00124 &   0.00097 &     11 &    0.7 &     no \\
  251 &   0.01440 &   0.01140 &   0.01290 &    0.00292 &   0.00086 &   0.00098 &     22 &    2.2 &     no \\
  257 &   0.05420 &   0.05790 &   0.05600 &   --0.00369 &   0.00143 &   0.00201 &     --6 &    1.5 &     no \\
  258 &   0.01540 &   0.01080 &   0.01310 &    0.00462 &   0.00083 &   0.00102 &     35 &    3.5 &    yes \\
  261 &   0.01040 &   0.00753 &   0.00896 &    0.00287 &   0.00074 &   0.00113 &     31 &    2.1 &     no \\
  264 &   0.00674 &   0.00773 &   0.00723 &   --0.00099 &   0.00078 &   0.00112 &    --13 &    0.7 &     no \\
  267 &   0.00823 &   0.00956 &   0.00890 &   --0.00133 &   0.00070 &   0.00223 &    --14 &    0.6 &     no \\
  268 &   0.00338 &   0.00656 &   0.00497 &   --0.00318 &   0.00050 &   0.00161 &    --63 &    1.9 &     no \\
  294 &   0.00267 &   0.00385 &   0.00326 &   --0.00118 &   0.00053 &   0.00096 &    --36 &    1.1 &     no \\
  299 &   0.00366 &   0.00833 &   0.00600 &   --0.00467 &   0.00057 &   0.00187 &    --77 &    2.4 &     no \\
  300 &   0.01960 &   0.02180 &   0.02070 &   --0.00211 &   0.00082 &   0.00143 &    --10 &    1.3 &     no \\
  306 &   0.00465 &   0.00376 &   0.00421 &    0.00089 &   0.00054 &   0.00084 &     21 &    0.9 &     no \\
  311 &   0.00359 &   0.01200 &   0.00779 &   --0.00840 &   0.00073 &   0.00192 &   --107 &    4.1 &    yes \\
  328 &   0.00546 &   0.02300 &   0.01420 &   --0.01750 &   0.00057 &   0.00180 &   --123 &    9.3 &    yes \\
  329 &   0.02790 &   0.09390 &   0.06090 &   --0.06610 &   0.00100 &   0.00309 &   --108 &   20.4 &    yes \\
  333 &   0.03530 &   0.12500 &   0.07990 &   --0.08930 &   0.00112 &   0.00336 &   --111 &   25.2 &    yes \\
  339 &   0.00551 &   0.01830 &   0.01190 &   --0.01280 &   0.00056 &   0.00370 &   --107 &    3.4 &    yes \\
\hline\hline
\end{longtable}

\twocolumn

%% file: var34.tex
\onecolumn

\begin{longtable}{rrrrrrrrrr}
\caption{Same as Table 
A.5 for Field 2 (OBS3 \& OBS4).
}
\label{tab:var_north}
\\
\hline\hline
ID & Rate3 & Rate4 & Mean Rate & Difference & Error3 & Error4 & Variability & Signicance & Significantly \\
& \multicolumn{2}{c}{[ct s$^{-1}$]} & [ct s$^{-1}$] & [ct s$^{-1}$] & \multicolumn{2}{c}{[ct s$^{-1}$]} & [\%] & & variable? \\ \hline \noalign{\smallskip}
\endfirsthead

\multicolumn{10}{c}%
{\tablename\ \thetable{} -- continued from previous page} \\
\hline
ID & Rate3 & Rate4 & Mean Rate & Difference & Error3 & Error4 & Variability & Signicance & Significantly \\
& \multicolumn{2}{c}{[ct s$^{-1}$]} & [ct s$^{-1}$] & [ct s$^{-1}$] & \multicolumn{2}{c}{[ct s$^{-1}$]} & [\%] & & variable? \\ \hline \noalign{\smallskip}
\hline
\endhead

\hline \multicolumn{10}{l}{{\footnotesize Continued on next page}} \\
\endfoot

\hline \\
\endlastfoot

   40 &   0.02130 &   0.00473 &   0.01300 &    0.01650 &   0.00102 &   0.00202 &    127 &    7.3 &    yes \\
   52 &   0.00532 &   0.00234 &   0.00383 &    0.00298 &   0.00067 &   0.00049 &     77 &    3.6 &    yes \\
   61 &   0.00313 &   0.00523 &   0.00418 &   --0.00209 &   0.00053 &   0.00117 &    --50 &    1.6 &     no \\
   73 &   0.00911 &   0.01400 &   0.01160 &   --0.00491 &   0.00067 &   0.00095 &    --42 &    4.2 &    yes \\
   79 &   0.00169 &   0.00264 &   0.00216 &   --0.00095 &   0.00038 &   0.00059 &    --43 &    1.3 &     no \\
   80 &   0.00182 &   0.00206 &   0.00194 &   --0.00024 &   0.00043 &   0.00055 &    --12 &    0.4 &     no \\
   85 &   0.00467 &   0.00685 &   0.00576 &   --0.00218 &   0.00057 &   0.00092 &    --37 &    2.0 &     no \\
   86 &   0.01200 &   0.01280 &   0.01240 &   --0.00083 &   0.00082 &   0.00118 &     --6 &    0.6 &     no \\
   93 &   0.00794 &   0.00399 &   0.00597 &    0.00395 &   0.00088 &   0.00059 &     66 &    3.7 &    yes \\
   96 &   0.01240 &   0.00462 &   0.00852 &    0.00781 &   0.00130 &   0.00064 &     91 &    5.4 &    yes \\
  101 &   0.00278 &   0.00263 &   0.00271 &    0.00016 &   0.00047 &   0.00054 &      5 &    0.2 &     no \\
  102 &   0.00236 &   0.00192 &   0.00214 &    0.00044 &   0.00037 &   0.00044 &     20 &    0.8 &     no \\
  104 &   0.00372 &   0.00237 &   0.00304 &    0.00135 &   0.00047 &   0.00064 &     44 &    1.7 &     no \\
  106 &   0.00198 &   0.00321 &   0.00260 &   --0.00123 &   0.00038 &   0.00065 &    --47 &    1.6 &     no \\
  108 &   0.01260 &   0.02000 &   0.01630 &   --0.00738 &   0.00066 &   0.00104 &    --45 &    6.0 &    yes \\
  113 &   0.00529 &   0.00180 &   0.00354 &    0.00349 &   0.00093 &   0.00044 &     98 &    3.4 &    yes \\
  116 &   0.01740 &   0.01860 &   0.01800 &   --0.00115 &   0.00127 &   0.00118 &     --6 &    0.7 &     no \\
  117 &   0.00200 &   0.00302 &   0.00251 &   --0.00101 &   0.00034 &   0.00067 &    --40 &    1.3 &     no \\
  121 &   0.00175 &   0.00452 &   0.00314 &   --0.00277 &   0.00049 &   0.00109 &    --88 &    2.3 &     no \\
  124 &   0.00725 &   0.00648 &   0.00686 &    0.00077 &   0.00056 &   0.00076 &     11 &    0.8 &     no \\
  128 &   0.00407 &   0.00226 &   0.00316 &    0.00182 &   0.00080 &   0.00058 &     57 &    1.9 &     no \\
  130 &   0.00454 &   0.00307 &   0.00380 &    0.00147 &   0.00096 &   0.00047 &     38 &    1.4 &     no \\
  133 &   0.00610 &   0.01170 &   0.00891 &   --0.00563 &   0.00053 &   0.00117 &    --63 &    4.4 &    yes \\
  135 &   0.00450 &   0.00826 &   0.00638 &   --0.00376 &   0.00049 &   0.00177 &    --58 &    2.0 &     no \\
  141 &   0.01070 &   0.00441 &   0.00757 &    0.00632 &   0.00051 &   0.00049 &     83 &    9.0 &    yes \\
  142 &   0.00725 &   0.00778 &   0.00751 &   --0.00053 &   0.00046 &   0.00084 &     --7 &    0.6 &     no \\
  146 &   0.20600 &   0.02640 &   0.11600 &    0.18000 &   0.00339 &   0.00112 &    154 &   50.4 &    yes \\
  148 &   0.00615 &   0.00555 &   0.00585 &    0.00060 &   0.00047 &   0.00069 &     10 &    0.7 &     no \\
  151 &   0.07890 &   0.08970 &   0.08430 &   --0.01080 &   0.00140 &   0.00172 &    --12 &    4.9 &    yes \\
  156 &   0.01070 &   0.01600 &   0.01330 &   --0.00528 &   0.00064 &   0.00104 &    --39 &    4.3 &    yes \\
  160 &   0.01450 &   0.00537 &   0.00993 &    0.00912 &   0.00100 &   0.00079 &     91 &    7.2 &    yes \\
  161 &   0.00419 &   0.00184 &   0.00302 &    0.00235 &   0.00061 &   0.00050 &     77 &    3.0 &     no \\
  167 &   0.00240 &   0.00207 &   0.00223 &    0.00032 &   0.00039 &   0.00041 &     14 &    0.6 &     no \\
  168 &   0.00402 &   0.00291 &   0.00347 &    0.00111 &   0.00032 &   0.00043 &     31 &    2.1 &     no \\
  169 &   0.00279 &   0.00250 &   0.00264 &    0.00030 &   0.00033 &   0.00047 &     11 &    0.5 &     no \\
  170 &   0.00155 &   0.00217 &   0.00186 &   --0.00062 &   0.00025 &   0.00034 &    --33 &    1.5 &     no \\
  177 &   0.00461 &   0.00284 &   0.00373 &    0.00177 &   0.00058 &   0.00053 &     47 &    2.3 &     no \\
  179 &   0.00559 &   0.00509 &   0.00534 &    0.00050 &   0.00037 &   0.00098 &      9 &    0.5 &     no \\
  180 &   0.00337 &   0.00223 &   0.00280 &    0.00114 &   0.00032 &   0.00042 &     40 &    2.2 &     no \\
  183 &   0.00317 &   0.00193 &   0.00255 &    0.00124 &   0.00040 &   0.00037 &     48 &    2.3 &     no \\
  186 &   0.00123 &   0.00158 &   0.00140 &   --0.00035 &   0.00026 &   0.00037 &    --25 &    0.8 &     no \\
  193 &   0.00648 &   0.00316 &   0.00482 &    0.00332 &   0.00040 &   0.00042 &     68 &    5.8 &    yes \\
  202 &   0.00315 &   0.00242 &   0.00279 &    0.00073 &   0.00055 &   0.00056 &     26 &    0.9 &     no \\
  207 &   0.00266 &   0.00468 &   0.00367 &   --0.00202 &   0.00032 &   0.00102 &    --54 &    1.9 &     no \\
  209 &   0.00285 &   0.00234 &   0.00260 &    0.00051 &   0.00035 &   0.00039 &     20 &    1.0 &     no \\
  210 &   0.01360 &   0.02400 &   0.01880 &   --0.01040 &   0.00052 &   0.00093 &    --55 &    9.8 &    yes \\
  212 &   0.00196 &   0.00183 &   0.00189 &    0.00013 &   0.00024 &   0.00036 &      7 &    0.3 &     no \\
  214 &   0.00461 &   0.00145 &   0.00303 &    0.00316 &   0.00059 &   0.00039 &    104 &    4.4 &    yes \\
  219 &   0.00500 &   0.00438 &   0.00469 &    0.00062 &   0.00034 &   0.00044 &     13 &    1.1 &     no \\
  222 &   0.00148 &   0.00171 &   0.00159 &   --0.00023 &   0.00023 &   0.00033 &    --14 &    0.6 &     no \\
  225 &   0.00347 &   0.00380 &   0.00363 &   --0.00032 &   0.00053 &   0.00049 &     --8 &    0.4 &     no \\
  228 &   0.00136 &   0.00350 &   0.00243 &   --0.00214 &   0.00028 &   0.00044 &    --88 &    4.1 &    yes \\
  230 &   0.00269 &   0.00222 &   0.00245 &    0.00047 &   0.00030 &   0.00039 &     19 &    0.9 &     no \\
  232 &   0.00172 &   0.00253 &   0.00213 &   --0.00081 &   0.00029 &   0.00058 &    --38 &    1.3 &     no \\
  233 &   0.00155 &   0.00304 &   0.00229 &   --0.00149 &   0.00032 &   0.00062 &    --65 &    2.1 &     no \\
  235 &   0.00509 &   0.00723 &   0.00616 &   --0.00213 &   0.00036 &   0.00048 &    --34 &    3.6 &    yes \\
  237 &   0.00280 &   0.00233 &   0.00256 &    0.00047 &   0.00027 &   0.00035 &     18 &    1.1 &     no \\
  246 &   0.00237 &   0.00192 &   0.00215 &    0.00045 &   0.00027 &   0.00041 &     20 &    0.9 &     no \\
  247 &   0.00249 &   0.00308 &   0.00278 &   --0.00059 &   0.00029 &   0.00038 &    --21 &    1.2 &     no \\
  250 &   0.00332 &   0.00195 &   0.00264 &    0.00137 &   0.00101 &   0.00050 &     52 &    1.2 &     no \\
  251 &   0.01040 &   0.00237 &   0.00639 &    0.00804 &   0.00119 &   0.00051 &    125 &    6.2 &    yes \\
  252 &   0.00241 &   0.00257 &   0.00249 &   --0.00016 &   0.00027 &   0.00044 &     --6 &    0.3 &     no \\
  254 &   0.00205 &   0.00245 &   0.00225 &   --0.00040 &   0.00025 &   0.00034 &    --17 &    0.9 &     no \\
  255 &   0.04250 &   0.02540 &   0.03390 &    0.01710 &   0.00110 &   0.00116 &     50 &   10.7 &    yes \\
  259 &   0.00531 &   0.00232 &   0.00381 &    0.00299 &   0.00046 &   0.00038 &     78 &    5.0 &    yes \\
  260 &   0.00941 &   0.00711 &   0.00826 &    0.00230 &   0.00047 &   0.00050 &     27 &    3.4 &    yes \\
  261 &   0.00845 &   0.00482 &   0.00663 &    0.00363 &   0.00074 &   0.00059 &     54 &    3.8 &    yes \\
  268 &   0.00356 &   0.00440 &   0.00398 &   --0.00084 &   0.00041 &   0.00049 &    --21 &    1.3 &     no \\
  273 &   0.00174 &   0.00380 &   0.00277 &   --0.00206 &   0.00027 &   0.00067 &    --74 &    2.9 &     no \\
  277 &   0.00215 &   0.00229 &   0.00222 &   --0.00014 &   0.00028 &   0.00055 &     --6 &    0.2 &     no \\
  285 &   0.00434 &   0.02220 &   0.01330 &   --0.01790 &   0.00040 &   0.00147 &   --134 &   11.7 &    yes \\
  290 &   0.01300 &   0.01920 &   0.01610 &   --0.00616 &   0.00078 &   0.00077 &    --38 &    5.6 &    yes \\
  300 &   0.01310 &   0.00330 &   0.00822 &    0.00983 &   0.00111 &   0.00048 &    119 &    8.1 &    yes \\
  304 &   0.00184 &   0.00239 &   0.00211 &   --0.00055 &   0.00036 &   0.00049 &    --26 &    0.9 &     no \\
  305 &   0.00586 &   0.00148 &   0.00367 &    0.00437 &   0.00067 &   0.00036 &    119 &    5.7 &    yes \\
  308 &   0.00522 &   0.00160 &   0.00341 &    0.00362 &   0.00077 &   0.00035 &    106 &    4.3 &    yes \\
  311 &   0.01040 &   0.01110 &   0.01070 &   --0.00075 &   0.00125 &   0.00095 &     --6 &    0.5 &     no \\
  312 &   0.00268 &   0.00207 &   0.00238 &    0.00060 &   0.00032 &   0.00043 &     25 &    1.1 &     no \\
  314 &   0.00381 &   0.00320 &   0.00351 &    0.00061 &   0.00071 &   0.00067 &     17 &    0.6 &     no \\
  317 &   0.00145 &   0.00325 &   0.00235 &   --0.00179 &   0.00030 &   0.00060 &    --76 &    2.7 &     no \\
  318 &   0.00572 &   0.00540 &   0.00556 &    0.00032 &   0.00045 &   0.00053 &      5 &    0.5 &     no \\
  321 &   0.00338 &   0.00357 &   0.00347 &   --0.00019 &   0.00040 &   0.00100 &     --5 &    0.2 &     no \\
  322 &   0.00290 &   0.00387 &   0.00339 &   --0.00097 &   0.00049 &   0.00086 &    --28 &    1.0 &     no \\
  324 &   0.00636 &   0.00353 &   0.00494 &    0.00283 &   0.00047 &   0.00061 &     57 &    3.7 &    yes \\
  325 &   0.00591 &   0.00731 &   0.00661 &   --0.00140 &   0.00057 &   0.00095 &    --21 &    1.3 &     no \\
  326 &   0.00674 &   0.00856 &   0.00765 &   --0.00183 &   0.00063 &   0.00168 &    --23 &    1.0 &     no \\
  332 &   0.00388 &   0.00590 &   0.00489 &   --0.00202 &   0.00045 &   0.00065 &    --41 &    2.6 &     no \\
  341 &   0.00181 &   0.00460 &   0.00321 &   --0.00279 &   0.00033 &   0.00066 &    --87 &    3.8 &    yes \\
  343 &   0.00335 &   0.00373 &   0.00354 &   --0.00038 &   0.00044 &   0.00061 &    --10 &    0.5 &     no \\
  344 &   0.00521 &   0.00592 &   0.00556 &   --0.00072 &   0.00081 &   0.00078 &    --12 &    0.6 &     no \\
  346 &   0.00189 &   0.00206 &   0.00197 &   --0.00017 &   0.00037 &   0.00043 &     --8 &    0.3 &     no \\
  359 &   0.00290 &   0.00276 &   0.00283 &    0.00014 &   0.00049 &   0.00062 &      5 &    0.2 &     no \\
  360 &   0.00208 &   0.00412 &   0.00310 &   --0.00204 &   0.00040 &   0.00087 &    --65 &    2.1 &     no \\
  363 &   0.00151 &   0.00331 &   0.00241 &   --0.00180 &   0.00035 &   0.00070 &    --74 &    2.3 &     no \\
  369 &   0.00825 &   0.01080 &   0.00954 &   --0.00257 &   0.00075 &   0.00103 &    --26 &    2.0 &     no \\
  371 &   0.01080 &   0.03380 &   0.02230 &   --0.02290 &   0.00082 &   0.00205 &   --102 &   10.4 &    yes \\
  372 &   0.00201 &   0.00154 &   0.00178 &    0.00047 &   0.00044 &   0.00040 &     26 &    0.8 &     no \\
  375 &   0.00486 &   0.00593 &   0.00540 &   --0.00107 &   0.00049 &   0.00109 &    --19 &    0.9 &     no \\
  380 &   0.00456 &   0.00494 &   0.00475 &   --0.00039 &   0.00058 &   0.00095 &     --8 &    0.3 &     no \\
  382 &   0.03720 &   0.07570 &   0.05640 &   --0.03850 &   0.00135 &   0.00233 &    --68 &   14.3 &    yes \\
  385 &   0.00062 &   0.00383 &   0.00222 &   --0.00321 &   0.00021 &   0.00110 &   --144 &    2.9 &     no \\
\hline\hline
\end{longtable}

\twocolumn

%% file: onlyonce.tex
\onecolumn

\captionsetup{width=\textwidth}

\begin{longtable}{rrr}
\caption{Sources, which have only beed detected once even though they were observed twice.}
\label{unmatched}
\\
\hline\hline
ID & \xmm\ coordinates & Obs. ID \\
\hline
\endfirsthead

\multicolumn{3}{c}%
{\tablename\ \thetable{} -- continued from previous page} \\
\hline
ID & \xmm\ coordinates & Obs. ID \\
\hline
\endhead

\hline \multicolumn{3}{l}{{\footnotesize Continued on next page}} \\
\endfoot

\hline
\endlastfoot

  5 & J004308.5+413022 & OBS1 \\
 10 & J004319.8+413640 & OBS1 \\
 11 & J004320.1+413619 & OBS1 \\
 12 & J004320.5+412616 & OBS1 \\
 15 & J004322.2+412347 & OBS1 \\
 21 & J004326.7+413542 & OBS1 \\
 22 & J004327.1+413554 & OBS2 \\
 26 & J004330.8+412100 & OBS2 \\
 28 & J004331.4+413605 & OBS2 \\
 29 & J004332.3+412058 & OBS1 \\
 32 & J004335.4+413127 & OBS1 \\
 34 & J004337.1+414048 & OBS2 \\
 35 & J004338.2+413736 & OBS1 \\
 44 & J004342.9+413811 & OBS1 \\
 45 & J004344.2+414736 & OBS3 \\
 46 & J004344.5+412410 & OBS2 \\
 49 & J004346.8+413839 & OBS1 \\
 55 & J004349.0+415204 & OBS3 \\
 56 & J004349.2+415533 & OBS3 \\
 59 & J004350.6+412729 & OBS1 \\
 62 & J004351.6+414239 & OBS3 \\
 65 & J004353.1+415459 & OBS3 \\
 66 & J004353.4+414711 & OBS3 \\
 70 & J004354.3+413102 & OBS1 \\
 71 & J004354.8+414043 & OBS1 \\
 78 & J004357.1+414821 & OBS3 \\
 81 & J004357.3+413842 & OBS1 \\
 82 & J004357.4+413449 & OBS2 \\
 88 & J004400.0+411855 & OBS2 \\
 90 & J004401.0+414646 & OBS4 \\
 91 & J004401.1+413946 & OBS1 \\
 92 & J004401.3+415636 & OBS4 \\
 95 & J004402.5+411711 & OBS1 \\
 97 & J004403.2+411805 & OBS1 \\
 98 & J004403.3+412741 & OBS1 \\
 99 & J004403.5+413413 & OBS1 \\
107 & J004405.0+415235 & OBS3 \\
109 & J004406.1+414925 & OBS4 \\
110 & J004406.4+415735 & OBS3 \\
111 & J004406.5+412823 & OBS1 \\
112 & J004406.5+413608 & OBS1 \\
115 & J004407.6+412501 & OBS2 \\
121 & J004409.9+413345 & OBS1 \\
123 & J004410.6+415014 & OBS3 \\
125 & J004411.4+412525 & OBS1 \\
126 & J004411.9+413216 & OBS1 \\
129 & J004412.3+413731 & OBS1 \\
132 & J004413.3+412728 & OBS1 \\
136 & J004414.8+415750 & OBS3 \\
137 & J004415.9+415942 & OBS4 \\
139 & J004416.4+413223 & OBS2 \\
143 & J004418.3+412517 & OBS1 \\
150 & J004422.3+413850 & OBS3 \\
153 & J004423.0+415536 & OBS3 \\
154 & J004423.2+414009 & OBS4 \\
155 & J004423.5+414916 & OBS3 \\
158 & J004424.8+413730 & OBS4 \\
159 & J004424.8+414807 & OBS3 \\
162 & J004425.8+413525 & OBS1 \\
164 & J004426.4+414342 & OBS1 \\
165 & J004426.4+415820 & OBS3 \\
172 & J004429.7+411743 & OBS1 \\
173 & J004429.7+415314 & OBS3 \\
174 & J004430.0+415500 & OBS3 \\
178 & J004431.1+414905 & OBS4 \\
181 & J004431.9+415629 & OBS3 \\
185 & J004432.1+415505 & OBS3 \\
188 & J004433.8+412153 & OBS2 \\
189 & J004433.8+415548 & OBS3 \\
190 & J004434.6+413144 & OBS1 \\
191 & J004434.8+415427 & OBS3 \\
192 & J004434.9+412513 & OBS1 \\
194 & J004435.2+412141 & OBS1 \\
195 & J004435.4+414534 & OBS4 \\
196 & J004435.6+412511 & OBS2 \\
200 & J004437.9+414355 & OBS1 \\
201 & J004438.1+412523 & OBS2 \\
208 & J004441.9+414026 & OBS3 \\
213 & J004443.4+412656 & OBS2 \\
215 & J004443.6+420158 & OBS4 \\
216 & J004444.6+412821 & OBS2 \\
220 & J004445.8+412641 & OBS1 \\
221 & J004445.8+412147 & OBS1 \\
224 & J004446.2+413609 & OBS1 \\
231 & J004448.7+412857 & OBS1 \\
234 & J004448.9+420026 & OBS4 \\
236 & J004449.4+414526 & OBS3 \\
238 & J004449.8+415307 & OBS3 \\
239 & J004449.9+415242 & OBS3 \\
240 & J004450.5+415421 & OBS3 \\
241 & J004451.0+415458 & OBS4 \\
248 & J004453.4+420214 & OBS3 \\
253 & J004455.7+415656 & OBS3 \\
256 & J004456.8+415352 & OBS3 \\
262 & J004500.1+413948 & OBS1 \\
263 & J004500.1+412517 & OBS1 \\
265 & J004500.3+420116 & OBS3 \\
266 & J004500.3+413128 & OBS1 \\
269 & J004501.2+412812 & OBS1 \\
271 & J004502.2+415408 & OBS3 \\
274 & J004502.7+412230 & OBS2 \\
275 & J004503.8+415819 & OBS4 \\
276 & J004504.2+414011 & OBS1 \\
278 & J004504.2+412809 & OBS1 \\
279 & J004504.8+412229 & OBS1 \\
280 & J004505.4+412504 & OBS1 \\
281 & J004506.7+415325 & OBS3 \\
282 & J004507.5+415357 & OBS3 \\
284 & J004508.4+413512 & OBS3 \\
286 & J004509.4+415002 & OBS3 \\
287 & J004509.6+415812 & OBS3 \\
288 & J004510.9+415901 & OBS4 \\
289 & J004510.9+413246 & OBS1 \\
291 & J004511.2+412228 & OBS1 \\
292 & J004511.2+415854 & OBS3 \\
295 & J004512.2+420019 & OBS3 \\
296 & J004512.7+412053 & OBS2 \\
297 & J004513.4+412714 & OBS1 \\
298 & J004513.5+420149 & OBS4 \\
301 & J004514.4+414150 & OBS4 \\
302 & J004514.5+420057 & OBS3 \\
307 & J004516.1+412302 & OBS1 \\
309 & J004517.9+415315 & OBS3 \\
310 & J004518.1+412441 & OBS1 \\
313 & J004520.4+412731 & OBS1 \\
315 & J004522.0+415841 & OBS3 \\
316 & J004522.4+412759 & OBS1 \\
319 & J004522.6+414613 & OBS4 \\
320 & J004522.9+413818 & OBS3 \\
327 & J004526.8+420016 & OBS3 \\
330 & J004527.7+415110 & OBS3 \\
334 & J004528.8+415015 & OBS3 \\
335 & J004529.0+414101 & OBS3 \\
337 & J004529.3+413525 & OBS2 \\
338 & J004529.4+415104 & OBS3 \\
342 & J004531.8+412759 & OBS1 \\
345 & J004532.9+415702 & OBS3 \\
347 & J004533.2+414329 & OBS4 \\
350 & J004534.8+415607 & OBS3 \\
351 & J004534.9+414951 & OBS3 \\
354 & J004535.8+413322 & OBS1 \\
355 & J004535.9+413901 & OBS3 \\
356 & J004536.2+414707 & OBS4 \\
357 & J004537.2+414658 & OBS3 \\
361 & J004538.6+414206 & OBS3 \\
364 & J004540.2+415225 & OBS3 \\
367 & J004542.3+414833 & OBS4 \\
368 & J004542.4+414712 & OBS3 \\
374 & J004545.9+415009 & OBS3 \\
376 & J004548.7+414803 & OBS3 \\
377 & J004552.1+415351 & OBS3 \\
378 & J004553.4+414513 & OBS3 \\
381 & J004555.8+414556 & OBS3 \\
383 & J004559.7+414827 & OBS4 \\
\hline\hline
\end{longtable}

\twocolumn